\documentclass[useAMS,usenatbib]{mnras}
\usepackage[T1]{fontenc}

\usepackage{graphicx}
\usepackage{subcaption}
\usepackage{float}
\usepackage{amsmath}
\usepackage{amssymb}
\usepackage{braket}
\usepackage[dvipsnames]{xcolor}

\usepackage{multirow}
\usepackage{hhline}

\usepackage{bm}
\newcommand{\eagle}{{\sc eagle}}
\newcommand{\tng}{{\sc tng}}

\newcommand{\nbodykit}{{\sc nbodykit}}
\newcommand{\ha}{H$\alpha$}

\usepackage{siunitx}

\newcommand{\massunit}{\unit{M_\odot h^{-1}}}
\newcommand{\sfrunit}{\unit{M_\odot yr^{-1}}}
\newcommand{\kunit}{\unit{h\ Mpc^{-1}}}
\newcommand{\pkunit}{\unit{h^{-3}\ Mpc^3}}

\newcommand{\lenunit}{\unit{Mpc\ h^{-1}}}

\newcommand{\mvir}{$M_{\mathrm{vir}}$}

\renewcommand{\lq}{{`}}
\renewcommand{\rq}{{'}}
\renewcommand{\lim}{{\sc lim}}
\newcommand{\hod}{{\sc hod}}
\newcommand{\sfr}{{\sc sfr}}
\newcommand{\ssfr}{s{\sc sfr}}

\newcommand{\illustris}{{\sc illustrisTNG}}
\newcommand{\agn}{{\sc agn}}
\newcommand{\nfw}{{\sc nfw}}
\newcommand{\h}{h}
\newcommand{\hprime}{h'}
\newcommand{\gj}{{g}}
\newcommand{\gl}{g'}
\newcommand{\subj}{_g}
\newcommand{\subl}{_{g'}}
\newcommand{\subh}{_h}
\newcommand{\subhprime}{_{h'}}

\newcommand{\sj}{s}
\newcommand{\sprime}{{s'}}

\newcommand{\uobserved}{\hat{v}_h^{(s)}(\bm{k})}

\newcommand{\uparent}{\hat{u}}
\newcommand{\uparenth}{\hat{u}_h}

\newcommand{\usatp}{\hat{u}_{\mathrm{sat}}}

\newcommand{\sfrd}{{\sc csfrd}}
\newcommand{\lf}{{\sc lf}}
\newcommand{\halofit}{{\sc halofit}}

\newcommand{\usatpprime}{\hat{u}_{\mathrm{sat},\hprime}}

\usepackage{widetext}

\usepackage{scalerel}
\usepackage{tikz}
\usetikzlibrary{svg.path}
\definecolor{orcidlogocol}{HTML}{A6CE39}
\tikzset{
  orcidlogo/.pic={
    \fill[orcidlogocol] svg{M256,128c0,70.7-57.3,128-128,128C57.3,256,0,198.7,0,128C0,57.3,57.3,0,128,0C198.7,0,256,57.3,256,128z};
    \fill[white] svg{M86.3,186.2H70.9V79.1h15.4v48.4V186.2z}
                 svg{M108.9,79.1h41.6c39.6,0,57,28.3,57,53.6c0,27.5-21.5,53.6-56.8,53.6h-41.8V79.1z M124.3,172.4h24.5c34.9,0,42.9-26.5,42.9-39.7c0-21.5-13.7-39.7-43.7-39.7h-23.7V172.4z}
                 svg{M88.7,56.8c0,5.5-4.5,10.1-10.1,10.1c-5.6,0-10.1-4.6-10.1-10.1c0-5.6,4.5-10.1,10.1-10.1C84.2,46.7,88.7,51.3,88.7,56.8z};
  }
}

\newcommand\orcidicon[1]{\href{https://orcid.org/#1}{\mbox{\scalerel*{
\begin{tikzpicture}[yscale=-1,transform shape]
\pic{orcidlogo};
\end{tikzpicture}
}{|}}}}

\usepackage{comment}
\usepackage{xr}
\usepackage{subfiles}

\usepackage[capitalise]{cleveref}
\crefformat{equation}{Eq. (#2#1#3)}
\crefrangeformat{equation}{Eqs. #3#1#4-#5#2#6}

\crefrangeformat{figure}{Figs. #3#1#4-#5#2#6}

\Crefformat{equation}{Equation (#2#1#3)}
\Crefname{equation}{Equation}{Equations}

\title[Power spectrum of galaxies: a \lim\ perspective ]{The power spectrum of galaxies from large to small scales: a line-intensity mapping perspective}
\author[R. L. Jun et al.]{Rui Lan Jun$^{1 \orcidicon{0009-0006-7907-1283}}$, Tom Theuns$^{2\, \orcidicon{0000-0002-3790-9520}}$, Kana Moriwaki$^{1,3 \, \orcidicon{0000-0003-3349-4070}}$, Sownak Bose$^{2\, \orcidicon{0000-0002-0974-5266}}$\\
$^{1}$ Department of Physics, Graduate School of Science, The University of Tokyo, 7-3-1 Hongo, Bunkyo, Tokyo 133-0033, Japan\\
$^{2}$ Institute for Computational Cosmology, Durham University, South Road, Durham DH1 3LE, UK\\
$^{3}$ Research Center for the Early Universe, Graduate School of Science, The University of Tokyo, 7-3-1 Hongo, Bunkyo, Tokyo 113-0033, Japan}

\date{Accepted XXX. Received YYY; in original form ZZZ}
\pubyear{2024}

\begin{document}

\maketitle

    \begin{abstract}

We present a model for the power spectrum of the density field of galaxies weighted by their star formation rate. This weighting is relevant in line-intensity mapping (\lim) when the observed line luminosity is strongly correlated with star formation, as is the case for the H$\alpha$ line. Our model reproduces the measured power spectrum in the \illustris\ simulation to within a few per cent across all scales, with fitting parameters that have clear physical interpretations. On scales of tens of megaparsecs, the model accounts for the weighted non-linear bias of galaxies as well as halo exclusion (2-halo term). On smaller scales, it incorporates the weighted distribution of satellite galaxies within haloes (1-halo term). The random sampling of satellite galaxies introduces a galaxy shot noise term to the power spectrum on small scales, and their confinement to haloes introduces a halo shot noise term on large scales. Omitting satellite galaxies from the analysis results in an underestimation of both the large-scale bias and the mean intensity by $\sim 30$ per cent each at $z \sim 1.5$. Assigning the intensity of satellites to the centre of their respective haloes affects the power spectrum on scales $k \gtrsim 0.3\ \kunit$. Our fitting function provides a well-motivated parametrisation that can be used to interpret data from upcoming \lim~surveys.

\end{abstract}


\section{Introduction}

The distribution of galaxies in the Universe forms a web-like structure, known as the cosmic web, which closely matches the patterns predicted by cosmological simulations that assume galaxies reside in dark matter haloes \cite[e.g.][]{Millennium_2005}. The presence of such haloes is further supported by gravitational lensing measurements (see, e.g., \citealt{Jaelani_2020} and references therein). In the simulations, the pattern emerges through the gravitational amplification of initially small density perturbations imprinted in the initial conditions.

A statistical description of these density fluctuations as a function of scale can be provided by the power spectrum (see, e.g., \citealt{Peebles80}).
On large scales, the power spectrum follows linear theory, enabling us to extract information about the primordial universe (e.g. \citealt{Guth81}; \citealt{Linde82}; \citealt{Meerburg_2019}). 
The power spectrum on smaller scales provides information about non-linear structure formation but may also hold valuable information about the cosmological model. For instance, structure on these smaller scales may be suppressed due to the free-streaming of massive neutrinos \citep[e.g.][]{Hu_1998} or due to deviations from cold dark matter models \citep[e.g.][]{Murgia_2017}. 

Line-intensity mapping (hereafter \lim) is an emerging observational technique with the potential to map the universe across a wide range of scales and redshifts (for more details on \lim\ and reviews of upcoming surveys, see, e.g., \citealt{Kovetz_2017}). Rather than targeting galaxies individually, \lim\ aims to map the spatial fluctuations in an emission line. By operating with lower angular and spectral resolution, \lim\ trades detailed knowledge of individual galaxies for more efficient mapping of larger volumes.
Unlike traditional galaxy surveys (e.g. {\sc euclid}; \citealt{Ballardini24}), which resolve individual galaxies above a threshold (e.g. flux), \lim\ captures all line emission at the target wavelength. Contributions from galaxies with weak emission lines are included in the map, even when their signals fall below the noise level.

While \lim\ surveys are not yet competitive in scale with current large galaxy surveys, near-future \lim\ surveys have the potential to complement galaxy surveys on smaller scales by measuring the contribution from faint galaxies that were previously excluded. \lim\ surveys such as {\sc copssII} \citep{COPSSII_2016} have already provided preliminary results on scales $k = 0.5 - 2$ \kunit\ (where $k$ is the comoving wavenumber), albeit with low signal-to-noise ratio. Current and upcoming \lim\ surveys such as {\sc tim} \citep{TIM_2020}, {\sc SPHEREx} \citep{SPHEREx_2018}, {\sc spt-slim} \citep{SPT-SLIM_2022}, {\sc comap} \citep{Cleary_2022}, {\sc hetdex} \citep{HETDEX_2021} and {\sc mmIME} \citep{mmIME_2020}
will further probe these non-linear scales and could provide valuable insights into cosmology, as well as galaxy formation and evolution. 
As an example, the use of the small-scale \lim\ power spectrum to probe neutrino properties is discussed by \cite{Dizgah_2022_neutrino}. 

Since galaxies are not detected individually in \lim, the power spectrum is determined by flux rather than number density, meaning galaxies contribute weighted by their flux rather than equally. 
A common assumption for several emission lines, such as \ha, is that the line luminosity (and therefore flux for galaxies at a given redshift) is proportional to the galaxy's star formation rate (hereafter \sfr; \citealt{Kennicutt98,DeLooze_2014}). 
Therefore, an accurate calculation of the \sfr-weighted power spectrum is a fundamental ingredient in modelling the flux-weighted power spectrum across different emission lines. 
In turn, this requires a model that connects the luminosities (\sfr s) of galaxies to their host haloes.

Models such as abundance matching \citep{Vale_2004} and halo occupation distribution (\hod) \citep{Berlind02} are commonly used to \lq paint\rq\ galaxies onto dark matter haloes in $N$-body simulations \citep[e.g.][]{Millennium_2005}. 
Applications of these methods to \lim\ are discussed by \citet{Wyithe_2010, Sun_2019, Wolz_2019}. For \lim, it is also common to assume some relation between luminosity and halo mass, often with no distinction made between central and satellite galaxies \citep[e.g.][]{Fonseca_2017, Gong_2017, Silva+2017}.
If satellites are accounted for, then their spatial distribution in a halo is often modelled using the profile of Navarro, Frenk \& White (\nfw, \citealt{nfw_1997}) or modifications of it (see, e.g., \citealt{Padmanabhan_2017, Wolz_2019, Schaan+21-multi}), but the suitability of this is unclear.

The halo model \citep{Cooray2002,Asgari2023} is also widely used, particularly for increased flexibility in exploring a range of parameters or cosmological models. This model approximates the matter distribution in $N$-body simulations, and is combined with models connecting galaxies to haloes, such as \hod, to interpret observations \citep[e.g.][]{Cacciato_2012,VanDenBosch13,Wolz_2019, Zacharegkas_2022}.
However, the standard model underestimates the strength of clustering in the non-linear transition regime between the 2-halo and 1-halo terms \citep{Mead_2015}.
While the nonlinearity of galaxy clustering within haloes is captured in the 1-halo term, additional non-linearities due to halo exclusion and non-linear halo clustering outside the scale of the size of haloes have also been found. Such effects need to be modelled accurately to understand the power spectrum in this intermediate regime, where linear theory does not apply.
Deviations from the linear bias model \citep[e.g.][]{Smith_2007,Hand17,Mead_2021} and effects of halo exclusion \citep[e.g.][]{Baldauf_2013, VanDenBosch13, Garcia_2019} have been actively studied in the context of galaxy surveys. 
\citet{MoradinezhadDizgah_precision_tests} discuss such effects in the context of CO and [\ion{C}{II}] \lim\ using halo models. 

The omission of non-linear effects typically affect scales close to the size of haloes, degrading the robustness in the so-called \lq full-shape\rq\ analysis of the power spectrum, where both large and small scales are used for constraining cosmological and astrophysical models. In addition, in some \lim\ surveys, such as SPHEREx, the spectral resolution is low, which means that the observations obtained are 2D projections in wavelength slices. The 2D power spectrum at a given wavenumber receives contributions from smaller scales in three dimensions \citep[e.g.][]{Kaiser91}, further motivating the importance of understanding the 3D power spectrum in the non-linear regime.

To investigate the effects of satellites and nonlinearity, one could use cosmological hydrodynamical simulations, which model galaxy properties directly by solving equations of gravity and hydrodynamics, simulating processes such as radiative cooling, star formation, and stellar and black hole feedback on top of $N$-body simulations (e.g., \citealt{Vogelsberger2014, Dubois2014, Schaye15}; see \citealt{Vogelsberger2020} for a review). 
The $N$-body framework naturally accounts for non-linear evolution, while the inclusion of baryonic physics provides a physically motivated connection between galaxy properties and the underlying dark matter distribution.
Although computationally expensive and thus limited in volume and number, hydrodynamical simulations remain a valuable tool for understanding the shape of the \sfr-weighted power spectrum.

This paper aims to provide a model that captures the \lq full-shape\rq\ power spectrum of \lim\ to extract maximal information from upcoming \lim\ surveys.
We place particular emphasis on accounting for non-linear effects, which can be explored well using 
the \illustris\ hydrodynamical simulations \citep[][also see \cref{sec:sfr} for a brief description of \illustris]{Nelson_2018,IllustrisTNG_Springel_2018,Pillepich_2018, IllustrisTNG_Marinacci_2018,IllustrisTNG_release}. 
We describe the measured power spectrum with a model that includes a bias of galaxies weighted by their \sfr\ relative to a matter power spectrum on large scales (a 2-halo term), and a model for the distribution of satellite galaxies in haloes on smaller scales (a 1-halo term). The theoretical description is an extension of the halo model for the matter power spectrum \citep{Cooray2002}. Our model accounts for scale-dependent shot noise, non-linear clustering and halo exclusion, motivated by \citet{Baldauf_2013}.

This paper is organised as follows: \cref{sec:weighted_ps} presents the theory behind the model for the power spectrum; \cref{sec:sfr} briefly describes the \illustris\ simulations, with \cref{sec:sat} quantifying the impact of satellites on the power spectrum. In \cref{sec:ps_components}, we compare the power spectrum model to that obtained from the simulations, and provide fits for the 2-halo and 1-halo terms.
Finally, \cref{sec:conclusion} summarises our findings.

In this paper, we consider H$\alpha$ emission from galaxies at $z\sim 1.5$ as an example, but most of our results will also apply to other emission lines, provided the line luminosity is proportional to the galaxy's \sfr. We focus on emission from star formation, and not from active galactic nuclei (\agn).
For simplicity, we ignore contamination from continuum emission and from galaxies at other redshifts whose emission lines redshift to the same observed wavelength as the target line \citep[line interlopers; see, e.g.,][]{Gong_2020}. We use snapshot data and compute the power spectra in real space for simplicity.
We compute the power spectrum using the {\sc python} package \nbodykit\ \citep{nbodykit}. See \cref{app:interpolation} for further details on the computation of the power spectrum and a discussion on how our results depend on the choice of interpolation scheme.

\ifSubfilesClassLoaded{%
  \bibliography{bibliography}%
}{}

\end{document}

\section{Weighted power spectrum: theory}\label{sec:weighted_ps}

\begin{figure*}
\centering
    \includegraphics[width=13cm]{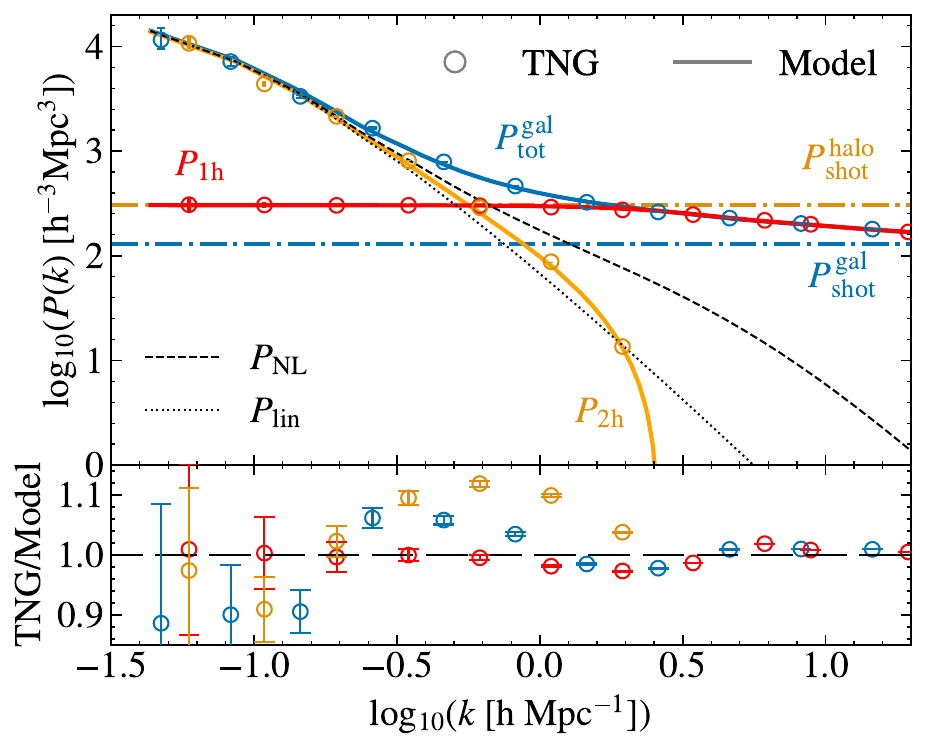}
\caption{\textbf{\textit{Upper panel}}: 
The galaxy power spectrum at $z=1.5$ ({\em blue}), which is the sum of a 2-halo term ({\em orange}) and 1-halo term ({\em red}), as in Eq.~\ref{eq:galaxy_ps_all}. The {\em open circles} are measured from the \illustris\ simulation, while {\em solid lines} are the model proposed in this paper.
The {\em orange} and {\em blue dash-dotted} lines show the Poisson shot noise in the model, computed using the luminosities of haloes and galaxies, respectively. The model's 1-halo term tends to the halo shot noise on large scales and the galaxy shot noise on small scales. The non-linear ({\em black dashed}) and linear ({\em black dotted}) matter power spectra are plotted for reference. For $-0.5 \lesssim \log (k\ [\kunit]) \lesssim 0$, neither the 1-halo term nor the 2-halo term dominates. \textbf{\textit{Lower panel}}: The ratio of the \tng\ power spectra to that of the model. The model proposed in this paper agrees with the \tng\ power spectrum to within a few per cent on all scales. On the largest scales, the \tng\ power spectrum is affected by sample variance, causing it to deviate from the fit. The fit is significantly better on scales $\log (k\ [\kunit]) \gtrsim -0.5$. 
In both panels, the error bars from \illustris\ are estimated using the number of modes within each $k$-bin, following $P(k)/\sqrt{N_{\mathrm{modes}}}$, where $N_{\mathrm{modes}} = V k_{\mathrm{mean}}^2 \Delta k / 2\pi^2$. Here, bins are constant in d$\log k = 0.25$.}
\label{fig:components_ps}
\end{figure*}

This section outlines the components of the weighted galaxy power spectrum, including the mathematical foundation for the model we propose. 
\citet{Baldauf_2013} demonstrated that the commonly assumed constant Poisson shot noise does not fully capture the stochasticity.
Instead there are contributions from the shot noise of galaxies and haloes, as well as from halo exclusion, where haloes are separated by a minimum distance due to their finite size. Additionally, the non-linear clustering of haloes outside the exclusion radius must also be taken into account. In this section, we revisit the results of \citet{Baldauf_2013} and extend them to the case where galaxies are weighted.

In \lim, galaxies contribute to the power spectrum with a weight, $W$, that is proportional to the line flux. 
While assigning equal weights is more common in galaxy survey analyses, alternative weighting schemes have also been proposed to enhance the information content of galaxy survey data \citep[e.g.,][]{Feldman_1994, Seljak_2009, Pearson_2016, Smith_2016}. Since weighted power spectra are not unique to \lim, we keep the weight $W$ general in this section.

Here, we provide an early glimpse of the final result we obtain in this section. The weighted 3D galaxy power spectrum is a sum of the 2-halo term and the 1-halo term:
\begin{align} \label{eq:galaxy_ps_all}
    P^\mathrm{gal}_{\mathrm{tot}}(k)
    &= \underbrace{P_{2h}(k, \uobserved)}_{\hbox{2-halo term}} 
    + \underbrace{U(k)^2(P_{\rm shot}^{\rm halo} - P^{\mathrm{gal}}_{\rm shot}) + P^{\mathrm{gal}}_{\rm shot}}_{\hbox{\rm 1-halo term}}\,.
\end{align}
Both $\uobserved$ and $U(k)$ depend on the distribution of weights within haloes. $k$ = $\left|\bm{k}\right|$ denotes the comoving wavenumber and is given in units of \kunit\ throughout this paper. Comoving units are assumed for all lengths and inverse lengths.
\Cref{fig:components_ps} shows the total galaxy power spectrum and both terms 
of Eq.~\ref{eq:galaxy_ps_all} measured from the hydrodynamical simulation \illustris\  (open circles) compared to the model (solid lines), which is described in more detail in \cref{sec:ps_components}. Both the simulation and the model are shown at redshift $z=1.5$.
The commonly used linear (dotted black) and non-linear (dashed black) power spectra are also shown in \cref{fig:components_ps} for reference, illustrating that these models do not provide good approximations for the 2-halo term on smaller scales.
The bottom panel of \cref{fig:components_ps} shows that our proposed model agrees with the \illustris\ power spectrum to within a few per cent on scales where sample variance does not dominate.

\begin{figure}
\includegraphics[width=\linewidth]{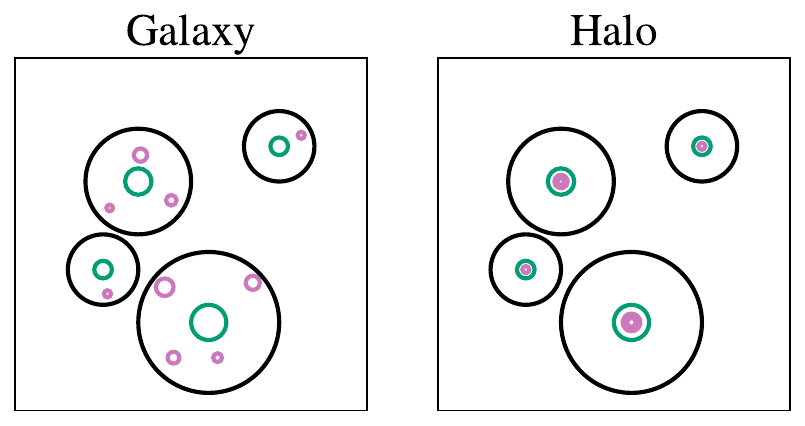}
\caption{Schematic comparing two different ways to assign weights to galaxies in a dark matter halo. The
\textit{green  circles} are each halo's central galaxy and the \textit{pink circles} denote satellites, with the radii of the circles being a measure of the weight assigned to the galaxy; the \textit{black circles} show the virial radii of the haloes. \textbf{\textit{Left panel}}: The resolved \lq galaxy\rq\ case, in which central and satellites are at different locations. \textbf{\textit{Right panel}}: The \lq halo\rq\ case: the weights of central and satellites are added to the position of the central galaxy. In \lim, the individual weights are the line fluxes.
}
\label{fig:distribution_schematic}
\end{figure}

In the following subsections, we derive \cref{eq:galaxy_ps_all} following closely the derivation by \citet{Baldauf_2013}, but for the case where galaxies and haloes are weighted. Figure \ref{fig:distribution_schematic} illustrates schematically some of the steps involved. 
We first consider the power spectrum of a weighted distribution in general in \cref{sec:cf_ps}, and discuss the concept of shot noise in \cref{sec:shot_noise}.
We start by considering haloes only (where the weights of satellite and central galaxies are summed and placed at the position of the central galaxy, illustrated in the right panel of \cref{fig:distribution_schematic}). We describe the 2-halo term and how this is modified by halo exclusion in \cref{sec:halo_ps}. In \cref{sec:gal_tracer}, we additionally consider the satellite distribution (as in the left panel of \cref{fig:distribution_schematic}), to describe all the terms in \cref{eq:galaxy_ps_all}.
We refer to the power spectrum computed for the case of the left panel of \cref{fig:distribution_schematic} as the `galaxy power spectrum', and that for the right panel as the `halo power spectrum'.

In this paper, we use a Fourier convention where the Fourier transform pair $f$ and $\hat f$ are related by
\begin{align}
\hat f({\bm k}) &= \frac{1}{V}\,\int d{\bm r}\,\exp(-i{\bm k}\cdot{\bm r})\,f({\bm r})\nonumber\\
f({\bm r}) &= \frac{V}{(2\pi)^3}\int d{\bm k}\,\exp(i{\bm k}\cdot{\bm r})\,\hat f({\bm k})\,.
\end{align}
$f$ and its Fourier transform $\hat f$ have the same dimension. The power spectrum is $P(k) = V\langle |\hat f(k)|^2\rangle\,$, where the angular brackets $\langle\cdot\rangle$ denote an ensemble average. In these expressions, $V=L^3$, where $L$ is a comoving length on which $f$ is periodic. In the case of simulations, $L$ is the linear extent of the simulation volume.

\subsection{Power spectrum of a weighted distribution}\label{sec:cf_ps}
The weight distribution of discrete, point-like objects (e.g. galaxies or haloes) can be written as
\begin{align}\label{eq:w_r}
    w(\bm{r}) = \sum_i  W_i  \delta^{(\rm{D})} (\bm{r}-\bm{r}_i ),
\end{align}
where $W_i $ and $\bm{r}_i $ are the weights and positions of each object, and $\delta^{\text{(D)}}$ is the Dirac delta function. 
It is common to define the overdensity as
\begin{align}   
    \delta_w(\bm{r}) = \frac{w(\bm{r})}{\overline{w}} - 1,
\end{align}
where $\overline{w} = \sum_i {W_i}/V$.

The Fourier transform of \cref{eq:w_r} is
\begin{align}\label{eq:w_k}
	\hat{w}(\bm{k}) &= \frac{1}{V} \int d\bm{r}\,\,w(\bm{r}) \exp(-i\bm{k}\cdot \bm{r}) \nonumber\\
    &= \frac{1}{V}\sum_i  W_i  \exp (-i\bm{k}\cdot \bm{r}_i ),
\end{align}
giving the Fourier transform of $\delta_w(\bm{r})$ as
\begin{align}
    \hat{\delta}_w(\bm{k}) = \frac{\hat{w}(\bm{k})}{\overline{w}} - \hat{\delta}^{(D)}(\bm{k}).
\end{align}
The delta function in Fourier space, $\hat{\delta}^{(D)}(\bm{k})$, is non-zero only for $k = 0$, where the power spectrum is 0. We will ignore this term in what follows.
The power spectrum of $\delta_w$ is
\begin{align}\label{eq:intensity_ps}
P_{\mathrm{tot}}(k) &= V \langle \hat{\delta}_w(\bm{k})\hat{\delta}_w(\bm{-k}) \rangle.
\end{align}

For \lim, $W_i$ would be the flux $F_i$ of an object at position $\bm{r}_i $. If all the objects are at the same redshift, then flux weighting is equivalent to luminosity weighting, except for the differences in the overall scaling factor. As we only consider one redshift at a time, we will consider luminosity in the rest of this paper. 
In \lim, $\overline{w}$ is given by the \emph{specific} mean intensity, 
\begin{equation}\label{eq:mean_intensity_discrete}
\bar{I} = \frac{\sum_i {F_i }}{{d\Omega}d\nu_{\mathrm{obs}}} = \frac{\sum_i\,L_i/(4\pi D_L^2)}{d\Omega d\nu_{\mathrm{obs}}},
\end{equation}
where $F_i $ is the flux of each object, $d\Omega$ is the solid angle of the survey, and $d\nu_{\mathrm{obs}}$ is the frequency corresponding to the depth considered. $L_i$ is the luminosity of each object and $D_L$ is the luminosity distance.

It is common in \lim\ to not divide by the mean intensity when computing the power spectrum, since the mean intensity may be difficult to measure. In this case, the power spectrum can be written as 
\begin{align}
    P_{\mathrm{tot},w}(k) = V\langle \hat{w}(\bm{k})\hat{w}(\bm{-k}) \rangle.
\end{align} In this paper, we use the subscript $w$ (or $I$ for intensity) to indicate the case of a field that is {\em not} normalised by its mean. 
Using \cref{eq:w_r}, this power spectrum can be written as the sum of all the contributions:
\begin{align} \label{eq:tot_ps_discrete}
P_{\mathrm{tot},w}(k) 
 &= \frac{1}{V} \braket{\sum_{i,j}  W_i  W_j  \exp[-i\bm{k}\cdot(\bm{r}_i -\bm{r}_j )] },
\end{align}
where $i$ and $j$ run over all objects.
\footnote{Although \lim\ observes voxel intensities, the power spectrum analysis considers $W_i$ as the flux of the underlying sources (e.g., galaxies) rather than the voxel intensity. \cref{sec:shot_noise} explains that shot noise arises from stochastic sampling, which does not apply to voxels.
The difference between the power spectra of galaxies and voxels is visible only on scales close to the voxel size.}

The total power spectrum can be separated into contributions from pairs between distinct objects and \lq self-pairs\rq:
\begin{align}\label{eq:ps_2terms}
P_{\mathrm{tot},w}(k) 
 &= \frac{1}{V} \braket{\sum_{i \neq j}  W_i  W_j  \exp[-i\bm{k}\cdot(\bm{r}_i -\bm{r}_j )] } \nonumber \\
    &+ \frac{1}{V} \braket{\sum_{i = j} W_i  W_j  \exp[-i\bm{k}\cdot(\bm{r}_i -\bm{r}_j )]}.
\end{align}
In the simplest case, the first and second terms describe clustering and shot noise, respectively.

\subsection{Self-pairs term of the power spectrum} \label{sec:shot_noise}
The self-pair term corresponds to the shot noise in the case of a Poisson-sampled distribution, and can be simplified to
\begin{align}\label{eq:shotnoise_discrete_weight}
P_{\mathrm{shot},w} &= \frac{1}{V} \sum_{i}  W_i ^2 = V \overline{w}^2 \frac{\sum_{i}  W_i ^2}{(\sum_i  W_i )^2}.
\end{align}
For a standard galaxy survey, all selected galaxies are weighted equally, such that $W_i  = 1$ for all objects, and the shot noise tends to the inverse number density of galaxies, $1/\bar{n}$.
The shot noise in the weighted case can be rewritten using the number density, $\bar{n}$, as follows:
\begin{align}
     P_{\mathrm{shot,} w} &= \bar{w}^2 V \frac{(\sum_i {W_i ^2})/N}{(\sum_i {W_i })^2/N}. \nonumber \\
     &= \bar{w}^2V\frac{\langle W^2 \rangle}{N\langle W \rangle ^2}\nonumber \\
     %&= V\frac{Var(L) + \langle L \rangle^2}{N_{\mathrm{gal}}\langle L \rangle ^2} \nonumber\\
     &= \frac{\bar{w}^2}{\bar{n}}\left(\frac{{\rm Var}(W) }{\langle W \rangle ^2} + 1\right).\label{eq:shotnoise_var}
\end{align}
We see that this depends on the variance of the weights (i.e. luminosities in \lim).
In standard galaxy surveys, the shot noise decreases when the number density of galaxies measured increases. In \lim, the shot noise additionally depends on the variance of the luminosities. When the power spectrum is not multiplied by the mean weight, the shot noise depends on the ratio of the variance of the weights over the square of their mean.

Shot noise refers to the contribution to the power spectrum from fluctuations inherent in the stochastic sampling of discrete objects. % from a probability distribution.
On the other hand, the contribution from distinct pairs (the first term of Eq.~\ref{eq:ps_2terms}) describes how objects are distributed relative to each other beyond what would be expected if they were randomly distributed.
If objects (e.g. galaxies) are independently sampled from an underlying probability distribution, 
then the only difference between the galaxy power spectrum and that of the underlying distribution is the shot noise. However, additional constraints, such as halo exclusion (\cref{sec:exclusion}) or the restriction of galaxies to reside within haloes (\cref{sec:gal_tracer}), mean that the objects (haloes or galaxies) are not sampled independently. In such cases, the resulting power spectrum differs not only due to shot noise but also due to additional stochastic effects introduced by these constraints.

\subsection{Halo power spectrum}\label{sec:halo_ps}
In this subsection, we consider the power spectrum when using haloes as tracers of the matter distribution. 
In this paper, we take the halo power spectrum to be the case where the signal, i.e. the weight, is solely at halo centres (right panel of \cref{fig:distribution_schematic}). This allows us to discuss the effects of non-linear bias and halo exclusion separately from how galaxies are distributed within individual haloes. 

The halo power spectrum is the sum of a term due to the clustering of pairs of haloes
(the 2-halo term, $P^{\mathrm{halo}}_{2h}$), and a shot noise term ($P^{\mathrm{halo}}_{\mathrm{shot}}$),
\begin{align} \label{eq:halo_ps}
    P^{\mathrm{halo}}_{\mathrm{tot}} = P^{\mathrm{halo}}_{2h} + P^{\mathrm{halo}}_{\mathrm{shot}}.
\end{align}
This corresponds to the two terms in \cref{eq:ps_2terms}.

\subsubsection{Non-linear halo bias}\label{sec:non-linear}

Haloes are biased tracers of the matter density, and their power spectrum can be written as
\begin{align}
    P^{\mathrm{halo}}_{\mathrm{tot}} = b_1^2 P_\mathrm{m} + P^{\mathrm{halo}}_{\mathrm{shot}}\,,
\end{align}
where $b_1$ is the bias factor. The matter power spectrum,  $P_\mathrm{m}$, can be measured from $N$-body simulations that evolve an initially linear density field into the non-linear regime. Several fitting functions for $P_\mathrm{m}$ exist \citep[e.g. \halofit;][]{Takahashi_2012}, but there is no analytical form that works on all scales. The non-linear power spectrum shown in \cref{fig:components_ps} is given by the \halofit\ function.

The matter power spectrum deviates from linear theory on small scales and at low redshifts, but the linear power spectrum is nevertheless useful since it can be described analytically.
Therefore it is common to define halo bias with respect to the linear power spectrum $P_{\rm lin}$ (dotted line in \cref{fig:components_ps}):
\begin{align}
    P^{\mathrm{halo}}_{\mathrm{tot}} = b_2^2 P_{\mathrm{lin}} + P^{\mathrm{halo}}_{\mathrm{shot}}.
\end{align}
In this case, the nonlinearities in the halo bias and in the matter power spectrum are both included in $b_2$. \citet{Mo_1996} showed that the halo bias is scale-independent on sufficiently large scales. However, on scales comparable to the radii of haloes, $b_2$ becomes scale-dependent \citep{Sheth_1999, Jose_2016}. One reason is that the matter density field itself becomes non-linear. This scale dependence of the bias $b_2$ has been found to be more significant at higher redshifts and weakens at lower redshifts \citep{Sheth_1999}.

\subsubsection{Halo exclusion} \label{sec:exclusion}

In addition to nonlinearity, the standard halo model also fails to take into account the fact that haloes cannot be within a certain distance of each other (halo exclusion). In the context of galaxy surveys, corrections to the power spectrum due to halo exclusion have been discussed by, e.g., \citet{Casas-Miranda_2002,Smith_2007,Baldauf_2013}.

The effect of halo exclusion can be understood more intuitively by looking at the correlation function, the Fourier transform of the power spectrum.
For the weight distribution, the correlation function is defined as
\begin{align}\label{eq:cf_delta}
    \xi(r) &= \langle \delta_w(\bm{r}_1)\delta_w(\bm{r}_2) \rangle,
\end{align}
where $r = \lvert \bm{r_1} - \bm{r_2}  \rvert$. Note that the correlation function here excludes self-pairs as we focus on the 2-halo term.

For the simple case where all haloes have the same radius $R$, the minimum distance between them is $D = 2R$, and the correlation function is $\xi(r) = -1$ for $r < D$. See \cref{app:halo_exclusion} for further explanation of the single distance case. For the more realistic case, where there is a distribution of exclusion distances, the correlation function is given by
\begin{align}\label{eq:cf_multiple_rexc}
	\xi(r) = F(r)(1 + \xi'(r)) - 1\,,
\end{align}  
where $f(r)$ represents the probability of finding a pair of haloes with separation less than $r$, and $\xi'(r)$ can be thought of as the \emph{hypothetical} correlation function in the absence of halo exclusion. We explore in more detail how $f(r)$ is related to the distribution of halo radii in \cref{sec:2halo_fit}.
While in the case of a single exclusion distance, the function $\xi'(r)$ only needs to be defined above the exclusion distance, in the case where there is more than a single exclusion distance, $\xi'(r)$ needs to be defined down to at least the smallest exclusion distance.
Note that although this means that $\xi'(r)$ is defined below the scale of the size of larger haloes, this does not mean the halo profile is being probed. The small-scale contribution is due to smaller haloes being able to reside closer to other haloes.

Fourier transforming the correlation function given by \cref{eq:cf_multiple_rexc}, we obtain the 2-halo term  of the power spectrum as
\begin{align} \label{eq:ps_exclusion_distribution}
	P_{2h}(k) &= 4\pi\int_0^\infty [F(r)(1 + \xi'(r)) - 1] \frac{\sin(kr)}{kr} r^2 {\rm d}r.
\end{align}

The general form for the total halo power spectrum accounting for halo exclusion is then
\begin{align} \label{eq:exclusion_summary}
	P^{\mathrm{halo}}_{\mathrm{tot}} &= 4\pi\int_0^\infty [F(r)(1 + \xi'(r)) - 1] \frac{\sin(kr)}{kr} r^2 {\rm d}r  + P^{\mathrm{halo}}_{\mathrm{shot}},
\end{align}
where $F(r)$ describes the probability distribution of finding a pair of haloes whose separation is larger than $r$ in the absence of clustering.
The non-linear bias discussed in \cref{sec:non-linear} should be included in the $\xi'(r)$ function, while the distribution of exclusion distances is described by $F(r)$.

In \cref{sec:shot_noise}, we introduced shot noise as the contribution to the power spectrum arising from statistical fluctuations due to sampling from a probability distribution. 
As in \citet{Baldauf_2013}, we use the term stochasticity to refer more generally to statistical fluctuations due to random processes, contributing to the non-deterministic relation between the distribution of the tracers and the underlying probability distribution.
Due to halo exclusion, haloes are not independently sampled from the probability distribution, as they must satisfy the additional condition of avoiding each other.
Therefore, halo exclusion introduces an additional contribution to the stochasticity, in addition to the Poisson shot noise.

\subsection{Galaxy power spectrum}\label{sec:gal_tracer}
Our formulation in the previous subsection has only considered haloes, but haloes are not observable. Often, tracers such as galaxies, which are hosted by haloes, are used. 

\Cref{eq:tot_ps_discrete} can be decomposed into the contributions from pairs of galaxies in different haloes and pairs in the same halo as follows:
\begin{align}
    P^{\mathrm{gal}}_{\text{tot},w}(k) &= P^{\mathrm{gal}}_{2h,w}(k) + P^{\mathrm{gal}}_{1h,w}(k)  \nonumber\\
	&= \frac{1}{V} \braket{\sum_{\substack{\h,h' \\ h \neq h'}} \sum_{\gj\in \h} \sum_{\gl \in h'} W\subj W\subl \exp[-i\bm{k}\cdot(\bm{r}\subj -\bm{r}\subl)]} \nonumber\\
    &+ \frac{1}{V} \braket{\sum_{\h} \sum_{\gj\in \h} \sum_{\gl \in \h} W\subj W\subl \exp[-i\bm{k}\cdot(\bm{r}\subj -\bm{r}\subl)]},
    \label{eq:1and2halo}
\end{align}
where $\h$ and $h'$ are indices running over all the haloes, while $\gj$ and $\gl$ run over all the galaxies in haloes. This decomposition is different from \cref{eq:ps_2terms}, where we decomposed the total power spectrum into contributions from pairs between distinct objects and self-pairs.
In this paper, we take the first term of \cref{eq:1and2halo} to be the 2-halo term and the second term to be the 1-halo term. This means the 1-halo term here includes the galaxy shot noise (self-pairs of galaxies).

To relate the Fourier transform of a density distribution, given by \cref{eq:w_k}, to the distribution of galaxies within a halo, we write the transform as
\begin{align}
     \hat{w}(\bm{k}) &= \frac{1}{V}  \exp(-i\bm{k}\cdot\bm{r}_h)\sum_i  W_i  \exp [-i\bm{k}\cdot (\bm{r}_i - \bm{r}_h) ]\nonumber\\
     &\equiv \exp(-i\bm{k}\cdot\bm{r}_h)\,\hat{w}_h(\bm{k}),
\end{align}
     where $\bm{r}_h$ denotes the halo centre, and we define $\hat{w}_h(\bm{k})$ as the Fourier transform of the distribution of galaxies within halo $h$:
\begin{align}
\hat{w}_h(\bm{k}) &= \frac{1}{V}\sum_i  W_i  \exp [-i\bm{k}\cdot (\bm{r}_i - \bm{r}_h) ].
\end{align}
We additionally define 
\begin{align}\label{eq:u_k_discrete}
    \hat{v}_h^{(s)}(\bm{k})  &=  \frac{\sum_{\gj \in h} W\subj  \exp [-i\bm{k}\cdot(\bm{r}\subj -\bm{r}_h)]}{\sum_{\gj \in h}W\subj}\,,
\end{align}
which is the normalised Fourier transform of the weighted distribution of sampled galaxies. We use the superscript `$(s)$' (for \lq sampled\rq) to distinguish between the Fourier transform of a discrete, sampled distribution of galaxies, and the Fourier transform $\uparent (k)$ of the underlying function from which the galaxies are sampled. Their power spectra differ by a shot noise term.

\subsubsection{2-halo term} \label{sec:2halo_sec2}
The 2-halo term, $P_{2h}(k)$, in the galaxy power spectrum is largely the same as in the halo power spectrum, except that it is influenced by $\uobserved$ on scales close to the size of haloes, where the variations in the distances between galaxies in a pair of haloes become noticeable. The 2-halo term is shown by the orange line in \cref{fig:components_ps}.

The expression for $\uobserved$ from \cref{eq:u_k_discrete} can be substituted into the first term in \cref{eq:1and2halo} to give
\begin{align}\label{eq:2halo}
    & P^{\mathrm{gal}}_{2h,w}(k) \nonumber\\
	&= \frac{1}{V} \braket{\sum_{\substack{\h,h' \\ \h\neq h'}} \sum_{\gj\in \h} \sum_{\gl \in h'} W\subj W\subl \exp[-i\bm{k}\cdot(\bm{r}\subj -\bm{r}\subl)]} \nonumber\\
     &= \frac{1}{V}  \braket{\sum_{\substack{\h,h' \\ \h\neq h'}} \sum_{\gj\in \h} \sum_{\gl \in h'} W\subj  W\subl \uobserved \hat{v}_{h'}^{(s)}(\bm{k}) \exp[-i\bm{k}\cdot(\bm{r}\subh -\bm{r}\subhprime )]} \nonumber\\
     &= \frac{1}{V}  \braket{\sum_{\substack{\h,h' \\ \h\neq h'}} W_{\h} W_{h'} \uobserved \hat{v}_{h'}^{(s)}(\bm{k})\exp[-i\bm{k}\cdot(\bm{r}\subh -\bm{r}\subhprime )]},
\end{align}
where $W_{\h} = \sum_{\gj\in \h} W\subj $ is the combined weight of all galaxies $g$ in halo $h$.
On large scales, $\uobserved \to 1$, and the 2-halo term tends to the halo power spectrum, $P^{\mathrm{gal}}_{2h,w}(k)\to P^{\rm halo}_{\mathrm{tot},w}(k)$. 

The positions of the weights $W$ are affected by halo exclusion and non-linear bias, as discussed in \cref{sec:non-linear} and \cref{sec:exclusion}, causing the yellow line in \cref{fig:components_ps} to lie in-between the linear and non-linear power spectra. It is also affected by how satellites are distributed inside their host haloes. This is described by the 1-halo term, which we turn to next.

\subsubsection{1-halo term, including scale-dependent shot noise} \label{sec:1halo_sec2}

The power spectrum of galaxies within the same halo (1-halo term, $P_{1h}(k)$) can be written in terms of the observed weight distribution of halos, $\uobserved$ (Eq.~\ref{eq:u_k_discrete}), as 
\begin{align}\label{eq:1halo_discrete_u}
	P_{1h,w}(k) = &\frac{1}{V} \braket{\sum_{\h} \sum_{\gj\in \h} \sum_{\gl \in \h} W\subj W\subl \, \exp[-i\bm{k}\cdot(\bm{r}\subj -\bm{r}\subl)]} \nonumber\\
        =&\frac{1}{V}  \braket{\sum_{\h} \sum_{\gj\in \h} \sum_{\gl \in \h} W\subj W\subl \, \uobserved \hat{v}_{h}^{(s)}(-\bm{k}) } \nonumber\\
         = &\frac{1}{V}  \Braket{\sum_{\h} \left(\sum_{\gj\in \h}W\subj  \right) \left(\sum_{\gl \in \h}W\subl \right) \,  \lvert \uobserved\rvert^2} \nonumber\\
         = &\frac{1}{V}  \Braket{\sum_{\h} W_{\h}^2\,  \lvert \uobserved\rvert^2}.
\end{align}
$P_{1h,w}(k)$ tends to the halo shot noise, Eq.~(\ref{eq:shotnoise_discrete_weight}),  when $k\gg 1/R$, with $R$ the virial radius of the largest halo, since 
$\uobserved\to 1$.

Galaxies sample the halo profile discretely, generating a shot noise term in $\lvert \uobserved\rvert^2$. This shot noise term corresponds to the part of the double sum with $g=g'$ in the above expression, with the $g\neq g'$ term describing the underlying halo profile. We describe the latter by defining 
\begin{align}\label{eq:uparent}
    |\uparenth(k)|^2 
    &= \frac{\Braket{\sum_{\gj \in \h} \sum_{\gl \neq \gj \in \h} W\subj W\subl \, \exp[-i\bm{k}\cdot(\bm{r}\subj -\bm{r}\subl)]}}{\sum_{\gj \in \h}\sum_{\gl \neq \gj \in \h} W\subj W\subl}.
\end{align}

The 1-halo term can also be separated into self-pairs and distinct pairs, yielding (see \cref{app:halo_profile} for details)
\begin{align} \label{eq:galaxy_ps_1halo}
    P^{\mathrm{gal}}_{1h}(k) 
    &= U(k)^2(P_{\rm shot}^{\rm halo} - P^{\mathrm{gal}}_{\rm shot}) \nonumber \\
    &+ P^{\mathrm{gal}}_{\rm shot},
\end{align}
where $U(k)^2$ is defined as
\begin{align}\label{eq:big_uk2}
    U(k)^2 = \frac{\sum_{\h} (W_{\h}^2 - \sum_{\gj \in \h} W\subj ^2) |\uparenth (k)|^2} {\sum_{\h} (W_{\h}^2 - \sum_{\gj \in \h} W\subj ^2)}.
\end{align}
This gives us the second term of \cref{eq:galaxy_ps_all} (the red line in \cref{fig:components_ps}).
Since each $|\uparenth (k)|^2 \to 1$ as $k \to 0$, and $|\uparenth (k)|^2 \to 0$ as $k \to \infty$ (given that $u_h (r)$ are integrable functions), $U(k)$ follows the same trends.
This means that the 1-halo term, $P^{\mathrm{gal}}_{1h}(k)$, tends to the halo shot noise (orange dash-dotted line) on large scales and the galaxy shot noise (blue dash-dotted line) on small scales. Similar to halo exclusion, it would be possible to account for galaxy exclusion, but since sizes of galaxies are small we will not do so here.

If all galaxies have the same weight, then $\uparenth(k)$ (defined in Eq.~\ref{eq:uparent}) corresponds to the Fourier transform of the density profile of the halo. In \cref{app:profile_sampled_from}, we discuss the physical interpretation of $\uparenth(k)$ when the weights are not the same.

The location of the central galaxy is not sampled from a density distribution -- it is fixed at the \lq centre\rq -- hence central and satellites should be treated differently. In \cref{app:cent_sat}, we show the formalism for the case where $\usatp (k)$ describes the distribution of satellite galaxies only.

\ifSubfilesClassLoaded{%
  \bibliography{bibliography}%
}{}

\end{document}

\section{The star formation rate in IllustrisTNG galaxies}\label{sec:sfr}

In the previous section, we provided the equations for the power spectrum of galaxies when they are weighted with some weight, $W$. In the case of \lim, 
$W$ would be the flux (or equivalently luminosity if all objects were at the same redshift) of the emission line, which for many emission lines is related to the
galaxy's \sfr. Hydrodynamical simulations provide physically motivated values of the \sfr. In this section, we briefly introduce the \illustris\ simulation that we analyse, and examine the \sfr s it predicts. In particular, we show that the simulation yields an H$\alpha$ luminosity function (\lf) that agrees well with observational data. We also examine how much central and satellite galaxies contribute to the total \sfr\ in individual haloes.

\subsection{Details of IllustrisTNG}

In this work, we primarily use the TNG300-1 (hereafter \tng) output of the \illustris\ hydrodynamical simulation \citep{Nelson_2018,IllustrisTNG_Pillepich_2018,IllustrisTNG_Marinacci_2018,IllustrisTNG_Naiman_2018,IllustrisTNG_Springel_2018}. \tng\ is performed in a cubic volume with comoving side length of 205 \lenunit\ at a dark matter resolution of $m_{\mathrm{DM}} \sim 3.98 \times$ $10^7$ \massunit, and gas resolution of $m_{\mathrm{gas}} \sim 7.44 \times$ $10^6$ \massunit. \tng\ uses the {\sc arepo} moving-mesh code \citep{Springel_2010} to solve for gravity and magneto-hydrodynamics. The simulation includes detailed recipes for the subgrid aspects of galaxy formation: gas cooling and photo-heating, star formation, stellar evolution, black hole seeding and growth, and the feedback from stars and accreting black holes (see \citealt{Weinberger_2017} and \citealt{Pillepich_2018} for details). Cosmological parameters are taken from {\sc Planck15} \citep{Planck15}. 

Dark matter haloes are identified from the dark matter particle distribution using the Friends-of-Friends ({\sc fof}) algorithm with the usual value of the linking length ($b=0.2$) \citep{FOF}. Substructures within {\sc fof} haloes are identified by the {\sc subfind} algorithm \citep{Subfind}, which works as follows: the local density for each particle within a halo is estimated, and density peaks are identified. Particles are assigned to density peaks in order of decreasing density. An unbinding procedure is applied, where unbound particles are removed to identify self-bound substructures. If the number of particles remaining after the unbinding procedure is greater than a predefined threshold number (20), they are labelled as constituting a \lq subhalo\rq. An {\sc fof} halo may contain many subhaloes, each of which contains either zero or one galaxy, consisting of stars, gas and black holes. The subhalo at the minimum of the gravitational potential of the {\sc fof} halo is called the central subhalo (below we give in brackets the name of the variable in the \tng\ database\footnote{\href{https://www.illustris-project.org/data/}{https://www.illustris-project.org/data}}; for the central subhalo this is \texttt{GroupFirstSub}), and all remaining subhaloes are satellite subhaloes, regardless of whether or not they are within the virial radius of the main halo. 
A central subhalo can host a central galaxy and a satellite subhalo can host a satellite galaxy. Categorising galaxies as central or satellite is not consistent across simulations; it is also challenging in observations, particularly as the dark matter halo is not directly observed. 
Nevertheless, the definition used in \tng\ offers a useful guide for the approximate effect of satellite galaxies. 

We use the virial mass (\mvir, \texttt{Group\_M\_TopHat200}) for the mass of the halo, defined such that the mean density within a sphere that encloses this mass is $\Delta_c$ times the critical density, where the dimensionless number $\Delta_c$ follows from the solution of the collapse of a spherical top-hat perturbation. The fitting formula provided by \citet{Bryan_1998} is used to evaluate $\Delta_c$.
We define the \sfr\ of a galaxy as the sum of the instantaneous \sfr s of all gas cells bound to its host subhalo (\texttt{SubhaloSFR}), and for a halo, as the total \sfr\ of gas cells in the {\sc fof} halo (\texttt{GroupSFR}).

We do not apply any aperture correction and do not apply any cuts in \sfr, since all \sfr\ should be detected in \lim. We show in \cref{sec:sat_contribution}, that the contribution to the signal from low halo masses (which may not be well-resolved) is subdominant. However, the simulations are not yet fully converged in resolution \citep{Pillepich_2018}, and we will see in \cref{sec:sec5_shot_noise} that quantities such as the shot noise are sensitive to numerical resolution.

Some substructures identified in the simulation may not qualify as separate galaxies but instead represent fragments or clumps within already formed galaxies (this is signalled by the value of the \texttt{SubhaloFlag}), making their exclusion more appropriate for certain analyses \citep{IllustrisTNG_release}.
However, \lim\ includes the contribution from all sources of \sfr, therefore we also include structures with \texttt{SubhaloFlag}=0. Excluding such subhaloes at $z=1.5$ decreases the \sfr\ density by 4 per cent and the amplitude of the \lim\ power spectrum by 5 per cent.

We focus mostly on $z \sim 1.5$ with H$\alpha$ \lim\ surveys such as 
{\sc SPHEREx} \citep{SPHEREx_2018} in mind. This redshift will also be observed
in CO \citep[by {\sc spt-slim};][]{SPT-SLIM_2022}, and [\ion{C}{II}] \citep[by {\sc tim};][]{TIM_2020}. We briefly show results for other redshifts as well for completeness.

The results we obtain in this study could be dependent on the galaxy formation model adopted in the hydrodynamical simulation. Therefore we also 
compare to the \eagle\ RefL100N1504 simulation \citep{Schaye15, Crain15}, which has a comoving side length of 67.77 \lenunit. The identification of \eagle\ galaxies is also based on {\sc subfind}, and we use the database described by \cite{McAlpine16}.

\subsection{SFR-luminosity relation} \label{sec:sfr-L}

\begin{figure}
    \centering
    \includegraphics[width=\linewidth]{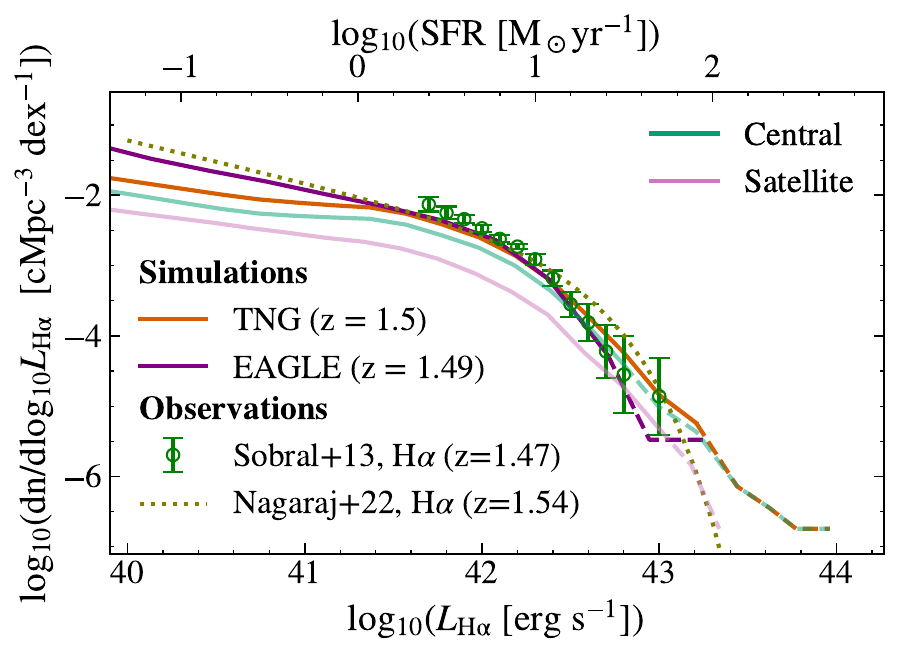}
    \caption{H$\alpha$ luminosity functions ({\lf}s) at redshift $z \sim 1.5$.  
    The {\em solid lines} correspond to {\lf}s computed from the \sfr\ in the simulations after applying dust attenuation ($A_{{\rm H}\alpha} = 1$, see Eq.~\ref{eq:Ldust}). The {\em orange lines} correspond to \tng\ at $z = 1.5$ and {\em purple lines} correspond to \eagle\ at $z = 1.49$; lines are {\em dashed} when there are fewer than 10 haloes in a luminosity bin. The {\em light green line} and the {\em light violet line} show the \tng\ {\lf}s for central and satellite galaxies only. 
    The {\em open circles} are the observed \lf\ from the {\sc HiZELS} survey at $z = 1.47$ \protect\citep{Sobral+13}. 
    The {\em dotted line} is the fit for the observed {\lf} from the {\sc 3D-HST} survey \protect\citep{Nagaraj_2023}.
    Both simulations generally agree well with each other and with the observations.}
    \label{fig:L_Ha_sfr}
\end{figure}

The luminosity of many observed emission lines is proportional to the galaxy's \sfr. Examples include CO, which is emitted by the cold molecular gas 
from which stars form, and H$\alpha$ and \ion{O}{III}, which are emitted in star-forming regions (e.g. \citealt{Bernal22}; we remind the reader that we do not include continuum radiation or emission from {\sc agn}s). It is common to adopt a linear relationship between the line flux and the \sfr, possibly after applying a correction for absorption (by dust and/or gas). This is most appropriate for \ha, as its emission is due to
the recombination of \ion{H}{II} in regions photo-ionised by ultraviolet photons emitted by young and massive stars. Other lines may be less direct tracers of star formation and more affected by metallicity or other environmental factors.

We compute the intrinsic \ha\ luminosity, $L_\alpha$, using
\begin{equation}\label{eq:L-sfr}
    \frac{L_{\alpha}}{\unit{erg\ s^{-1}}} = K \frac{\hbox{\sfr}}{\unit{M_\odot\ yr^{-1}}}\,,
\end{equation}
where $K =  2.0 \times 10^{41}$ \citep{Kennicutt98}
[we applied a $\sim 0.63$ factor conversion \citep{Madau_2014} from the \citet{Salpeter_1955} stellar initial mass function ({\sc imf}) assumed by \citet{Kennicutt98} to the more recent \citet{Chabrier2003} {\sc imf}]. We add the effect of dust attenuation in the galaxy's interstellar medium ({\sc ism}) by reducing the intrinsic \ha\ luminosity to
\begin{equation}\label{eq:Ldust}
    L^{\text{dust}}_{\alpha} = 10^{-A_{{\rm H}\alpha}/2.5} L^{\text{no dust}}_{\alpha}\,,
\end{equation}
setting $A_{{\rm H}\alpha} = 1$, following \citet{Garn10,Sobral12}.

In \cref{fig:L_Ha_sfr}, we compare the \ha\ {\lf} computed using \cref{eq:Ldust} from \tng\ and \eagle\ with each other, and with the observations from {\sc HiZels} \citep{Sobral+13} and {\sc 3D-HST} \citep{Nagaraj_2023}.
There is generally good agreement between both simulations and the observational data. At the faint end not probed by the observations, the \tng\ 
{\lf} is lower than the one from \eagle. This difference may be a consequence of the lower resolution of \tng. Indeed,
fig.~11 in \citet{Hirschmann_2023} shows that the {\lf} of the 300~Mpc \tng\ simulation falls below the {\lf} of the higher resolution simulations of the \illustris\ project. 
The contributions from centrals and satellites are shown separately in \cref{fig:L_Ha_sfr} (light green and light purple lines):
30 per cent of the {\lf} is due to satellites for $\log L_{\rm H\alpha} \gtrsim 41.5$.

The good agreement between the simulated and observed {\lf} in \cref{fig:L_Ha_sfr} provides some confidence that
combining \cref{eq:Ldust} with the \sfr\ from \tng\ is a good starting point to investigate the \lim\ power spectrum.
In the following subsections, we will investigate the contributions from central and satellite galaxies to the \sfr-\mvir\
relation, the evolution of the mean cosmic \sfr\ density (\sfrd), as well as the distribution of the \sfr\ inside haloes.

\subsection{The contribution of satellite galaxies to the cosmic star formation rate density} \label{sec:sat_contribution}

\begin{figure}
    \centering
    \includegraphics[width=\linewidth]{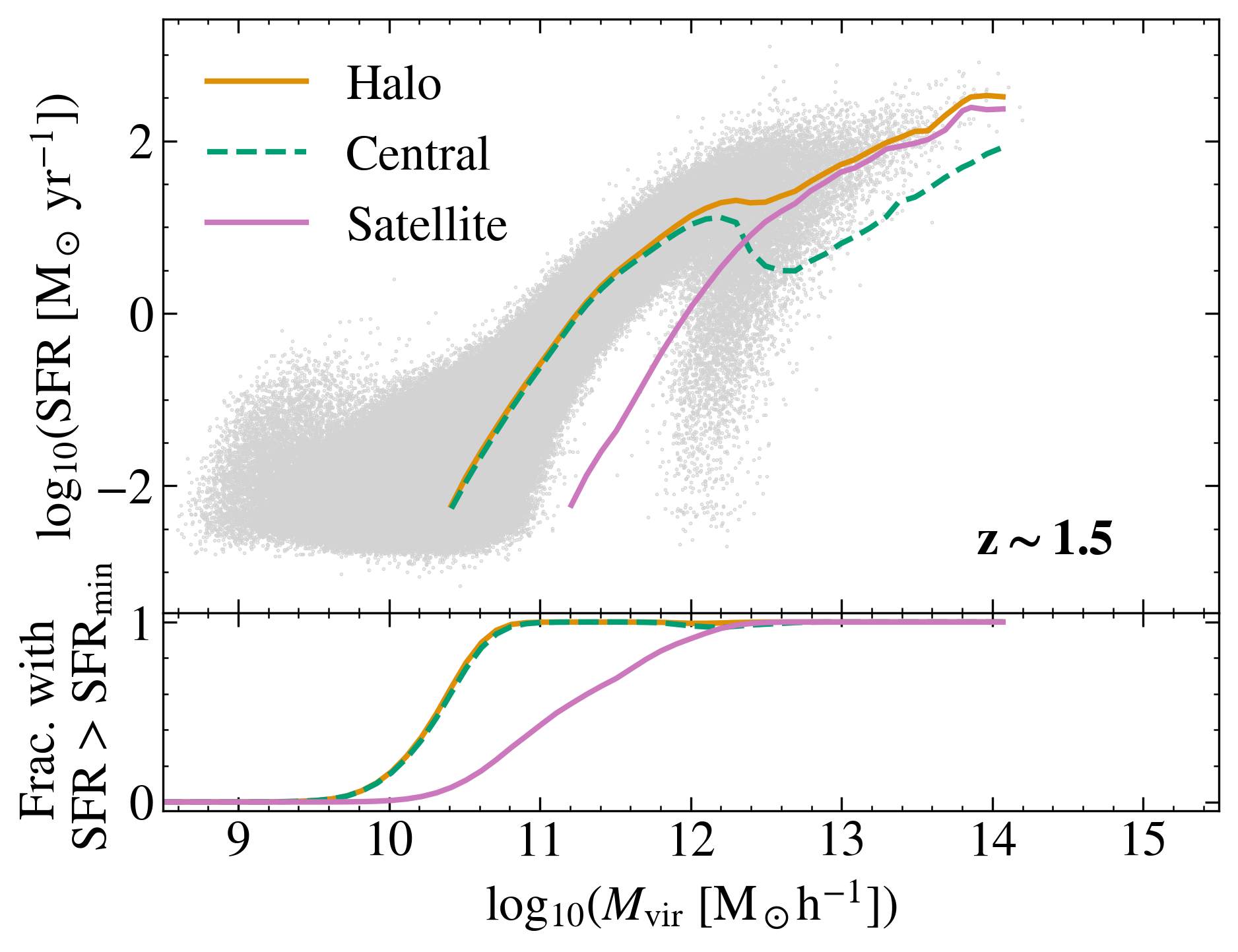}
    \caption{\textbf{\textit{Upper panel}}: \sfr-\mvir\ relation of \tng\ haloes at $z = 1.5$.
    The {\em dashed green line} is the median \sfr\ of the central galaxy only, the {\em solid violet} is the median of the sum of the \sfr s of all satellites in a halo, and the {\em solid orange} is the sum of these two contributions (i.e. the halo \sfr).
    The {\em grey dots} are the halo \sfr s of individual haloes. \textbf{\textit{Lower panel}}: Fraction of haloes for which the \sfr\ is numerically well resolved.
    }
    \label{fig:mvir_sfr_median}
\end{figure}

We plot the \sfr-\mvir\ relation of \tng\ galaxies at redshift $z=1.5$ in \cref{fig:mvir_sfr_median}. 
On average, the central galaxy contributes more to the halo's \sfr\ than all satellites combined for $\log M_{\mathrm{vir}} \lesssim 12$ (we will quote halo masses in units of \massunit\ unless specified otherwise), with satellites dominating the \sfr\ in more massive haloes. 

The lower panel of \cref{fig:mvir_sfr_median} shows the fraction of haloes with non-zero \sfr\ in the simulation.
Almost 100 per cent of haloes with $\log M_{\mathrm{vir}} \gtrsim 11$ have non-zero \sfr, compared to only 20 per cent for haloes with $\log M_{\mathrm{vir}} \sim 10$. The fraction of haloes with non-zero \sfr\ in their satellites at $\log M_{\mathrm{vir}} \sim 10$ is $\sim 0$ per cent, but reaches $\sim 100$ per cent for $\log M_{\mathrm{vir}} \gtrsim 12$.
These values may well depend on the details of the galaxy formation subgrid scheme adopted, but we expect that the general trends will not. In \eagle, for example, the decrease of the \sfr\ in the central galaxy sets in slightly earlier and is less sudden compared to \tng. The median fraction of star formation in central to satellites also does not reach as low a fraction as in \tng. The different subgrid physics implementations cause differences in the \sfr\ of both central and satellites, as we discuss in more detail in \cref{app:cent_frac_eagle}.

\begin{figure}
\includegraphics[width=\linewidth]{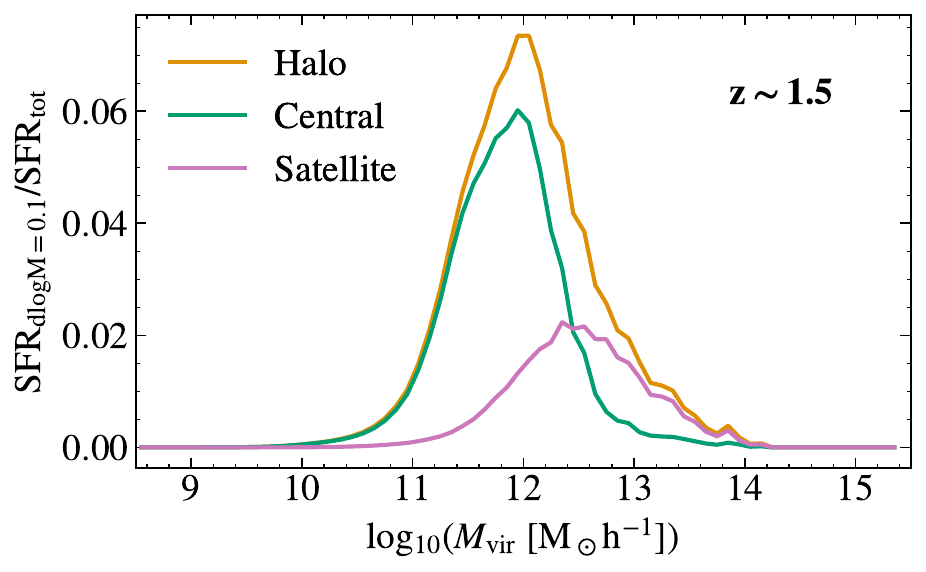}
\caption{The contribution from different halo masses to the total \sfr\ for haloes ({\em orange line}), central galaxies ({\em green line}), and satellite galaxies ({\em violet line}) in \tng\ at $z=1.5$. The amplitudes of the curves are the sum of the \sfr s of all haloes in a bin of \mvir\ with bin width $\Delta \log_{10} M_{\mathrm{vir}}=0.1$, divided by total \sfr\ in the computational volume. The contribution from satellite galaxies becomes larger than that from central galaxies beyond $\log M_{\mathrm{vir}} \sim 12.5$.}
\label{fig:contribution_mass_bins}
\end{figure}

\begin{figure}
\includegraphics[width=\linewidth]{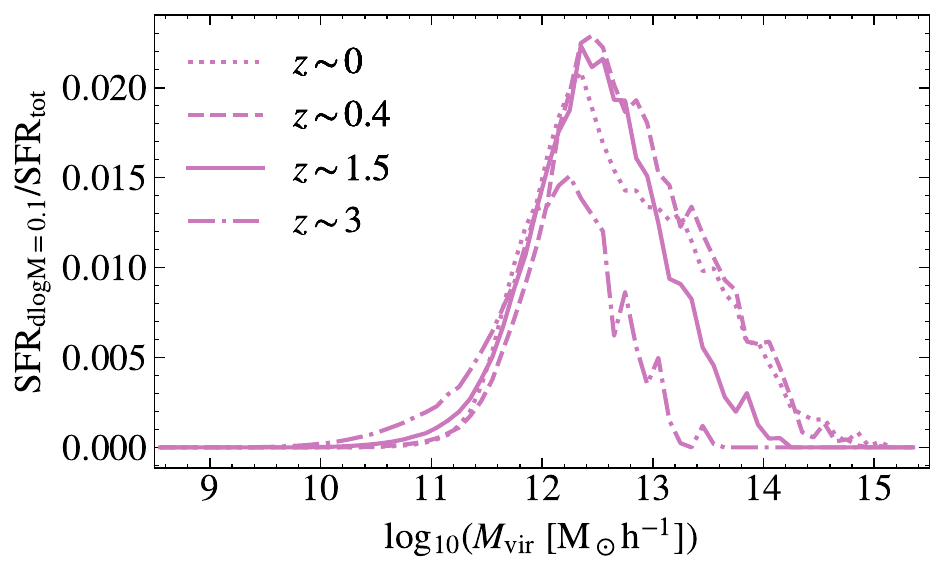}
\caption{The contribution from different halo masses to the total satellite \sfr\ in \tng\ at different redshifts as labelled in the panel. Comparing the curves for $z\sim 3$ ({\em dash-dotted line}), $z\sim1.5$ ({\em solid line}), and $z\sim0.4$ ({\em dashed line}) shows that the contribution to the \sfr\ from satellite galaxies shifts to more massive haloes at lower redshifts.}
\label{fig:contribution_mass_bins_z}
\end{figure}

To understand the observed \lim\ power spectrum, it is useful to quantify the extent to which haloes of a given mass contribute to the \sfr. \Cref{fig:contribution_mass_bins} shows the contribution to the total \sfr\ from haloes of different masses. Higher-mass haloes have higher \sfr s on average but then they are rarer than lower-mass haloes. As a result, intermediate-mass haloes, and in particular those with $\log M_{\mathrm{vir}} \sim 12$ contribute most as they are abundant and have high \sfr.  The central galaxy dominates the \sfr\ in such haloes because such haloes rarely have highly star-forming satellites. At $\log M_{\mathrm{vir}} \sim 12.5$, satellite galaxies start contributing more to a halo's \sfr\ than the central. This is both because the \sfr\ of central galaxies becomes quenched due to {\sc AGN} feedback, and because these haloes typically contain several highly star-forming satellites.
Observationally, verifying this prediction of the simulations is challenging due to the difficulty in determining the relative contribution from central and satellite galaxies to the \sfr\ (for a recent attempt using DESI data, see, e.g., \citealt{Gao_2023}).

\Cref{fig:contribution_mass_bins_z} shows the redshift evolution of the contribution of different halo masses to the total satellite \sfr. In general, the peak contribution comes from lower halo masses at higher redshifts, as haloes are generally less massive at earlier times. However, the contribution at $z \sim 0$ comes from slightly lower mass haloes than at $z \sim 0.4$, possibly due to satellite \sfr s becoming quenched.

\begin{figure}
    \centering
    \includegraphics[width=\linewidth]{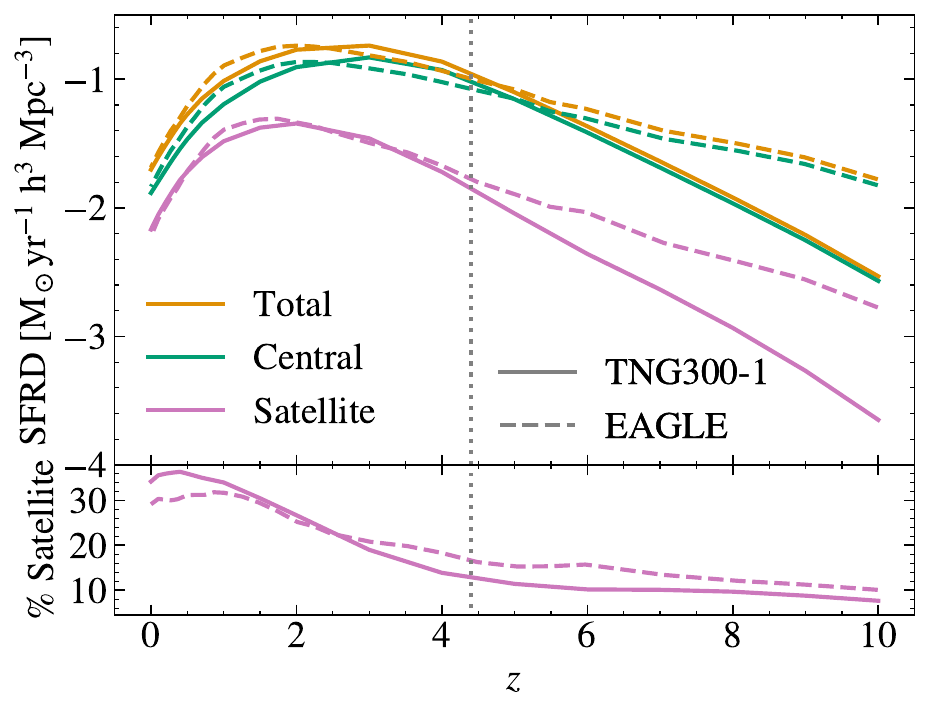}
    \caption{\textbf{\textit{Upper panel}}: Cosmic star formation rate density as a function of redshift, comparing \tng\ ({\em solid line}) to \eagle\ ({\em dashed line}). The total \sfr\ density is shown in {\em orange}, while the contribution from central and satellite galaxies separately are shown in {\em green} and {\em violet}. \textbf{\textit{Lower panel}}: Percentage contribution of the \sfr\ due to satellites. The {\em vertical dotted grey line} indicates the redshift at which the \sfr\ from haloes with fewer than 100 dark matter particles contributes 5\% to the total {\sfr} density in \tng. To the right of this line, lack of numerical resolution makes these results less reliable.}% - note that resolution also affects galaxy formation.}}
    \label{fig:sfrd}
\end{figure}

\Cref{fig:sfrd} shows the cosmic star formation rate density (\sfrd) as a function of redshift (orange lines), as well as the separate contributions from central (green lines) and satellite galaxies (violet lines) for \tng\ and \eagle. 
At any redshift, the \sfrd\ is dominated by central galaxies. For \tng, the contribution of satellites increases from $\sim 10$ per cent at $z\sim 4$ to $\sim 30$ per cent
below $z\sim 1.5$, reaching a maximum of $\sim 34$ per cent at $z\sim 0.4$. Redshift $z=4$ is close to the vertical dotted line, which indicates the redshift above which the \sfr\ from haloes with fewer than 100 dark matter particles contribute more than 5 per cent to the total \sfr\ in \tng. The lack of numerical resolution may well affect the satellite contribution close to and above this redshift. However, the higher contribution from satellites at low $z$ compared to higher $z$ seems numerically robust, and this trend is seen in both simulations. Below $z \sim 4$, the simulations show good agreement, but at higher redshifts, the \sfr\ in both satellites and centrals is significantly higher in \eagle\ compared to \tng. However, it should be noted that both simulations are increasingly affected by numerical resolution at these redshifts.

\subsection{The distribution of the SFR inside haloes} \label{sec:sfr-distr-halo}

\begin{figure*}
\centering
\begin{subfigure}[t]{.49\textwidth}
    \centering
    \includegraphics[width=\linewidth]{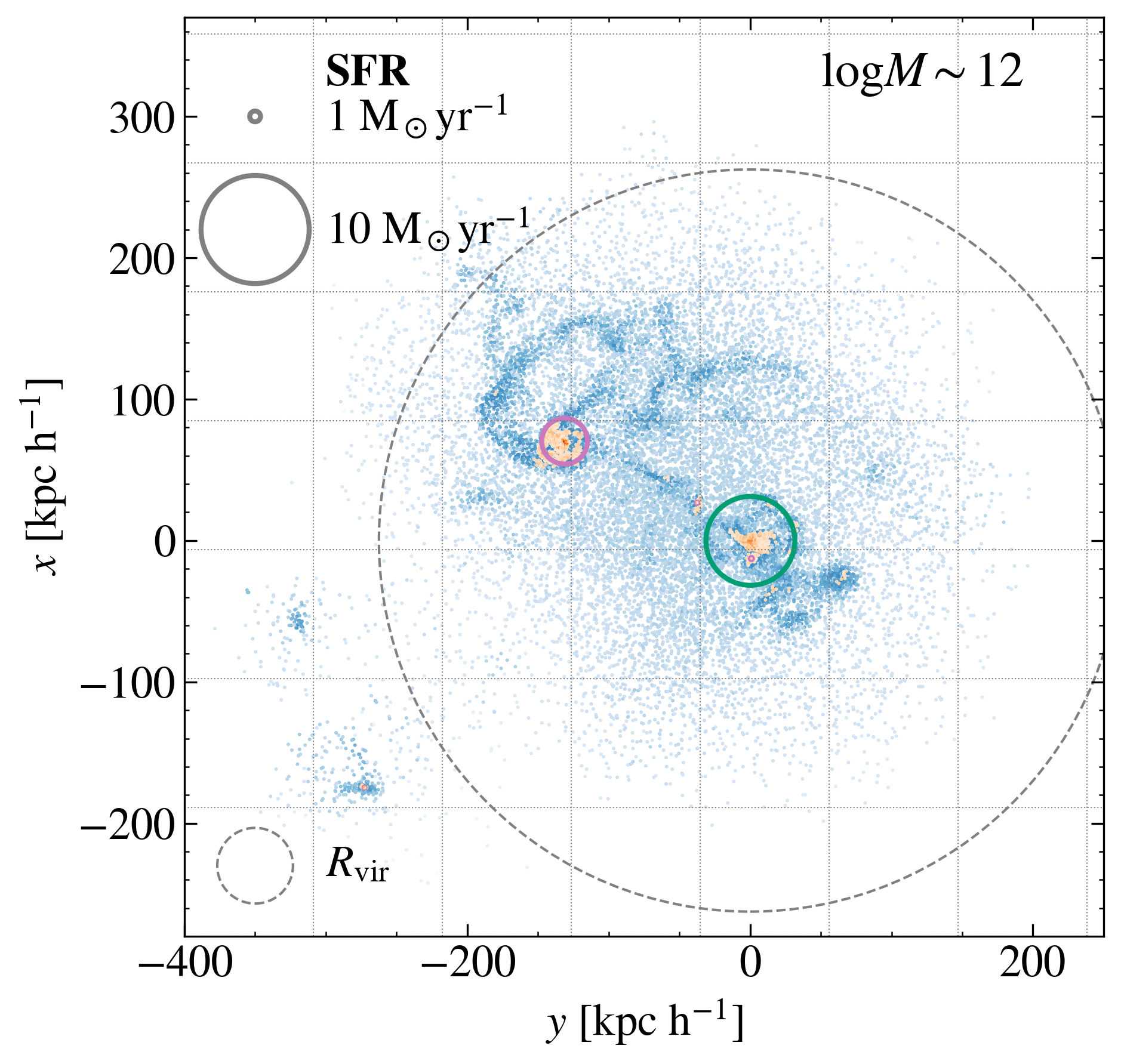}
\end{subfigure}
\begin{subfigure}[t]{.49\textwidth}
    \centering
    \includegraphics[width=\linewidth]{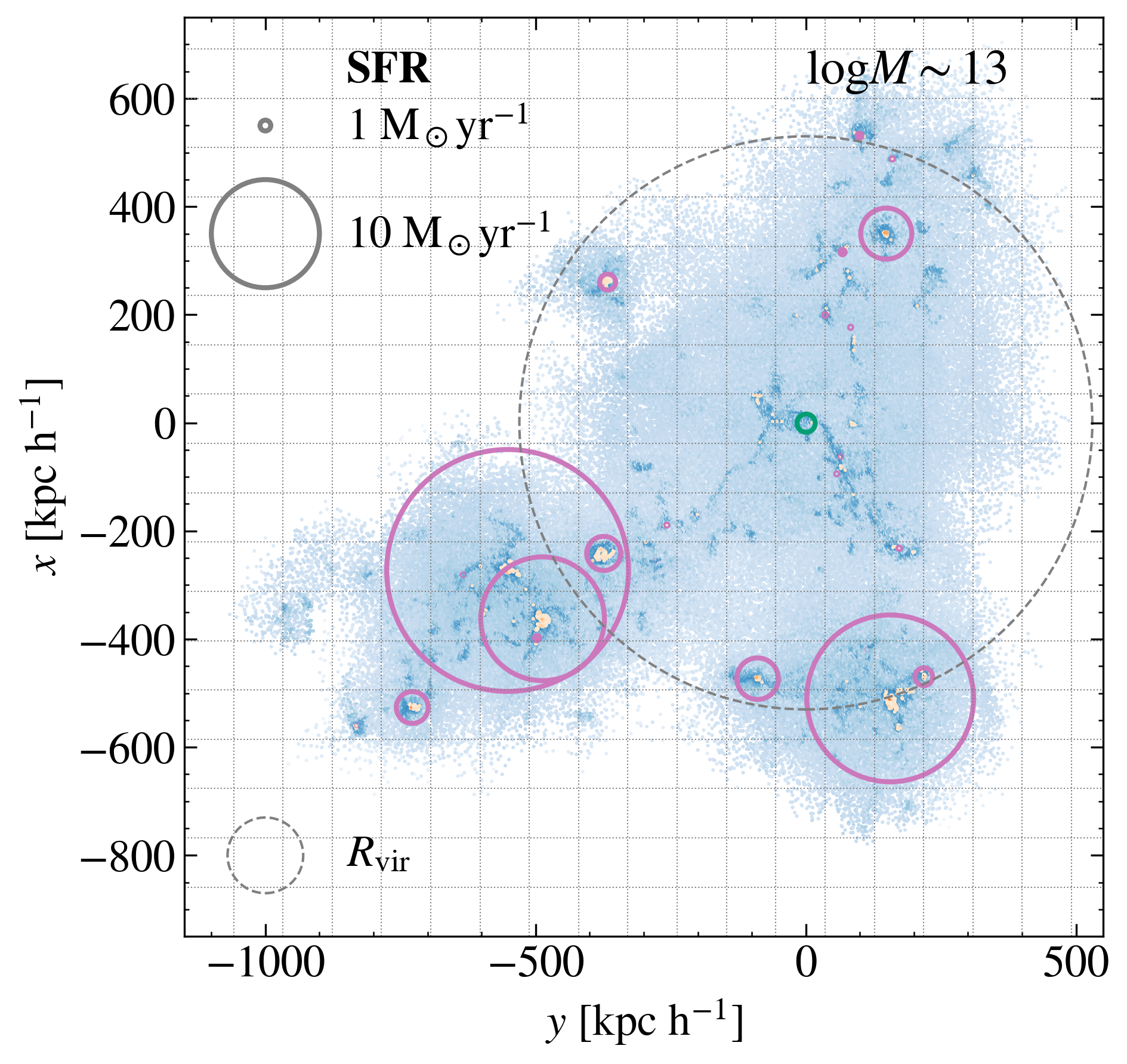}
\end{subfigure}
\caption{The distribution of star formation in 
\tng\ haloes. In both panels, the {\em dashed grey circle} indicates the virial radius of the halo, while galaxies are indicated by a {\em circle} with a radius proportional to the \sfr. The {\em green circles} correspond to the central galaxy, and {\em violet circles} correspond to satellite galaxies. The {\em dotted blue points} are gas particles, and {\em dotted orange points} are star-forming gas particles. 
\textbf{\textit{Left panel}}: Halo with mass $\log M_{\mathrm{vir}}\sim 12$ and \sfr\ $\sim 13$ \sfrunit. \textbf{\textit{Right panel}}: Halo with mass $\log M_{\mathrm{vir}}\sim 13$ and \sfr\ $\sim 72$ \sfrunit. (Please notice the difference in linear scale of the panels).
The {\em light grey dotted grid lines} mark the angular resolution (\SI{6.2}{\arcsecond}) for the {\sc SPHEREx} survey ($\sim 91$ kpc h$^{-1}$ at $z=1.5$). Both haloes would be spatially resolved in the data. (These haloes are index 564 and 17776 in the {\sc fof} halo catalogue in \tng).}
\label{fig:particles}
\end{figure*}

How star formation is distributed inside haloes impacts the small-scale \lim\ power spectrum if haloes are spatially resolved.  \Cref{fig:particles} shows two examples of the distribution of star-forming galaxies in \tng\ haloes of mass 
$\log M_{\mathrm{vir}}\sim 12$ and $\log M_{\mathrm{vir}}\sim 13$. Note that, in this paper, we refer to all galaxies that have non-zero \sfr\ in the simulation as star-forming galaxies.
Although the total gas is distributed widely, the {\it star-forming} gas is generally concentrated at the centres of subhaloes. The \sfr\ is not all at the centre of the halo, and some \lim\ observations have angular resolutions that resolve the distribution of the \sfr\ in sufficiently large haloes \citep[e.g.][]{COPSSII_2016, mmIME_2020}. Although not all \lim\ surveys will be able to resolve haloes, the upcoming {\sc SPHEREx} will spatially resolve haloes of mass 
$\log M_{\mathrm{vir}}\sim 12$ at $z\sim 1.5$. The resolution of this survey is indicated by the grey dotted grid in \cref{fig:particles}.

\ifSubfilesClassLoaded{%
  \bibliography{bibliography}%
}{}

\end{document}

\section{Impact of satellite galaxies on the power spectrum}\label{sec:sat}

\label{sect:satellites}
In the previous section, we illustrated the contribution of satellite galaxies to the total \sfr\ and how this \sfr\ is spatially distributed inside of their host halo.
Their net contribution and their spatial distribution are two ways in which satellites affect the galaxy power spectrum. In this section, we will consider these effects in detail by comparing three different ways of assigning \sfr s. Two of the schemes are illustrated in \cref{fig:distribution_schematic}.
The \lq galaxy\rq\ scheme is most similar to the real Universe: the \sfr\ of satellites is assigned to their subhaloes. To test whether the spatial distribution of the satellite \sfr\ matters, we consider the \lq halo\rq\ scheme, in which the \sfr\ of all galaxies in the same halo are summed and placed at the position of the central galaxy. The total \sfr\ of a halo is identical in both schemes. In the third and final case, we examine whether the \sfr\ of satellites matters at all by simply neglecting their contribution to the total \sfr\ altogether. We refer to the power spectrum corresponding to this case as the `central' power spectrum. 
We show the effect of removing satellite galaxies in \cref{sec:removing_sats} and the effect of how they are spatially distributed inside their host halo in \cref{sec:sat_dist}.

\ifSubfilesClassLoaded{%
  \bibliography{bibliography}%
}{}

\end{document}
\subsection{Impact of star formation in satellites on the power spectrum}\label{sec:removing_sats}

\begin{figure}
\centering
    \includegraphics[width=\linewidth]{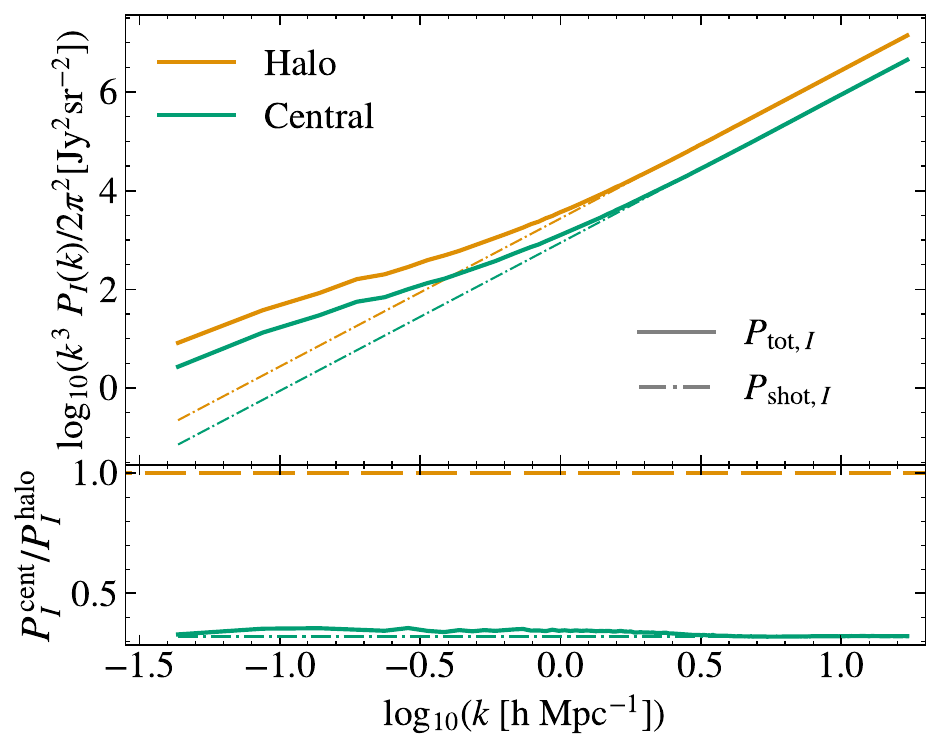}
\caption{\textbf{\textit{Upper panel}}: Power spectrum of galaxies weighted by their \sfr\ in \tng\ at redshift $z=1.5$, scaled to the intensity that would be measured in a \lim\ survey (i.e. the mean intensity has not been divided out). 
The \lq halo\rq\ power spectrum ({\em orange lines}) assigns the \sfr\ of galaxies to the centre of their host halo, while the \lq central\rq\ power spectrum neglects satellites. The {\em solid lines} represent the total power spectrum, while {\em dash-dotted lines} correspond to the shot noise contribution. 
\textbf{\textit{Lower panel}}: The ratio of the central to the halo power spectra, for total ({\em solid lines}) and shot-noise ({\em dash-dotted lines}).
The \lq central\rq\ power spectrum is less than $\sim 40$~per cent of the \lq halo\rq\ power spectrum, and the suppression is similar for the shot noise contribution.
We show results for \ha\ but these results should also hold for other lines.
}
\label{fig:halo_cent_ps}
\end{figure}

\begin{figure}
\centering
    \includegraphics[width=\linewidth]{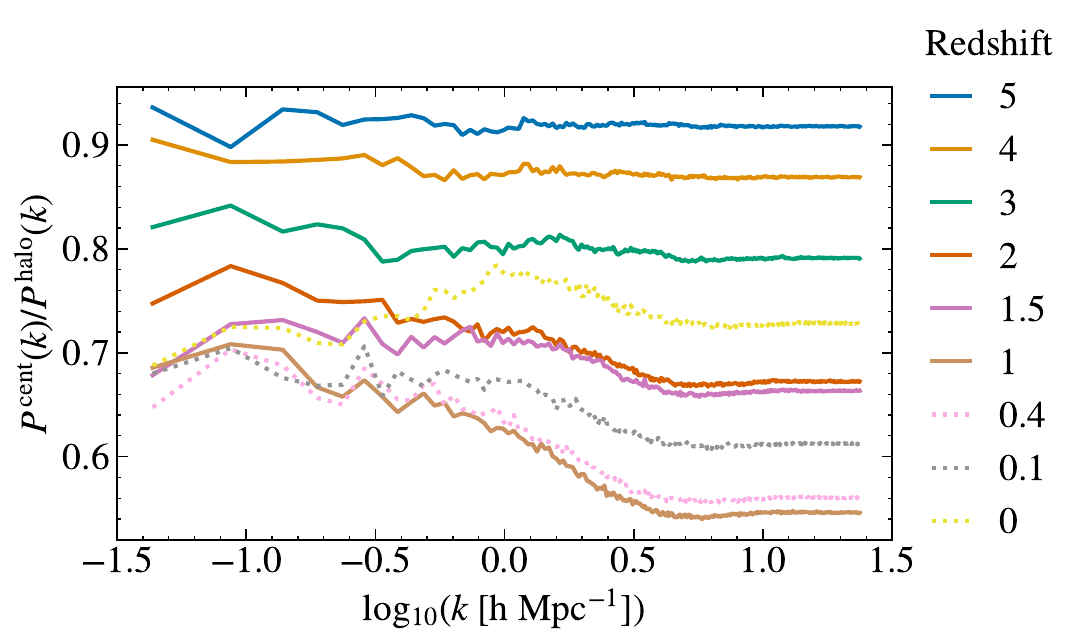}
\caption{The ratio of the central power spectra to halo power spectra, with both power spectra divided by their respective specific mean intensities. It is in effect the ratio $(b_{\mathrm{cent}}(k)/b_{\mathrm{halo}}(k))^2$.
This is shown for different redshifts, with line colours and styles indicated by the legend. The effect of satellites increases with decreasing redshift down to $z\sim 0.4$. The {\em dotted lines} are used for $z \sim 0.4,\ 0.1$ and 0 to emphasise the reversal of the trend from higher $z$.
The \sfr-weighted bias of central galaxies only is lower than that of the haloes (which includes the contribution from satellites) at all redshifts.
}
\label{fig:halo_cent_ps_z}
\end{figure}

We compute the power spectrum with galaxies weighted by their \sfr, in two ways: ({\em i}) the \lq central\rq\ power spectrum ($P_{\mathrm{cent},I}$),
which only accounts for central galaxies, and ({\em ii}) the \lq halo\rq\ power spectrum ($P_{\mathrm{halo},I}$, right panel of \cref{fig:distribution_schematic}). The latter is computed after adding the \sfr\ of satellites to the central galaxy. 
The total number density of sources is the same but the total \sfr\ is higher in the halo case.

The lower panel of \cref{fig:halo_cent_ps} shows the ratio of the central power spectrum to the halo power spectrum, while the upper panel shows the power spectra themselves. The power spectra have not been divided by the mean intensity.
We find that the amplitude of the central power spectrum is $\sim 35$ per cent of that of the halo power spectrum, indicating that removing satellite galaxies results in a significant underestimate of the power spectrum. 

The \lim\ 2-halo term can be written as $P_{2h,I}(k) = \bar{I}^2b(k)^2P_\mathrm{m}(k)$, showing that both the specific mean intensity and the bias affect the power spectrum.
The specific mean intensity is set by the \sfrd. 
In \cref{fig:sfrd}, we showed that satellite galaxies contribute $\sim$ 30\% to the \sfrd\ at $z =1.5$.  This scales with the specific mean intensity, $\bar{I}$, by a constant factor. 
The intensity power spectrum is proportional
to the mean intensity squared, so the difference due to the contribution from the \sfr\ of satellites would cause $P_{\mathrm{cent},I}$
to be $\sim 50$~per cent of $P_{\mathrm{halo},I}$, but we find $P_{\mathrm{cent},I}$ to be even lower than this. This suggests that the inclusion of satellite galaxies also affects the bias, $b(k)$.

\Cref{fig:halo_cent_ps_z} shows the ratio of the central power spectrum to the halo power spectrum but with the power spectra divided by the square of their respective specific mean intensities, such that the ratio corresponds to $(b_{\mathrm{cent}}(k)/b_{\mathrm{halo}}(k))^2$. The evolution with redshift is also displayed.
The violet line in \cref{fig:halo_cent_ps_z} shows that even when we consider the intensity-divided power spectrum, $P_{\mathrm{cent}}$ is $\sim 70$ per cent of $P_{\mathrm{halo}}$ at $z=1.5$. 
The reason for the difference is that satellite galaxies tend to reside in higher mass haloes.
Such haloes are more highly biased \citep{Tinker_2010}. 
When satellite galaxies are included, higher mass haloes have higher weight, causing the overall bias to be higher in the \lq halo\rq\ case.

The contribution of satellite galaxies to the \sfrd\ increases with decreasing redshift, at least until $z\sim 0.4$, as we showed in the previous section. As a consequence, satellites will also contribute more to the power spectrum at lower $z$. 

The amplitude of the power spectrum is set by the shot noise  on small scales. Although the number density of galaxies is the same in the \lq halo\rq\ and \lq central\rq\ case, the shot noise contribution is different. Indeed, the shot noise scales with $\text{{\rm Var}}(L)/\braket{L}^2$, where ${\text{\rm Var}}(L)$ is the variance in luminosity, as we demonstrated in \cref{eq:shotnoise_var}. Accounting for satellites in \tng\ increases the variance more than the square of the mean, resulting in a higher shot noise in the \lq halo\rq\ case.

\ifSubfilesClassLoaded{%
  \bibliography{bibliography}%
}{}

\end{document}

\subsection{Impact of the spatial distribution of satellites on the power spectrum}\label{sec:sat_dist}

\begin{figure}
\centering
    \includegraphics[width=\linewidth]{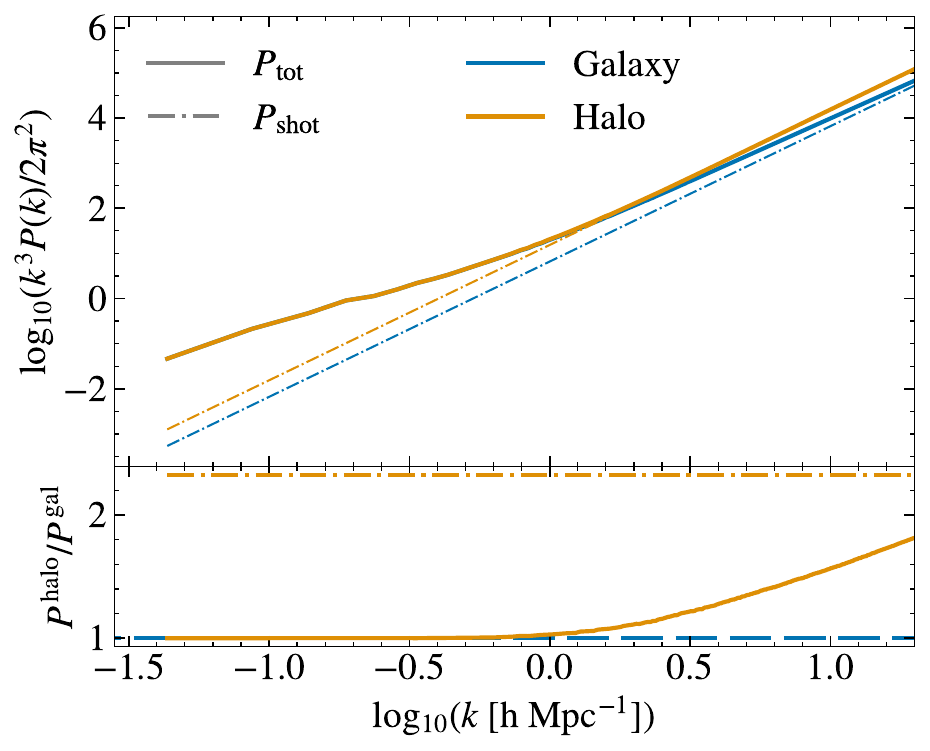}
\caption{\textbf{\textit{Upper panel}}: The halo power spectrum ({\em orange}) compared with the galaxy power spectrum ({\em blue}) at $z=1.5$. 
The {\em solid line} represents the total power spectrum and the {\em dash-dotted line} represents the Poisson shot noise power spectrum. 
\textbf{\textit{Lower panel}}: The ratio of the halo power spectrum relative to the galaxy power spectrum, respectively for the total and shot noise power spectrum. 
The galaxy power spectrum approaches the halo power spectrum on scales larger than the typical size of haloes ($k \sim 1$ \kunit).
}
\label{fig:halo_ps}
\end{figure}

\begin{figure}
\centering
    \includegraphics[width=\linewidth]{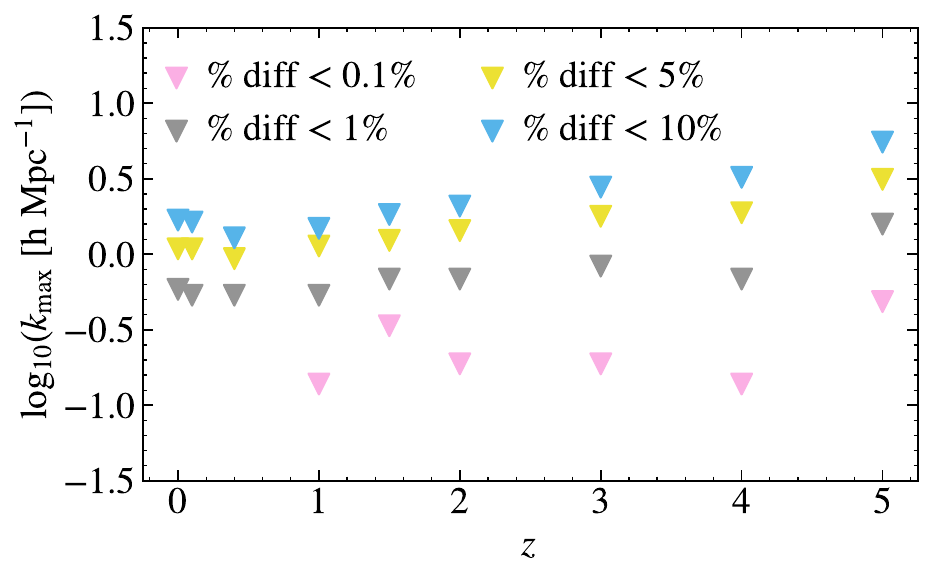}
\caption{The wavenumber up to which the galaxy and halo power spectra agree to within 0.1\% ({\em pink}), 1\% ({\em grey}), 5\% ({\em green}) and 10\% ({\em blue}) for several redshifts. The 0.1\% values for $z < 1$ are not shown as the variation in the ratio is larger than 0.1\% (the percentage error does not clearly stay below 0.1\%). 
}
\label{fig:gal_vs_halo_ps_z}
\end{figure}

In the previous section, we have only considered the case where all the \sfr\ is at the centre of haloes, yielding what we refer to as the halo power spectrum. This would be the power spectrum obtained from some fast simulation codes \citep[e.g.][]{Pinocchio}, which only have the mass and position of the {\sc fof} haloes and do not track the satellite subhaloes, in exchange for speed.
However, as we have seen in \cref{fig:particles}, the \sfr\ is not all at the centre, and the resolution of surveys like SPHEREx (\SI{6.2}{\arcsecond}) would be able to resolve the spatial distribution of the \sfr\ in a halo.

In this subsection, we provide a comparison of the halo power spectrum with the {\it galaxy} power spectrum, which more closely reflects the observed power spectrum. 
This helps us understand the contributions to the power spectrum and how the power spectrum computed from haloes only would differ from the observed power spectrum.

\Cref{fig:halo_ps} shows the halo and galaxy power spectra. They agree on large scales, but on small scales, the halo power spectrum tends to the {\it halo} Poisson shot noise, $P^{\text{halo}}_{\rm shot}$, whereas the galaxy power spectrum tends to the galaxy Poisson shot noise, $P^{\text{gal}}_{\rm shot}$. 
The halo Poisson shot noise is higher than that of galaxies as there are fewer haloes than there are galaxies, and the variance of luminosities is larger (see Eq.~\ref{eq:shotnoise_var}). 

The difference between the halo and galaxy power spectra becomes non-negligible when $U(k) \neq 1$ (see Eq.~\ref{eq:galaxy_ps_all}).
\Cref{fig:gal_vs_halo_ps_z} shows, for several redshifts, the $k$ values up to which the halo and galaxy power spectra agree to within a certain percentage. At lower redshifts, the sizes of haloes are generally larger, so the wavenumbers up to which the power spectra agree is smaller ($U(k) = 1$ beyond the size of the largest halo).
We find that a model that does not distribute the satellite galaxies is sufficient to reproduce the large-scale power spectrum to within 1 per cent on scales $\log k \lesssim -0.3$ for $z \lesssim 5$. The findings are similar when comparing the galaxy and halo power spectra computed for the \eagle\ simulation.

\ifSubfilesClassLoaded{%
  \bibliography{bibliography}%
}{}

\end{document}

\section{A model for the galaxy power spectrum} \label{sec:ps_components}

A \lim\ survey such as that proposed by {\sc SPHEREx} will sample the galaxy power spectrum from large linear scales down to smaller scales comparable to or even smaller than those of haloes. Here we present details of a model that captures this full range of scales. %We already showed some results of this model in \cref{sec:weighted_ps} (\cref{eq:galaxy_ps_all}). 
The model is based on the physically motivated components of the power spectrum explained in detail in \cref{sec:weighted_ps}.
In \cref{fig:components_ps}, we plotted the components of the model against the corresponding components in \tng, and showed that our model could reproduce the \tng\ power spectrum to within a few per cent. 

The model is motivated by the \lq halo model\rq\ \citep{Cooray2002} for the matter power spectrum, and consists of a 2-halo term -- arising from the clustering of haloes -- and a 1-halo term -- arising from the distribution of galaxies in a given halo.
In \cref{sec:2halo_fit}, we provide a fit for the 2-halo term, which accounts for halo exclusion and non-linear halo clustering outside the exclusion radius. 
In \cref{sec:1halo}, we provide a fit for the 1-halo term, which accounts for the different origins of shot noise on large and small scales, and is based on how galaxies are distributed inside a halo.
Finally, in \cref{sec:sec5_summary}, we summarise the parameters of the model.

\ifSubfilesClassLoaded{%
  \bibliography{bibliography}%
}{}

\end{document}

\subsection{Fitting the 2-halo term -- accounting for nonlinearity and halo exclusion} \label{sec:2halo_fit}

The 2-halo term imprints the clustering of haloes on the galaxy power spectrum. It is plotted for \tng\ as open orange circles in \cref{fig:components_ps}.
We expect the 2-halo term to be a biased version of the matter power spectrum on scales much larger than the radii of haloes. Nonlinearity will start to play a role on scales comparable to those radii.
Finally, on scales smaller than the radii of the largest haloes, we expect that halo exclusion will play an increasingly important role.
\citet{Mead_2021} introduce a correction to the halo model that incorporates both non-linear halo bias and halo exclusion within a single non-linear term, calibrated using simulations. However, this approach does not separately account for the individual effects of non-linear bias and exclusion. In this section, we present a fitting formula that explicitly distinguishes between these two contributions.

We compute the 2-halo term in the simulations by Fourier transforming the density field of haloes, where each halo is a point object with a weight equal to the total \sfr\ of all its galaxies. 
The 2-halo term is the power spectrum of the weighted halo density field after subtracting the halo shot noise term (Eq.~\ref{eq:halo_ps}). 
%As before, we use {\sc nbodykit} \citep{nbodykit} to perform this calculation, accounting for aliasing. 
Note that the 2-halo term in the galaxy power spectrum differs slightly from that in the halo power spectrum on scales smaller than the size of haloes (where $\uobserved \neq 1$ in Eq.~\ref{eq:2halo}). The distances between galaxies in a pair of haloes are not identical and will be noticeable on these scales. Nevertheless, we still use this 2-halo term as an approximation to the one in the galaxy power spectrum. 
In \cref{sec:sat_dist}, we showed that the galaxy power spectrum deviates from the halo power spectrum for $\log k \gtrsim -0.5$. This marks the scale where $\uobserved < 1$ for some haloes, in which case the 2-halo term in the galaxy power spectrum may be smaller than that in the halo power spectrum, and thus the 1-halo term would be larger than proposed (since the 1-halo term is derived by subtracting the 2-halo term of the halo power spectrum). 
At $\log k \sim -0.5$, only the largest haloes will have $\uobserved < 1$, therefore the difference between the galaxy and halo 2-halo term will be small, with the difference increasing at smaller scales where more haloes will have $\uobserved < 1$.

We will write the correlation function as $\xi(r) = [F(r)(1 + \xi'(r)) - 1]$ (see Eq.~\ref{eq:cf_multiple_rexc}), where the factor $F(r)$ accounts for halo exclusion, and $\xi'(r)$ is the non-linear correlation function in the absence of halo exclusion. These factors are computed as follows.

\vspace{2mm}
\noindent
$\bullet$ {\em The halo  exclusion factor.}

\noindent 
Halo exclusion imprints a feature in the correlation function $\xi(r)$. If all haloes had the same size, the correlation function $\xi(r)=-1$ for $r<D$ (where $D$ is the minimum distance between a pair of haloes before they overlap, i.e. $D$ is twice the halo's radius; see \cref{app:halo_exclusion}). Since they do not all have the same size, we expect this step-like feature in $\xi$ to be smeared out.
Therefore, we need to find the probability distribution of exclusion distances, ${\rm PDF}(d)$. \citet{Baldauf_2013} propose a lognormal distribution,
\begin{align}
	{\rm PDF}(d) = \frac{1}{(\ln10) d \sigma  \sqrt{2\pi}} \exp\left[ -\frac{(\log_{10}(d/d_0))^2}{2\sigma^2} \right]\,,
 \label{eq:pdf}
\end{align}
with parameters $\log_{10} d_0$ (the logarithmic mean value of $d$) and $\sigma$ (the logarithmic scatter around the mean). The probability of finding a pair of haloes whose separation is less than some distance $r$ is then 
\begin{align} \label{eq:erf}
	F(r) = \frac{1}{2}\left( 1 + {\rm erf}\left[ \frac{\log_{10} (r/d_0)}{\sqrt{2} \sigma}\right]\right)\,.
\end{align}
This is the function that appears in \cref{eq:cf_multiple_rexc}. 

We find that the measured probability distribution of halo separations in \tng\footnote{We estimate the {\sc fof} radius of haloes by multiplying the virial radius by $(M_{\mathrm{FOF}}/M_{\mathrm{vir}})^{1/3}$. Then, we find the probability distribution of all pairs of {\sc fof} radii.} for haloes in the range $\log M_{\mathrm{vir }}=12-12.5$ is described well by the lognormal model of \cref{eq:pdf}. \citet{Sullivan_2021} additionally test the use of an exponential function, but conclude that the lognormal function fits better.

\vspace{2mm}
\noindent
$\bullet$ {\em The non-linear correlation function, $\xi'(r)$.}

\noindent
Before we can compute the halo power spectrum by Fourier transforming $\xi$, we need to compute the correlation function on scales comparable to or even smaller than the typical halo radius -- indeed, some smaller haloes in the lognormal distribution can be closer together than larger haloes. We referred to this extrapolated correlation function
as $\xi'(r)$. % in  \cref{eq:cf_multiple_rexc} (see \cref{sec:non-linear}).
\citet{Baldauf_2013} attempt to take into account the non-linear enhancement of clustering outside the exclusion radius by adding a quadratic bias, but this results in an underestimation of the power spectrum measured in their simulation (see their fig.~5).  %In \cref{sec:2halo_fit}, we show that the physically motivated formalism for halo exclusion presented in this subsection can be included in a fit to reproduce the 2-halo term measured in simulations.

%We find that the ratio of $\xi'(r)$ to the linear matter correlation function depends on $r$: linear bias is not a good description for halo bias. However, w
We find that the ratio of $\xi'(r)$ to the {\em non-linear} matter correlation function is approximately constant at $z\sim 1.5$.
This finding agrees with the results of \citet{Sheth_1999}, who also found
that the non-linear bias of haloes is approximately constant (for $z\sim 2$ in their case). At both higher and lower $z$, the non-linear bias becomes scale-dependent, as also shown by \citet{Sheth_1999}. % (their fig.~1). A more detailed investigation of the scale-dependence of bias is beyond the scope of the present study but, in principle, the same exclusion formula could be combined with a better model for scale-dependent bias at other redshifts.

\begin{figure}
\centering
    \includegraphics[width=\linewidth]{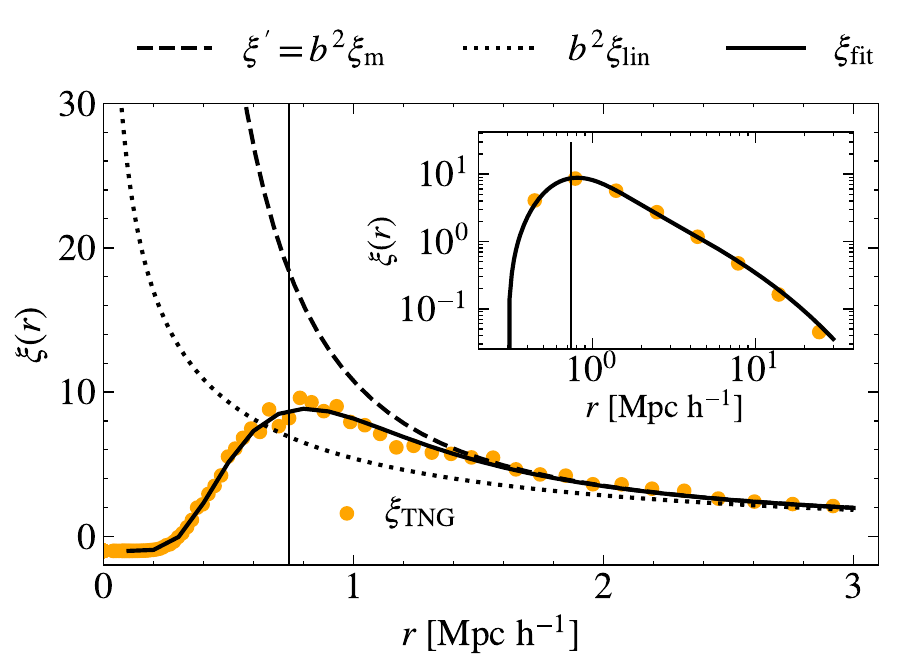}
\caption{The \sfr-weighted halo correlation function at $z=1.5$.
The correlation function measured from \tng\ is shown by {\em orange dots}, and the fit discussed in the text is plotted as a {\em solid black line}. The fit is described by three parameters: a constant bias, $b$, relative to the non-linear correlation function, and two parameters that characterise halo exclusion. This fit captures the non-linear enhancement outside the exclusion radius as well as the decrease at small radii caused by halo exclusion. The {\em vertical solid line} marks the mean exclusion radius. The {\em dashed black line} is $b^2$ times the non-linear correlation function, and the {\em dotted black line} is $b^2$ times the linear correlation function. The linear correlation function underestimates and the non-linear correlation function overestimates the measured correlation function on small scales.
The inset shows the correlation function on a log-log scale up to $r \sim 30$ \lenunit.}
\label{fig:exclusion_cf}
\end{figure}

We compute the non-linear matter correlation function, $\xi_\mathrm{m}$, with the {\sc halofit} model\footnote{We use {\halofit} -- rather than measuring the correlation function from \tng\ directly -- purely for computational ease.} of \citet{Takahashi_2012}.
The extrapolated correlation function $\xi'(r)$ is then taken to be $b^2 \xi_\mathrm{m}$. 

\vspace{2mm}
\noindent
$\bullet$ {Incorporating halo exclusion and nonlinearity.}

\noindent
We use \cref{eq:cf_multiple_rexc} as a fitting function for the \sfr-weighted correlation function, with $F(r)$ given by \cref{eq:erf} and $\xi'(r)$ given by the non-linear matter correlation function. The three parameters of $\xi$ are: $d_0$ and $\sigma$, which characterise exclusion (see Eq.~\ref{eq:erf}), and the bias parameter $b$.
Fitting to the halo correlation function in \tng, we obtain the best-fitting values of $d_0 = 0.74$ \lenunit, $\sigma = 0.17$ dex and $b=2.0$.

In \cref{fig:exclusion_cf}, the fit (solid black line) is compared to the halo correlation function measured in \tng\ (orange dots). On scales $r\gtrsim 2$ \lenunit, the \tng\ correlation function is well approximated by $b^2$ times the linear (dotted line) or non-linear correlation function (dashed line). On smaller scales, $1\lesssim r/\lenunit\lesssim 2$, nonlinearity increases the \tng\ correlation function above the biased linear correlation function: this increase is captured well by $b^2$ times the non-linear correlation function. On even smaller scales, halo exclusion is detected in \tng, and the rapid decrease in amplitude of the halo correlation function is captured well by our model.

We note that the halo exclusion multiplier $F(r) \leq 1$ for all $r$, hence $\xi(r) \leq \xi'(r)$ for all $r$. Therefore, the amplitude of $\xi'(r)$ needs to be higher than that of the halo correlation function at all $r$. Figure \ref{fig:exclusion_cf} shows that this is not true for the linear correlation function (dotted line) for $r\sim 1$ \lenunit\ -- which is why the non-linear correlation function is needed in the definition of $\xi$.

\begin{figure*}
\centering
\begin{subfigure}[t]{.5\textwidth}
    \centering
    \includegraphics[width=\linewidth]{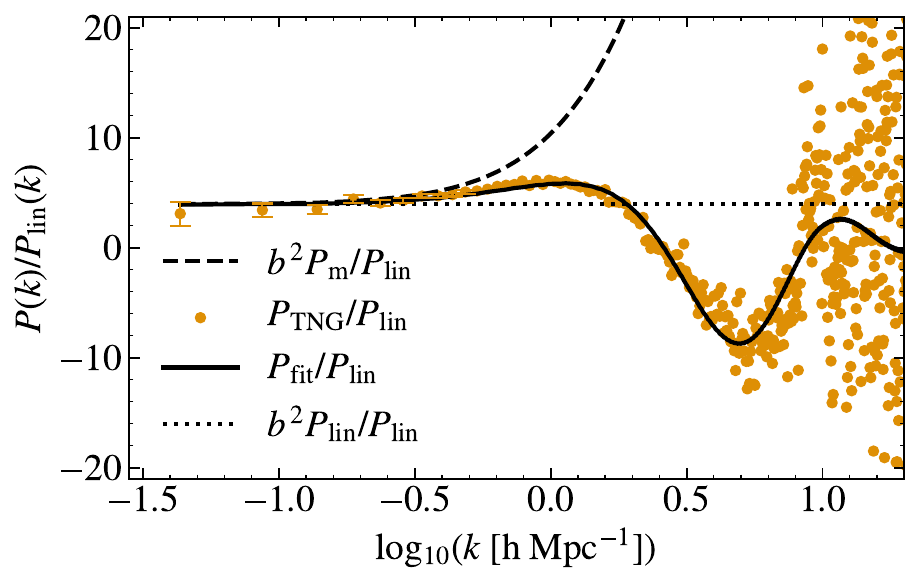}
\end{subfigure}%
\begin{subfigure}[t]{.5\textwidth}
    \centering
    \includegraphics[width=\linewidth]{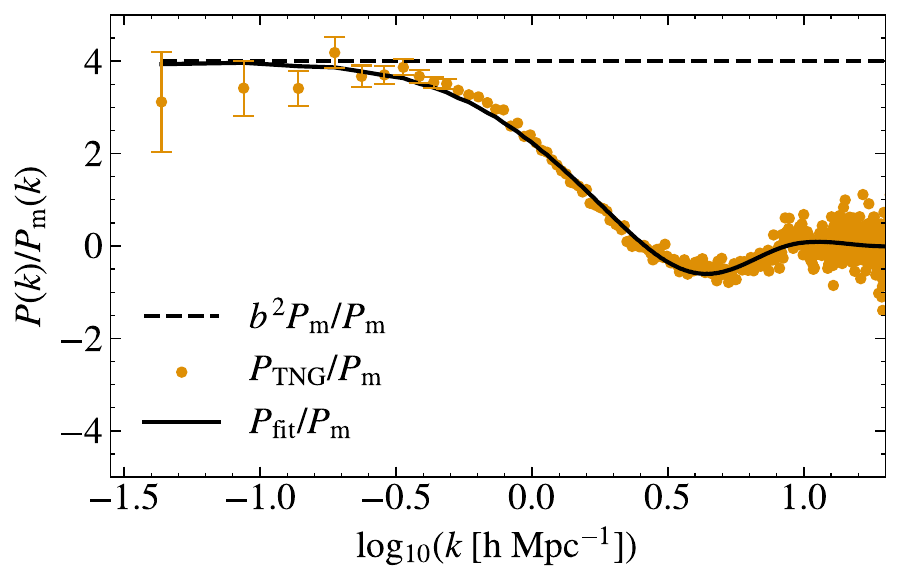}
\end{subfigure}
\caption{
The 2-halo power spectrum of our model ({\em black solid line}) compared to that measured in \tng\ ({\em orange dots}) at $z=1.5$. Here, the power spectrum is binned in linear bins of width d$k = 0.05$ \kunit. \textbf{\textit{Left panel}}: Both power spectra are divided by the linear power spectrum of \protect\citet{Eisenstein_1998}. The {\em dashed black line} is $b^2$ times the non-linear matter power spectrum divided by the linear power spectrum, where $b$ is the bias. The {\em dotted line} is the square of the constant bias value we use in our fit. 
\textbf{\textit{Right panel}}: Both power spectra are divided by the non-linear matter power spectrum. The {\em dashed black line} is the square of the constant bias value we use in our fit. 
Our fit to the 2-halo term -- which includes a bias with respect to the non-linear matter spectrum and a correction for halo exclusion -- reproduces the power spectrum measured in \tng.
}
\label{fig:bias2}
\end{figure*}

We Fourier transform the fit to the correlation function to get the model's power spectrum. The halo exclusion feature in the correlation function creates wiggles in the power spectrum (Fig.~\ref{fig:bias2}).
The figure also shows the scale-dependent bias with respect to the linear and non-linear power spectrum (left and right panel). The model presented here (solid black line), which uses the non-linear matter power spectrum and takes into account the halo exclusion effect, reproduces the power spectrum measured in \tng\ (orange dots).  We show in \cref{app:2halo_eagle} that a similar fit also reproduces the power spectrum measured in \eagle.

The left panel of Fig.~\ref{fig:bias2} shows that the linear power spectrum multiplied by a constant bias (dotted line) underestimates the \tng\ 2-halo term on intermediate scales of $\log k \sim 0$. On the other hand, multiplying the non-linear matter power spectrum by a constant bias -- without accounting for halo exclusion (dashed line) -- overestimates the \tng\ power spectrum on these scales. Our model does better than a constant bias model as it accounts for both non-linear halo bias and halo exclusion.

When using the non-linear matter power spectrum as $\xi'(r)$, it is important to clarify that this does not represent the halo clustering in the absence of exclusion. Such an interpretation would be incorrect as it includes the 1-halo term. We are simply using it as a tool to reproduce the observed 2-halo term, which includes halo exclusion.
By removing the power on scales smaller than the size of haloes due to halo exclusion, we naturally remove the effect of the 1-halo term. 

In \cref{fig:components_ps}, the largest difference between TNG and the model is on the largest scales -- this is where the impact of sample variance is greatest. To estimate the impact of sample variance, we compute the \tng\ matter power spectrum at the start of the simulation ($z \sim 127$), where the power spectrum is expected to be linear.
We then multiply the result by a scaling factor -- in this case, the power spectrum at $z=1.5$ agrees with the scaled power spectrum for $\log k \lesssim -0.8$, indicating the scale above which the power spectrum is still linear. This suggests that the error on scales $\log k \lesssim -0.8$ is due to sample variance.

In summary, the 2-halo contribution to the \tng\ power spectrum is not fit well by multiplying the linear power spectrum with a constant bias: this fails to capture the shape of the power spectrum on scales comparable to that of haloes ($\log k \sim 0$) and (unsurprisingly) is even worse at smaller scales.
Our model -- which uses a constant bias relative to the non-linear power spectrum and accounts for halo exclusion -- fits the \tng\ 2-halo term on all scales at $z \sim 1.5$. We turn to the 1-halo term next.

\ifSubfilesClassLoaded{%
  \bibliography{bibliography}%
}{}

\end{document}

\subsection{The 1-halo term -- halo profile and shot noise} \label{sec:1halo}

We compute the 1-halo term of a simulation power spectrum by subtracting the 2-halo term of the halo power spectrum from the total galaxy power spectrum (Eq.~\ref{eq:galaxy_ps_all}). The result is plotted as open red circles in \cref{fig:components_ps} for the \tng\ simulation. Whereas the 2-halo term results from the clustering of galaxies in different haloes, the 1-halo term arises from correlations of galaxies within the same halo, including with themselves. 

\Cref{eq:galaxy_ps_all} shows that the 1-halo term combines three contributions: ({\em i}) shot noise from haloes, ({\em ii}) shot noise from galaxies, and ({\em iii}) a term that reflects the distribution of galaxies in haloes. Determining these contributions can help constrain the galaxy formation model. 
Understanding the 1-halo term is also important on large scales, where shot noise contributes
to the measured power spectrum.

In \cref{sec:sec5_shot_noise}, we discuss the amplitude of the 1-halo term, which is set by galaxy shot noise on small scales and halo shot noise on large scales. In \cref{sec:halo_profile}, we discuss the shape of the 1-halo term, which depends on the distribution of luminosity in haloes.

\subsubsection{Galaxy and halo shot noise contributions to the 1-halo term} \label{sec:sec5_shot_noise}

\begin{figure}
\centering
    \includegraphics[width=\linewidth]{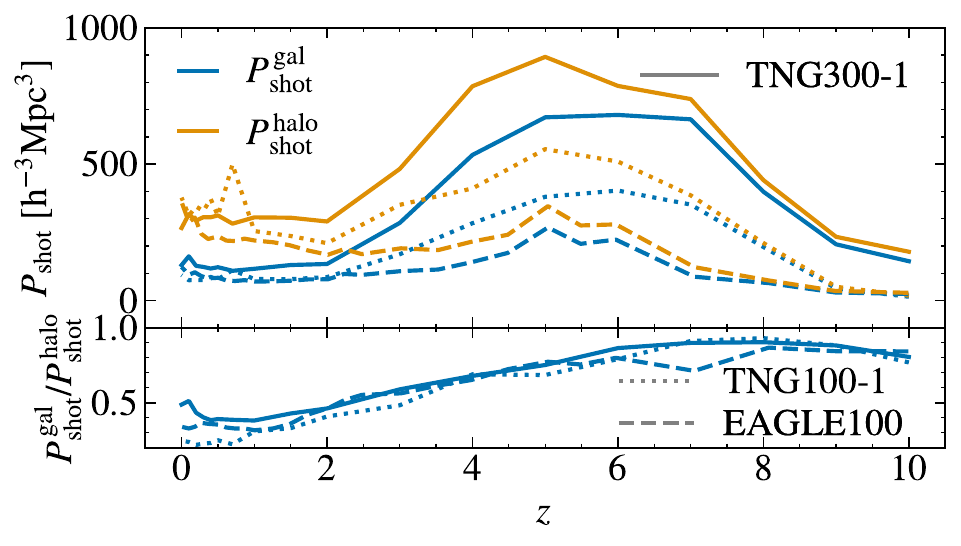}
\caption{\textbf{\textit{Upper panel}}: Amplitude of the shot noise term as a function of redshift.
The galaxy shot noise, $P^{\mathrm{gal}}_{\mathrm{shot}}$, is plotted in {\em blue}, and the halo shot noise, $P^{\mathrm{halo}}_{\mathrm{shot}}$, in {\em orange}. Results for \tng\ (TNG300-1) are shown by {\em solid lines}, for TNG100-1 as {\em dotted lines}, and for \eagle\ as {\em dashed lines}.
%The {\em upper panel} shows the amplitude of the shot noise terms, and the {\em lower panel} shows the 
\textbf{\textit{Lower panel}}: The ratio $P^{\mathrm{gal}}_{\mathrm{shot}}/P^{\mathrm{halo}}_{\mathrm{shot}}$. 
For all simulations, the difference between the galaxy and halo shot noise decreases with increasing redshift, where the contribution of satellites to the \sfrd\ is decreases.
There are significant amplitude differences between TNG300-1 and TNG100-1, likely due to numerical resolution. Despite similar resolutions, TNG100-1 and EAGLE still show noticeable differences in shot noise amplitude.
}
\label{fig:shot_redshift}
\end{figure}

We compute the shot noise using Eq.~(\ref{eq:shotnoise_discrete_weight}), summing over all simulated galaxies for the galaxy shot noise term, $P^{\mathrm{gal}}_{\mathrm{shot}}$, and over all haloes, with weight equal to the sum of their galaxies' \sfr s, for the halo shot noise term, $P^{\mathrm{halo}}_{\mathrm{shot}}$. By construction, these numbers are independent of scale. They are plotted as a function of redshift in the upper panel of \cref{fig:shot_redshift}. With increasing redshift, increasingly fewer haloes have highly star-forming satellites, and the values of the two shot noise terms becomes similar. The ratio of the two terms are shown in the lower panel of \cref{fig:shot_redshift}. The galaxy shot noise is $\sim 50\%$ of the halo shot noise for $z \lesssim 2$, rising to $75\%$ at $z \sim 4$.

We also show the shot noise computed using the TNG100-1 simulation (dotted lines) and the \eagle\
simulation (dashed lines) to show the effect of resolution and galaxy formation model; both have a simulation box of around 100 cMpc and comparable numerical resolution. Both of these are higher in resolution than the TNG300-1 simulation. The shot noise computed in the TNG100-1 simulation is lower than in the TNG300-1 simulation, both because star formation is better resolved, resulting in a higher volume density of galaxies with non-zero \sfr, and because the \sfr\ of a galaxy tends to be higher at higher resolution in the \illustris\ model. Indeed, while the hydrodynamical simulations have well-motivated physical processes, the simulations are not yet fully converged in resolution \cite[see e.g.][]{Pillepich_2018}. 
In particular, resolution affects the \sfr\ \citep[see fig. 11 of][]{Hirschmann_2023} and hence the amplitude of the shot noise term. The shot noise computed from the \eagle\ simulation is even lower than the one computed from TNG100-1. 
Although the variance of the \sfr\ is higher in the \eagle\ simulation, the mean is also higher, resulting in a lower shot noise in \eagle.

The surprisingly large differences between simulations with different resolutions (TNG300-1 versus TNG100-1), or between different implementations of galaxy formation (TNG100-1 versus \eagle) sound a note of caution when using simulations to estimate the shot noise term in an observed data set ($P^{\mathrm{halo}}_{\mathrm{shot}}$ cannot easily be measured in the data since haloes are not observable). Nevertheless, the qualitative trends are relatively independent of resolution: there is a difference between galaxy and halo shot noise at all redshifts, with the difference becoming less significant with increasing redshift.

\begin{figure}
\centering
    \includegraphics[width=\linewidth]{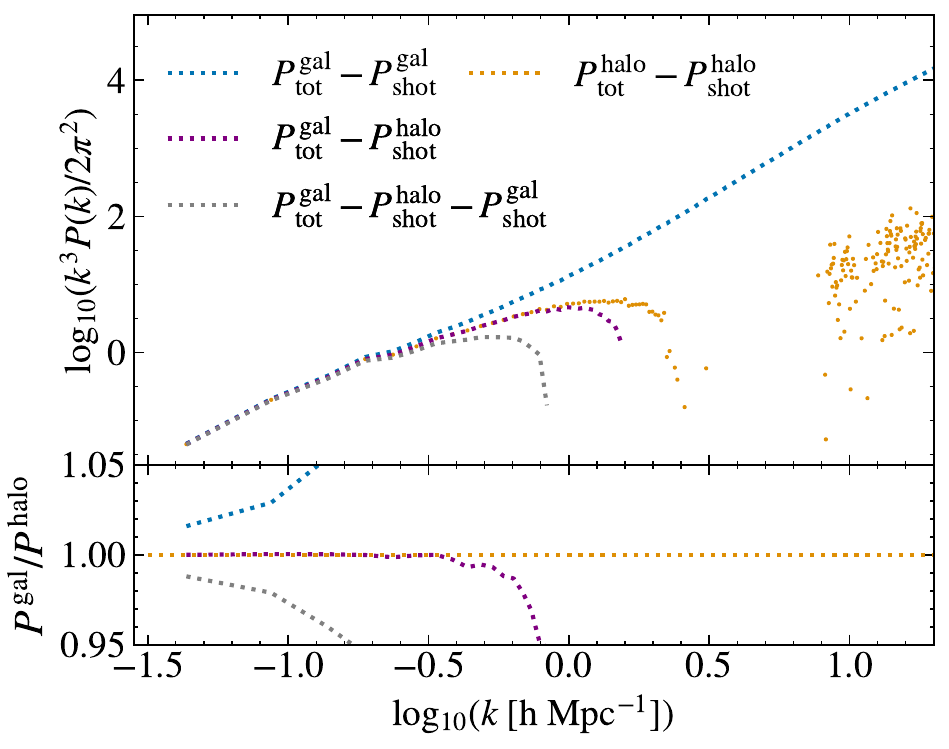}
\caption{\textbf{\textit{Upper panel}}: Subtracting various choices for the shot noise from the galaxy power spectrum of \tng\ at $z=1.5$. 
The shot noise subtracted is that of galaxies ({\em blue dotted}), haloes ({\em purple dotted}), and both galaxies and haloes ({\em grey dotted line}). The {\em yellow dotted line} corresponds to the 2-halo term of the halo power spectrum (obtained by subtracting the halo shot noise from the halo power spectrum). 
\textbf{\textit{Lower panel}}: The ratio of the shot noise-subtracted power spectra relative to the 2-halo term derived from the halo power spectrum ({\em yellow dotted line}). The {\em purple dotted line} agrees with the {\em yellow dotted line} for $\log k \lesssim -0.5$, indicating that the halo shot noise is the appropriate term to subtract on large scales to recover the 2-halo term.
}
\label{fig:subtract_shot_noises}
\end{figure}

We stress that $P^{\mathrm{gal}}_{\mathrm{shot}}$ -- which might conceivably be measured from the observed small-scale galaxy power spectrum -- is not the correct shot noise to be used on large scales.
If one subtracts $P^{\mathrm{gal}}_{\mathrm{shot}}$ rather than $P^{\mathrm{halo}}_{\mathrm{shot}}$
from the galaxy power spectrum, then the 2-halo term, and therefore the bias, will be estimated incorrectly. \Cref{fig:subtract_shot_noises} illustrates the errors incurred if one subtracts the wrong shot noise. The orange dotted line shows the halo power spectrum with the halo shot noise subtracted. This is our reference for the correct 2-halo term on large scales (where $U(k) = 1$; see Eq.~\ref{eq:galaxy_ps_all}). 

Several common assumptions are made when discussing shot noise, some of which are illustrated in
Fig.~\ref{fig:subtract_shot_noises}. The first common assumption is that the galaxy shot noise is independent of scale. The blue dotted line shows the case where only the galaxy shot noise is subtracted from the galaxy power spectrum. If we assume that subtracting the galaxy shot noise gives the 2-halo term then it is overestimated by more than 1~per cent on large scales. Another common assumption is that the 1-halo term tends to the halo shot noise on large scales and that the galaxy shot noise is constant on all scales. In this case, subtracting the sum of the galaxy shot noise and the halo shot noise should give the 2-halo term. However, the grey dotted line shows that doing this underestimates the 2-halo term by more than 1~per cent. Subtracting only the halo shot noise allows us to reproduce the 2-halo term to within 0.01\% for $\log k \lesssim -0.5$, confirming that the halo shot noise is the correct shot noise on large scales. The shot noise-subtracted galaxy and halo power spectra start to deviate for $k$ sufficiently large such that $U(k)$ -- the Fourier transform of the halo profile -- deviates significantly from 1. \Cref{fig:subtract_shot_noises} shows that on the largest scales in \tng, shot noise is not negligible and therefore the distinction between the galaxy and halo shot noise must be made in order to estimate the 2-halo term accurately.

While subtracting the correct shot noise term is important to obtain the correct 2-halo term on all -- including large -- scales, it is also worth investigating what information can be obtained from the term itself. By measuring the small scales where the galaxy Poisson shot noise is dominant, we can obtain information about the luminosities of galaxies (Eq.~\ref{eq:shotnoise_var}), providing a constraint on their luminosity function. We only have the total power spectrum from observations, therefore to determine if the galaxy shot noise is dominant, one should check that the slope of the power spectrum is constant on those scales. In \cref{fig:components_ps}, we plotted the power spectra to half the Nyquist frequency of the {\sc SPHEREx} pixel length at $z=1.5$. At these scales, the galaxy Poisson shot noise is not yet dominant.

Similarly, the halo shot noise could, in theory, provide a constraint on the luminosity-weighted abundance of haloes. However, while the galaxy shot noise can be measured on small enough scales, the halo shot noise is always either subdominant or multiplied by $U(k) \neq 1$, %not dominant on any scale, 
and it is therefore not straightforward to extract its value from the observation directly unless one does forward modelling. Nevertheless, the halo shot noise could be considered to be a free parameter that is a constant on scales where the 2-halo term dominates. The value of the halo shot noise 
is always larger than that of the galaxy shot noise by construction, thus the galaxy shot noise measured on small scales can provide a lower bound.

\subsubsection{Inferring the halo profile from the 1-halo term} \label{sec:halo_profile}

\begin{figure}
\centering
    \includegraphics[width=\linewidth]{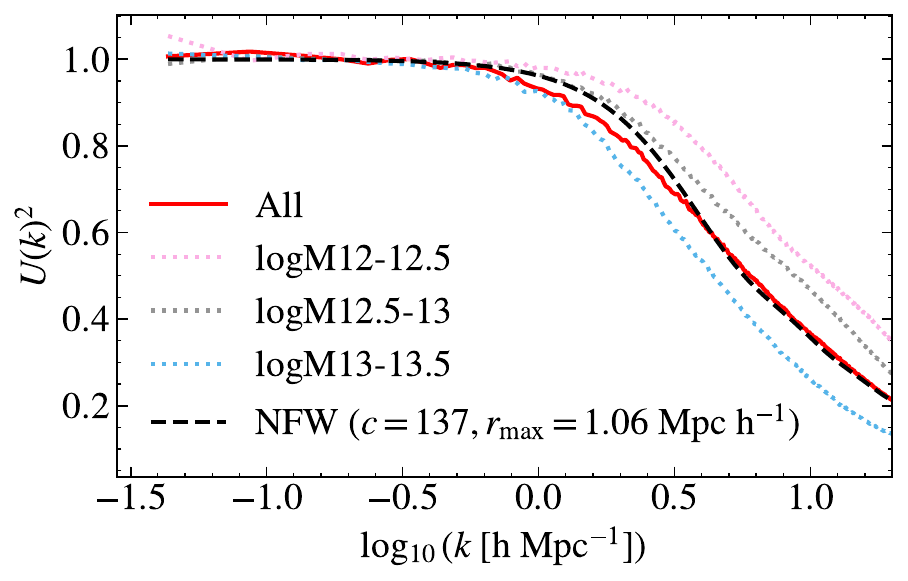}
\caption{Luminosity-weighted average of the square of the Fourier transform of the luminosity distribution within haloes at $z=1.5$, computed using \cref{eq:galaxy_ps_all}. The {\em solid red line} is computed from the power spectrum including all the haloes. The {\em pink}, {\em grey} and {\em blue dotted lines} show that computed using power spectra including haloes with mass in the ranges $\log M_{\mathrm{vir}}=12-12.5$, $\log M_{\mathrm{vir}}=12.5-13$ and $\log M_{\mathrm{vir}}=13-13.5$, respectively. The square of the Fourier transform of an NFW profile with parameters $c=137$, $r_{\mathrm{vir}}=0.5$ \lenunit\ and $r_{\mathrm{max}}=1.06$ \lenunit\ is shown by the {\em dashed black line}.
The $U(k)^2$ for the total power spectrum lies between that for $\log M_{\mathrm{vir}}=12.5-13$ and $\log M_{\mathrm{vir}}=13-13.5$.}
\label{fig:u_k}
\end{figure}

\begin{figure*}
\centering
\begin{subfigure}[t]{.5\textwidth}
    \centering
    \includegraphics[width=\linewidth]{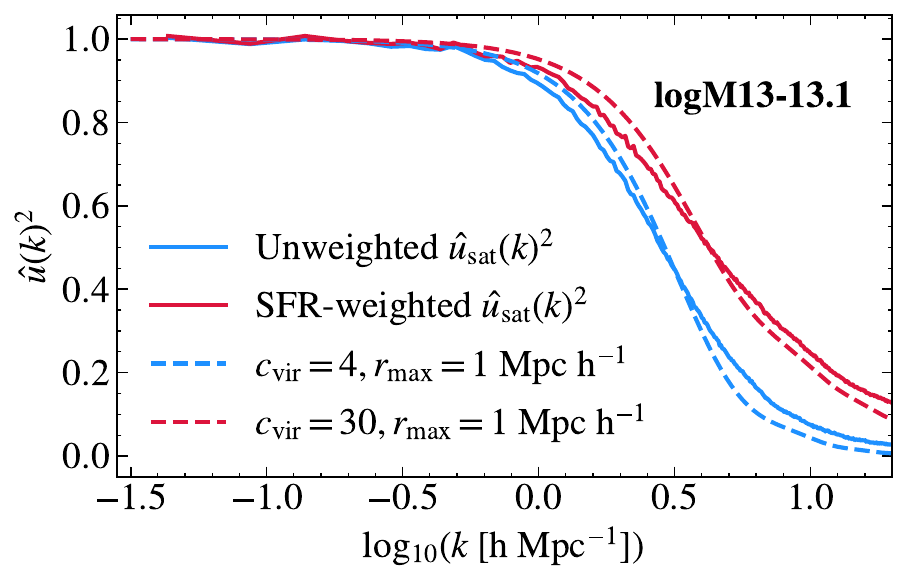}
\end{subfigure}%
\begin{subfigure}[t]{.5\textwidth}
    \centering
    \includegraphics[width=\linewidth]{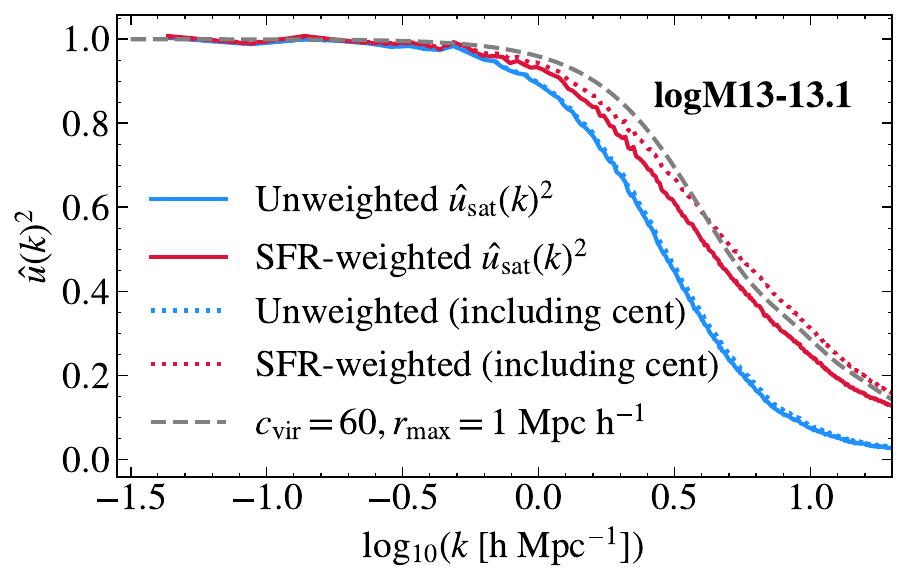}
\end{subfigure}
\caption{\textbf{\textit{Left panel}}: Square of the Fourier transform of the halo profiles, $\hat{u}_{\mathrm{sat}}(k)^2$, for satellites in haloes of mass $\log$ \mvir $\in [13, 13.1]$ in \tng\ at $z=1.5$ ({\em solid lines}). 
The {\em solid blue line} is the profile when satellites with non-zero \sfr\ are given equal weight, and the {\em solid red line} is for when satellites are weighted by their \sfr. The {\em dashed lines} are {\nfw} fits, with the legend stating the best-fitting {\nfw} parameters. The virial radius is given by $r_{\mathrm{vir}}=0.5$ \lenunit\ for all the \nfw\ fits. \textbf{\textit{Right panel}}: {\em Solid lines} are the same as in the left panel. The {\em dotted lines} show (the Fourier transform squared of) the halo profiles if the central galaxy is included as well. The {\em dashed grey line} is an {\nfw} fit to the {\em dotted red line} (the case where both the central galaxy and satellites are included and weighted by their \sfr). 
If an \nfw\ profile is used as a fitting function, a higher concentration than is typical for dark matter haloes of this mass is required to fit $\hat{u}_{\mathrm{sat}}(k)$ when galaxies are weighted by their \sfr\ ({\em red}), compared to the unweighted case ({\em blue}).
%Including the central galaxy has little effect in the unweighted case (i.e., the {\em dotted blue line} is nearly identical to the {\em blue solid line}), but the effect is greater in the weighted case (i.e., the {\em dotted red line} has a different shape from the {\em red solid line}).
}
\label{fig:u_k_logM13}
\end{figure*}

We demonstrated in \cref{sec:1halo_sec2} that the 1-halo term $P_{1h}^{\rm gal}(k)\to P^{\rm halo}_{\rm shot}$ for wavenumbers $k\to 0$, and $P_{1h}^{\rm gal}(k)\to P^{\rm gal}_{\rm shot}$ for wavenumbers $k\to \infty$. The shape of $P_{1h}^{\rm gal}(k)$ on intermediate scales depends on the average weight distribution of galaxies inside of the haloes. $P_{1h}^{\rm gal}(k)$ is plotted in red in
\cref{fig:components_ps}.

$\uparenth (k)$ for a given halo is related to the underlying halo profile that the galaxies in the halo are thought to be sampled from, as well as the correlation of weight with position (see \cref{app:profile_sampled_from} for explanation). 
Note that $\uparenth (k)$ differs from the actual distribution of \sfr s in a halo, which we define as $\uobserved$ in \cref{sec:1halo_sec2}.
$U(k)^2$ is a weighted average of the individual $\uparenth (k)^2$ profiles (Eq.~\ref{eq:big_uk2}).
To compute $U(k)$ from \cref{eq:galaxy_ps_all}, we use the values for $P^{\rm halo}_{\rm shot}$ and $P^{\rm gal}_{\rm shot}$ from the previous section, and the 2-halo term is computed by subtracting the halo shot noise from the halo power spectrum, as in \cref{sec:2halo_fit}. We plot
$U(k)^2$ measured in \tng\ at $z=1.5$ as the red line in \cref{fig:u_k}; 
$U(k)^2 \to 1$ for $k\to 0$ by construction.

$U(k)^2$ starts to deviate noticeably from 1 for $\log k \gtrsim -0.5$, signalling the scale below which the galaxy power spectrum starts to deviate from the halo power spectrum (Eq.~\ref{fig:subtract_shot_noises}). We can repeat the calculation, this time restricting it to galaxies in haloes within a narrow mass range. Although haloes of mass
$\log M_{\mathrm{vir}} \sim 12-12.5$ (pink dotted line) dominate the \sfrd, $U(k)^2$ computed for the case including all haloes is more similar to that for haloes with a slightly higher mass, $\log M_{\mathrm{vir}} \sim 13$.
This makes sense, since such haloes have significantly more star formation in satellites than the lower-mass haloes, as we showed in \cref{fig:contribution_mass_bins}.

As the 2-halo term for the galaxy power spectrum differs on small scales from the 2-halo term computed from the halo power spectrum, the shape of $U(k)^2$ may be less accurate in the range $ -0.5 \lesssim \log k \lesssim 0.5$. For $\log k \lesssim -0.5 $, $U(k)^2 = 1$, so they are equivalent. For $\log k \gtrsim 0.5 $, the 2-halo term is negligible due to halo exclusion, therefore the error has little impact on the total power spectrum on these scales.

$U(k)$ is a weighted average of the \sfr\ distribution in haloes. What does
this distribution look like for a single halo? The distribution of the {\em dark matter density} follows the {\nfw} profile \citep{nfw_1997},
\begin{align}
    \rho(r) = \frac{\rho_0}{r/r_s\,\left(1+r/r_s\right)^2}\,,
\end{align}
where $r_s = r_{\mathrm{vir}}/c$ is called the scale radius, $c$ the concentration of the halo, and $\rho_0$ is a normalisation that sets \mvir. As a first approximation, we might assume that the subhaloes that host satellites follow a similar profile. If, additionally, we ignore the fact that the contribution of a satellite is weighted by its \sfr, then the halo profile is simply the Fourier transform, $\uparent (k)$, of the \nfw\ profile -- apart from the overall normalisation. However, the distribution of subhaloes is not generally the same as that of the mass \cite[e.g.][]{Zavala19} and satellites {\em are} weighted by their \sfr.

The \sfr-weighted distribution of galaxies is not necessarily similar to the unweighted distribution of galaxies selected above a threshold in \sfr\ or specific \sfr\ (\ssfr; \sfr\ per unit stellar mass).
Galaxies with high \sfr\ or \ssfr\ tend to lie in the outskirts of haloes \citep{Orsi_2018,Rocher_2023}, thus deviating from the \nfw\ profile traced by the dark matter. \citet{Avila2020} suggest to account for this by using a lower concentration, while \citet{Reyes-Peraza_2024} provide a modified \nfw\ profile. It is unclear if a similar approach would work for the weighted distribution of \sfr\ in haloes, since weighting differs from galaxy number counts above a threshold.

Nevertheless, we start our investigation by assuming an {\nfw} profile, which has been used to model the distribution of galaxies in \lim\ \citep{Schaan+21-multi,MoradinezhadDizgah_precision_tests}. In addition to the parameters $r_s$ and $c$, there is a third parameter to consider: the radius to which we integrate when Fourier transforming the profile. This radius does not have to be the virial radius, $r_{\mathrm{vir}}$, and will be denoted by $r_{\mathrm{max}}$.

\Cref{eq:big_uk2} relates the halo profile of haloes,
$\uparent (k)$, to the function $U(k)^2$. Even if the $\uparent (k)$'s for all haloes were well-described by an {\nfw} profile, there is no reason that $U(k)$ would be well-described by that profile.  Nevertheless, we find fitting parameters $r_{\mathrm{vir}}=0.5$ \lenunit, $c=137$, and $r_{\mathrm{max}}=1.06$ \lenunit\ so that (the Fourier transform of) the {\nfw} profile roughly reproduces $U(k)$ from \tng: the fit is shown by the dashed black line in \cref{fig:u_k}.

The values for $r_{\mathrm{vir}}$ and $r_{\mathrm{max}}$ are not unexpected, but the best-fitting value of $c\sim 137$ for the concentration is unexpectedly high -- we would have expected a value of around 10 at most, based on typical halo mass-concentration relations (e.g. \citealt{Klypin_2016}). Why is such a high concentration required?
One factor is that the $\uparent (k)$ is contributed to by haloes of all masses, and therefore sizes, but we integrate the {\nfw} profile to $r_{\mathrm{max}}=1.06$ \lenunit. For many of the smaller haloes, there will be no star formation at larger radii, and this is instead accounted for by increasing the concentration.
Another factor is that the central galaxy is always at the \lq centre\rq\ of the halo. This galaxy contributes to
the galaxy power spectrum, $P^{\rm gal}_{\rm tot}$, and hence to $U(k)$, but it may be inappropriate to include it in the {\nfw} profile. Therefore, we next consider the profile $\usatp (k)$, which includes satellite galaxies only (Eq.~\ref{eq:1halo_usat} writes the 1-halo term in terms of $\usatp (k)$). See also \citet{McDonough_2022} for a comparison of the satellite distribution in \tng\ with {\nfw}-like profiles.

We compute the power spectrum of satellite galaxies in haloes in a narrow range of masses,
$\log M_{\rm vir}\in [13, 13.1]$ (at $z=1.5$), for both the case where the satellite galaxies are at their fiducial positions and for the case where their contribution is assigned to the centre of their respective haloes. We use these power spectra to compute $\usatp (k)$ (using Eq.~\ref{eq:galaxy_ps_all}). The $\usatp (k)$
for this case is plotted as solid lines in both panels of \cref{fig:u_k_logM13}: the blue line corresponds to weighting each satellite equally (provided the satellite's \sfr\ is non-zero), while the red line corresponds to weighting each satellite by its \sfr\ (the latter being relevant to \lim). The dashed lines in the left panel show {\nfw} profiles, with the legend including the best-fitting values of the {\nfw} parameters, and $r_{\mathrm{vir}}=0.5$ \lenunit\ for all the \nfw\ fits. The case of equal weighting (blue lines) yields a reasonable concentration of $c\sim 4$ (see, e.g.. \citealt{Klypin_2016, Child18} for measurements of the concentration-mass relation of dark matter haloes). However, weighting satellites by their \sfr\ (red lines) yields a much higher concentration of $c\sim 30$.

Note that we extrapolate the {\nfw} profile beyond the halo's virial radius
(we use $r_{\rm max}=1\ \lenunit$, which is much larger than the virial radius of the haloes in the mass range shown in Fig.~\ref{fig:u_k_logM13}).
If, instead, we Fourier transform the {\nfw} profile only up to the virial radius (0.5 \lenunit), we find that the resulting $\usatp (k)$ differs significantly from that measured from the simulation. This is because there are satellite galaxies with non-negligible \sfr\ outside the virial radii of haloes. 
Although the \nfw\ profile does not fit the distribution (of dark matter or galaxies) well outside the virial radius, integrating past the virial radius is still better than ignoring the contribution from outside the virial radius completely.
Including satellite galaxies outside the virial radius has similarly been found to reproduce the clustering of galaxies with high \ssfr\ in the {\sc desi} One-Percent survey better \citep{Rocher_2023}. Nevertheless, the radial distribution of galaxies with high \ssfr\ (or \sfr) (parametrised in terms of a modified \nfw\ profile; e.g., \citealt{Rocher_2023,Reyes-Peraza_2024}) differs from our fit for the distribution of galaxies weighted by their \sfr.
In terms of the dark matter profile, \cite{Zhou23, Zhou24} show that using the \cite{Einasto69} profile -- rather than the {\nfw} profile -- and extrapolating it beyond the halo's virial radius, improves the fit to the non-linear matter power spectrum obtained within the halo model. 
We further note that we used a single set of {\nfw} parameters for haloes of a given mass -- whereas it is known 
that there is scatter in concentration \citep{Bullock_2001} and some haloes may not follow the {\nfw} profile well \citep{Jing_2000}. Using the Einasto profile and accounting for scatter may improve the fits to the 1-halo term.

As mentioned previously, the best-fitting value of $c$ is significantly higher for the case where satellites
are weighted by their \sfr. This suggests that the \sfr s of satellites are not independent of their position in the halo. The increased concentration suggests that there is increased star formation occurring in galaxies near the centre of the halo compared to in its outskirts, relative to the dark matter. We also find that the \sfr s of distinct satellite galaxies are correlated with each other, a finding we discuss in \cref{app:profile_sampled_from}.

In the right panel of \cref{fig:u_k_logM13}, we plot both $\usatp (k)$, the halo profile of satellites only (solid lines), i.e. after removing the central galaxy, and $\uparent (k)$ (dotted lines), which includes the contribution of the central galaxy. Including the central galaxy does not change the profile much in the unweighted case (blue dotted compared to blue solid line), but the effect is more pronounced in the \sfr-weighted case (red dotted compared to red solid line in the right panel). Including the central galaxy increases the best-fitting value of the concentration from $c=30$ to $c=60$.
Since the central galaxy is always at the centre of the halo, it might be physically more meaningful to add a delta function to the profile to represent the central galaxy -- rather than increasing $c$.
We discuss this in \cref{app:profile_sampled_from}.

\ifSubfilesClassLoaded{%
  \bibliography{bibliography}%
}{}

\end{document} 

 \subsection{Model summary} \label{sec:sec5_summary}
 
We briefly summarise the total model for the galaxy power spectrum, $P^{\rm gal}_{\rm tot}$, that we introduced in Eq.~(\ref{eq:galaxy_ps_all}) and discussed in the previous sections. All equations are repeated in Table~\ref{table:summary}.

The galaxy power spectrum is the sum of a 2-halo term, $P^{\rm gal}_{2h}$, and a 1-halo term, $P^{\rm gal}_{1h}$. The 2-halo term, $P^{\rm gal}_{2h}$, captures the clustering of galaxies in distinct haloes. On large scales, it captures the clustering of the dark matter haloes themselves. We describe $P^{\rm gal}_{2h}$ as a biased version of the non-linear matter power spectrum, with bias $b$, on scales larger than that of haloes. On scales comparable to or smaller than the sizes of haloes, power is suppressed due to halo exclusion. Halo exclusion is modelled by assuming that the distribution of exclusion distances has a lognormal shape, parametrised by $\log_{10}(d_0)$ -- the logarithmic mean of the exclusion distance -- and $\sigma$ -- the logarithmic scatter around the mean. These two parameters define the function $F$, the probability of finding two haloes at distance less than $r$ (see Table~\ref{table:summary}).

The 1-halo term, $P^{\rm gal}_{1h}$, has three independent contributions. On scales much larger than $d_0$, $P^{\rm gal}_{1h}$ is constant, set by the {\em halo} shot noise. On scales much smaller than $d_0$, $P^{\rm gal}_{1h}$ is also constant, but set by the {\em galaxy} shot noise.
The smooth interpolation of $P^{\rm gal}_{1h}$ between its asymptotes is regulated by the mean profiles of satellites in haloes, when they are weighted by their \sfr. For haloes within a narrow range of masses, this profile can be fit with an {\sc nfw} profile, provided that ({\em i}) the profile is extrapolated to $r_{\rm max}$, a radius well beyond the virial radius of the halo, and ({\em ii}) the concentration $c$ of the {\sc nfw} profile is much higher than that of the matter density. 
As an alternative to introducing these constraints, we also present a more physically motivated fit in \cref{app:profile_sampled_from}, where we consider the position and weighting of satellites separately.
However, for simplicity, we adopt the {\sc nfw} profile (with the above constraints) in Table~\ref{table:summary}.
Note also that we use a single profile, despite the fact that the 1-halo term should depend on a weighted average of the individual profiles.

\renewcommand{\arraystretch}{1.5}
\begin{table*}
\centering
\begin{tabular}{ |p{4.5cm}||p{7cm}|p{1.5cm}|p{3cm}| }
 \hline
 \multicolumn{4}{|c|}{Model parameters for fitting the galaxy power spectrum, $P^\mathrm{gal}_{\mathrm{tot}}(k)$ } \\
 \hline
\multicolumn{4}{|c|}{$P^\mathrm{gal}_{\mathrm{tot}}(k) = P^\mathrm{gal}_{2h}(k, \uobserved) + U(k)^2(P_{\rm shot}^{\rm halo} - P^{\mathrm{gal}}_{\rm shot}) + P^{\mathrm{gal}}_{\rm shot}$ (Eq.~\ref{eq:galaxy_ps_all})}  \\ 
\hline
 2- and 1-halo terms & Component & Parameter & Value at $z\sim 1.5$ \\
 \hline
 2-halo term, $P^{\mathrm{gal}}_{2h}(k)$ &  
 \multirow{2}{\textwidth}{\shortstack[l]{\hbox{Halo exclusion}\\ $F(r)= \frac{1}{2}\left( 1 + {\rm erf}\left[ \frac{\log_{10} (r/d_0)}{\sqrt{2} \sigma}\right]\right)$ (Eq.~\ref{eq:erf}) }}
 
 &    $d_0$ & 0.74 \lenunit \\ 
$= 4\pi\int_0^\infty [F(r)(1 + \xi'(r)) -1] \frac{\sin(kr)}{kr} r^2 {\rm d}r$  &                & $\sigma$   & 0.17 dex\\
\cline{2-4}
(Eq.~\ref{eq:ps_exclusion_distribution})    &  
\multirow{2}{\textwidth}{\shortstack[l]{\hbox{Non-linear bias}\\ 
$\xi'(r) =  b^2 \xi_m (r)$
}}
&  $b$ & 2.0\\
    &                 & $\xi_m (r)$ & {\sc halofit}\\
\hline
 1-halo term, $P^{\mathrm{gal}}_{1h}(k)$   
 & \hbox{Halo shot noise}
 & $P_\mathrm{shot}^\mathrm{halo}$ &  304.5 \pkunit \\
$ 
    = U(k)^2(P_{\rm shot}^{\rm halo} - P^{\mathrm{gal}}_{\rm shot})
    + P^{\mathrm{gal}}_{\rm shot}$   
    &  \hbox{Galaxy shot noise}
    & $P_\mathrm{shot}^\mathrm{gal}$  & 130.5 \pkunit  \\
    \cline{2-4}
 (Eq.~\ref{eq:galaxy_ps_1halo})            & \multirow{2}{\textwidth}{\shortstack[l]{\hbox{Halo profile:} $U(k) = $\\ $\hat{u}_{\mathrm{NFW}}(k) \propto \int_0^{r_\mathrm{max}} ${\large $\frac{1}{(r/r_s)(1+r/r_s)^2}$} $\exp(-i\bm{k}\cdot \bm{r})\ \mathrm{d}^3r$}}   & $r_{\mathrm{max}}$   & 1.06 \lenunit\\
            &  &  $r_s$   & 0.0036 \lenunit \\
             &  &    & ($c = r_{\mathrm{vir}}/r_s = 137$ for $r_{\mathrm{vir}} = 0.5$ \lenunit) \\
 \hline
\end{tabular}
\caption{Parametrisation of the galaxy power spectrum, $P^\mathrm{gal}_{\mathrm{tot}}(k)$, in terms of a 2-halo and 1-halo term.
The 2-halo term is parametrised in terms of the bias, $b$, with respect to the non-linear matter correlation function, $\xi_m$ (which we compute using {\sc halofit}, \citealt{Takahashi_2012}), and the function $F(r)$ that captures halo exclusion.
The latter is parametrised by the mean exclusion distance, $d_0$, and the standard deviation $\sigma$ around the lognormal distribution of exclusion distances.
The 1-halo term depends on the halo and galaxy shot noise terms, and the function $u$ that describes how star formation is distributed in haloes, on average. A high concentration is required when fitting $u$ with an {\sc nfw} profile.
The values for all parameters in the last column are for the \tng\ simulation at redshift $z=1.5$.}
\label{table:summary}
\end{table*}
\renewcommand{\arraystretch}{1}

\ifSubfilesClassLoaded{%
  \bibliography{bibliography}%
}{}

\end{document} 

 \section{Summary and Conclusions}\label{sec:conclusion}

Line-intensity mapping (\lim) is an emerging technique for efficiently mapping the spatial distribution of galaxies
\citep[e.g.][]{Kovetz_2017}. \lim\ promises to measure this distribution on the very large scales that are still in the linear regime, enabling inferences about the primordial power spectrum and the nature of the initial perturbations. 
\lim\ surveys may also have sufficient spatial resolution to measure the distribution of galaxies on much smaller scales, and these measurements could be used to constrain theories of galaxy formation as well as some cosmology. 

In this paper, we investigate the power spectrum of galaxies when they are weighted by their star formation rate (\sfr). 
Our analysis is general but with a focus on redshift $z\sim 1.5$, an ideal target for \lim\ surveys based on the \ha\ line. We mostly analyse the 300 Mpc realisation of the \illustris\ simulation (\citealt{Pillepich_2018}; hereafter \tng), 
but also use the higher resolution 100 Mpc realisation as well as the \eagle\ simulation \citep{Schaye15}. 
Our aim is to fit the simulated power spectrum in terms of a set of parameters that have clear physical interpretations.

Galaxies in the simulations are labelled as centrals or satellites. This distinction is relevant because the galaxy power spectrum
depends on the spatial distribution of the \sfr\ of satellites in a halo on scales comparable to or smaller than that of haloes. 
We therefore write the total galaxy power spectrum as a sum of two terms: a 1-halo term -- quantifying how satellites are distributed in haloes -- and a 2-halo term -- which quantifies how haloes are clustered. This dissection is inspired by the halo model \citep{Cooray2002} for the total matter power spectrum.
We show that the 1-halo term can be written in terms of the galaxy and halo shot noises, and a parameter that describes the distribution of \sfr\ within haloes. The 2-halo term is written in terms of bias with respect to the non-linear matter distribution and a function describing halo exclusion.

Our main findings are as follows:
\begin{enumerate}
    \item The galaxy power spectrum, where galaxies are weighted by their \sfr, is fitted well by a model that incorporates halo exclusion and biased clustering of haloes relative to the non-linear matter power spectrum in the 2-halo term, and includes both shot noise and the spatial distribution of satellite galaxies within haloes in the 1-halo term.
    \item Satellite galaxies contribute $\sim$ 30 per cent to the total star formation rate density (\cref{fig:sfrd}) in \tng\ at $z\sim 1.5$. Neglecting their contribution
    leads to an underestimate of the galaxy power spectrum by $\sim$ 30 per cent on large scales (\cref{fig:halo_cent_ps}) -- even
    when the power spectrum is normalised by the mean \sfr. On average, the effect of satellite galaxies increases with decreasing redshift (\cref{fig:halo_cent_ps_z}). 
    \item Satellites dominate the \sfr\ in 
    haloes with virial mass greater than
    $\log M_{\mathrm{vir}}\ [\massunit]\sim 12.5$ (\cref{fig:contribution_mass_bins}). This is primarily due to two factors: the quenching of the \sfr\ of the central galaxy due to feedback from its accreting supermassive black hole ({\sc agn} feedback), and the increase in both the number and \sfr s of satellites with increasing halo mass.
    \item We test the impact of the spatial distribution of satellites on the power spectrum by summing the \sfr\ of all galaxies in a halo and assigning it to the position of the central galaxy.
    The resulting power spectrum differs by less than 1~per cent from the original power spectrum (in which haloes are resolved into central and satellite galaxies) on scales $\log_{10} (k\ [\kunit]) \lesssim -0.5$ (\cref{fig:gal_vs_halo_ps_z}). Therefore, 
    assigning a single luminosity to the centre of each halo is sufficient for reproducing the large-scale power spectrum,  
    unless one wants to probe the 1-halo term.    
    \item We fit the 2-halo term by taking into account halo exclusion and non-linear halo bias (\cref{fig:bias2}). Approximating the distribution of exclusion distances using a lognormal distribution works well.
    \item Shot noise -- arising from finite sampling of a distribution -- is scale-dependent: it is set by the \sfr s of galaxies on small scales, and the \sfr s of haloes on large scales (\cref{fig:components_ps}; see also \citealt{Baldauf_2013}). On the largest scales in the TNG300 box ($\log_{10} (k\ [\kunit])\sim -1.5$), the shot noise is still not negligible, therefore care should be taken when making the assumption that the shot noise is negligible on ``large scales''. Rather than treating shot noise as separate from the 1-halo term, we demonstrate that it is more physically intuitive to consider the amplitude of the 1-halo term as being set by shot noise. Specifically, the 1-halo term approaches the halo shot noise on large scales and the galaxy shot noise on small scales, with its shape determined by the distribution of star formation inside haloes.
    \item We investigated the effect of weighting satellites by their \sfr\ and including the \sfr\ of the central galaxy -- as appropriate for \lim\ modelling -- on the halo profile. Although the halo profile can be approximately fitted by an {\sc nfw} profile, that profile should be extrapolated beyond the halo's virial radius, and the best-fitting value of the concentration is much higher than that of the dark matter profile (\cref{fig:u_k_logM13}).
\end{enumerate}

\subsection{Limitations of modelling}

We have made several approximations that could be improved upon. We have only considered a linear relation between \sfr\ and line luminosity, yet the luminosity of emission lines can also depend on other properties, such as metallicity. For example, satellites can have higher metallicity than central galaxies \citep{Bahe_2017}, which can affect luminosities of emission lines such as [\ion{O}{III}], in which case satellites should be considered even more carefully. 
Nevertheless, the underlying components of the model should be similar for lines with luminosity not proportional to the \sfr, provided they are sourced from galaxies, as the backbone of the model is motivated by the clustering of galaxies and haloes.

{\sc agn} feedback may suppress the \sfr\ of central galaxies, but they themselves can also contribute to line luminosities \citep[e.g.][]{Favole_2024} -- an effect not accounted for in this study. It has been found, in galaxy surveys \citep{Sobral_2016} and in \tng\ \citep{Hirschmann_2023}, that the contribution from \agn{}s is most significant at the high-luminosity end of the luminosity function. This is also the range that will be well-probed by galaxy surveys such as {\sc euclid} \citep{Ballardini24} and the {\emph Roman Space Telescope} \citep{Wfirst_2019}. These surveys can therefore serve to identify bright \agn\ \citep{Silva+2017}.
The contribution from lower luminosity \agn{}s is thought to be less significant \citep{Hirschmann_2023}, but the impact of \agn{}s in \lim\ is worth investigating in more detail.

The observed power spectrum will be affected by redshift-space distortions ({\sc rsd}). For example, the motions of satellite galaxies in large haloes cause the so-called Fingers-of-God effect \citep{Jackson1972} on small scales. 
In this paper, we focus on understanding the contributions to the power spectrum from the \sfr\ and clustering of galaxies, excluding additional complexities due to {\sc rsd}. However, for interpreting observations, the effect of {\sc rsd} will need to be accounted for. For a model that incorporates {\sc rsd} in \lim, see \citet{Schaan+21-multi}.
Instrumental noise and line interlopers are other contributions to the power spectrum we have not considered here (see, e.g., \citealt{Fonseca_2017} for the effect of instrument noise on the error and for the contributions from different lines). Possible ways to mitigate them include enhancing instrument sensitivity or applying theoretical approaches \citep[e.g.][]{Cheng_2020,Moriwaki_2021_noise, Qezlou_2023}.

On the largest scales probed by TNG300 (linear extent of 205 \lenunit), sample variance limits our ability to confirm the existence of a constant bias.
Studies have suggested that the halo bias may be non-linear even on larger scales \citep{Penin_2018, MoradinezhadDizgah_precision_tests}, and could impact large-scale constraints, such as those from BAO. Therefore accounting for the scale-dependence of the bias will be important in future work.
Our parametrisation for the galaxy power spectrum can be applied to other cosmological hydrodynamical simulations as well as to other redshifts, in which case, a model for the scale-dependence of the halo bias relative to the matter power spectrum needs to be used.

\subsection{Future prospects}

This work provides a flexible and physically-motivated framework for modelling the weighted galaxy power spectrum, relevant for \lim, effectively capturing nonlinearities and enabling analysis of power spectrum data from upcoming surveys.

While we anticipate the planning of large-scale \lim\ surveys, we should maximise the value of the data from smaller scale surveys, which are more feasible in the near-term. 
The non-linear regime provides us with information about structure formation, and can provide tests for cosmological models. 
In {\sc COMAP} season 2 results \citep{Chung_2024}, for instance, the shot noise and bias of the CO power spectrum have been constrained by fitting the obtained upper limit of the power spectrum on scales $-1 \lesssim \log_{10} (k\ [\kunit]) \lesssim 0$. In such analyses, one can take into account the non-linear effects discussed in this study to better extract physical information from the data.

Additionally, for surveys such as SPHEREx, where the spectral resolution is low, the 2D power spectrum becomes relevant. The 2D power spectrum can be obtained by integrating the 3D power spectrum along the line of sight \citep[see, e.g.,][]{Kaiser91,Einasto_2020}. As this integration includes contributions from small scales in the 3D power spectrum, the small-scale effects discussed in this paper could also be important for large-scale 2D power spectrum predictions relevant for upcoming surveys.

\ifSubfilesClassLoaded{%
  \bibliography{bibliography}%
}{}

\end{document}

\section*{Acknowledgements}

RLJ has been supported by The University of Tokyo Fellowship. KM acknowledges JSPS KAKENHI Grant Number 23K03446, 23K20035, and 24H00004. SB is supported by the UK Research and Innovation (UKRI) Future Leaders Fellowship (grant number MR/V023381/1).
This work used the DiRAC@Durham facility managed by the Institute for Computational Cosmology on behalf of the STFC DiRAC HPC Facility (www.dirac.ac.uk). The equipment was funded by BEIS capital funding via STFC capital grants ST/K00042X/1, ST/P002293/1, ST/R002371/1 and ST/S002502/1, Durham University and STFC operations grant ST/R000832/1. DiRAC is part of the National e-Infrastructure. 
Python packages used include \texttt{matplotlib} \citep{Hunter_2007}, \texttt{numpy} \citep{numpy_2020} and \texttt{corrfunc} \citep{corrfunc_proceedings,corrfunc_mnras}.

\section*{Data Availability}
The \illustris\ data are publicly available at \url{https://www.tng-project.org/}.
The \eagle\ data are publicly available at \url{https://icc.dur.ac.uk/Eagle/}.
We will share the scripts used in this paper upon reasonable request.

\bibliography{bibliography}

\appendix

\crefalias{section}{appendix}

%\onecolumn

\section{Numerical details for calculating simulation power spectra}\label{app:interpolation}

We use \nbodykit\ \citep{nbodykit} with {\sc python} version 3.6 for computing the power spectrum of galaxies from a simulation. This involves interpolating the weight associated with discrete galaxies to a regular mesh and then computing the power spectrum using a discrete Fourier transform. The largest scale computed for the power spectrum is limited by the size of the simulation box, while the smallest scale depends on the resolution of the survey.

How the weight is assigned to the mesh may affect the power spectrum on scales of a few mesh cells. In this paper, we adopt the triangular-shaped cloud interpolation scheme and correct for interlacing. The fiducial resolution of the mesh is taken to be approximately the {\sc SPHEREx}
linear resolution at $z=1.5$ (corresponding to an angular resolution of 6.2 arcsec), which is 
$\sim$ 0.091 \lenunit. This corresponds to using a mesh with $N_{\rm mesh}^3$ cells, with $N_{\rm mesh}=2248$, for
the \tng\ simulation with linear extent 
$L_{\rm mesh}=205$~\lenunit. 
The Nyquist frequency is $k_{\mathrm{Nyq}} = \pi N_{\mathrm{mesh}}/L_{\mathrm{mesh}} \approx 34.5$ \kunit. In the main text, we show all plots to $\log_{10} (k\ [\kunit]) \sim 1.3$, which is approximately half the Nyquist frequency. 
The small-scale power in the observations will additionally depend on the point spread function (see, e.g., \citet{Symons_2021} for {\sc SPHEREx}).

\subsection{Aliasing effects and interlacing}
Aliasing effects become important close to the Nyquist frequency. The effect of aliasing is discussed in detail in \citet{Jing_2005}. We adopt the interlacing option provided by \nbodykit\ to correct for them. An explanation of this technique is detailed in Section 3.1 of \citet{Sefusatti_2016}. 

\begin{figure}
\includegraphics[width=\linewidth]{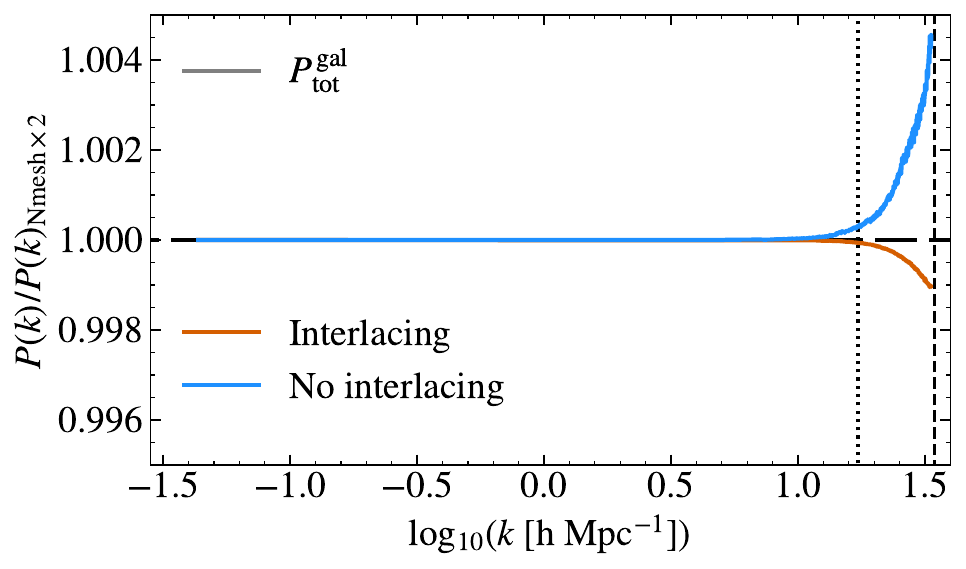}
\centering
\caption{Ratio of the power spectra computed on a mesh with the fiducial resolution relative to that on a mesh with twice the resolution. 
The {\em orange lines} represent the power spectrum where interlacing has been used, whereas for the {\em blue points} interlacing has not been used. The {\em dashed vertical line} represents the Nyquist frequency, $k_{\mathrm{Nyq}} = \pi N_{\mathrm{mesh}}/L_{\mathrm{box}}$, for the fiducial resolution and the dotted vertical line represents $k_{\mathrm{Nyq}}/2$.}
\label{fig:tsc_interlacing}
\end{figure}

\Cref{fig:tsc_interlacing} shows the effect of interlacing. 
The reference power spectrum, $P_{\mathrm{Nmesh \times 2}}$, uses $N_{\rm mesh}=2\times 2248$ -- {i.e.} twice our fiducial resolution.
The Nyquist frequency for the reference
mesh is higher, so aliasing effects are less significant at the Nyquist frequency of the fiducial mesh, therefore the reference power spectrum is closer to the \lq true\rq\ power spectrum.
The figure shows that the fiducial resolution reproduces the reference power spectrum better when the interlacing option is selected (orange line), than when it is not used (blue line). We therefore always use the interlacing option by default.

\subsection{Effect of the mass-assignment scheme}
There exists a hierarchy of schemes for assigning the mass (or weight) of point particles to a mesh \cite[see, {e.g.},][]{Hockney&Eastwood1981}. Lower-order schemes (for example {\bf N}earest{\bf G}rid {\bf P}oint, {\sc ngp}) are numerically faster than higher-order schemes (such as {\bf C}loud {\bf I}n {\bf C}ell, {\sc cic}, or {\bf T}riangular-{\bf S}haped {\bf C}loud, {\sc tsc}).
Higher order interpolation schemes provide more accurate results at the expense of higher computational cost
\citep{Hockney&Eastwood1981,nbodykit}.

\begin{figure}
\includegraphics[width=\linewidth]{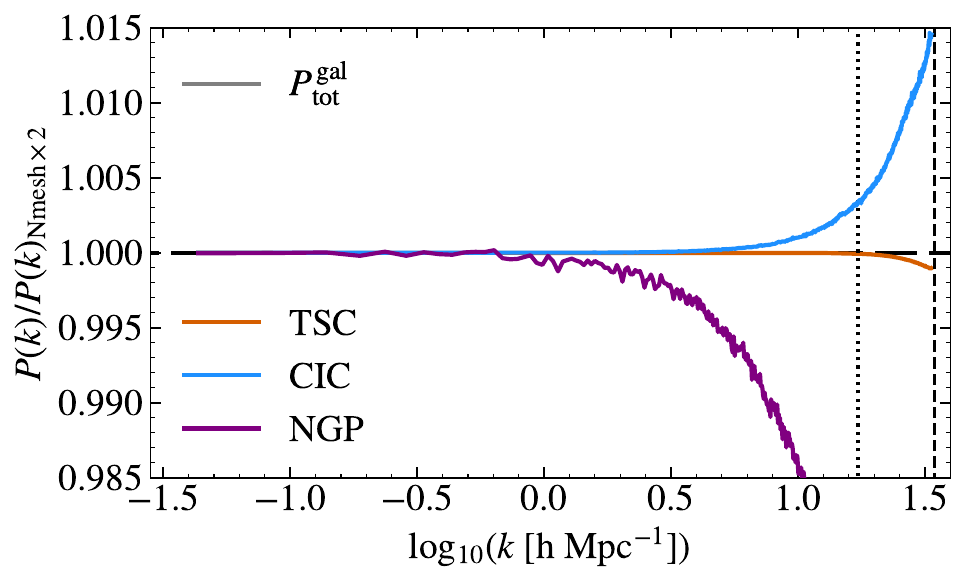}
\centering
\caption{Ratio of the power spectra computed on meshes with the fiducial resolution, interpolated using the {\sc tsc} ({\em orange}), {\sc cic} ({\em blue}) and {\sc ngp} ({\em purple}) assignment schemes relative to a reference power spectrum computed on a mesh with twice the resolution. 
Interlacing has been applied in the computation of all the power spectra. Except for the {\sc ngp} case, deconvolution has also been applied.
The vertical lines are the same as in \cref{fig:interp}.
In this paper we adopt the {\sc tsc} scheme with interlacing and kernel deconvolution.
}
\label{fig:interp}
\end{figure}

\Cref{fig:interp} shows the ratio of the power spectra obtained from the {\sc ngp}, {\sc cic} and {\sc tsc} schemes on meshes with the fiducial resolution
($N_{\rm mesh}=2248$) relative to the reference power spectrum ($N_{\rm mesh}=4496$) using {\sc tsc}. At $k_{\mathrm{Nyq}}$ of the fiducial mesh (vertical dotted black line), the {\sc cic} and {\sc tsc} schemes differ from the reference power spectrum by less than 1.5 and 0.1 per cent respectively.
The {\sc ngp} scheme shows a much larger difference of more than 40~per cent.
Even at $k_{\mathrm{Nyq}/2}$, the {\sc ngp} scheme still differs by $\sim 15$~per cent from the reference result, whereas {\sc cic} and {\sc tcs} differ by only 0.2 and 0.1~per cent.
In this paper, we adopt the {\sc tsc} interpolation scheme. Even higher order schemes such as the PieceWise Cubic Spline ({\sc pcs}) can be adopted if higher accuracy is required.

{\sc cic} or {\sc tsc} interpolation can be thought of as a convolution of the particle density field (a sum of Dirac delta functions) with a smooth assignment kernel. {\sc nbodykit} deconvolves the density field with this kernel when selecting the option 
\texttt{compensated = True}. We apply deconvolution throughout this paper.

\ifSubfilesClassLoaded{%
  \bibliography{bibliography}%
}{}

\end{document}

\section{Halo exclusion for a single exclusion distance}\label{app:halo_exclusion}

Here, we derive the power spectrum for the case where we assume all haloes are of the same size $R$, such that the minimum distance between them is $D=2R$ (see also, e.g., \citealt{Baldauf_2013}).
In this case, the correlation function is given by
\begin{equation}\label{eq:exclusion_xi_1R}
  \xi(r) =
    \begin{cases}
        -1 & \hbox{\rm when } r < D \\
        \xi'(r) & \hbox{\rm when } r\gtrsim D,
    \end{cases} 
\end{equation}
where 
$\xi'(r)$ describes the clustering of haloes for scales larger than the exclusion distance, $D$. 
The correlation function given by \cref{eq:exclusion_xi_1R} corresponds to the definition provided in \cref{eq:cf_delta}, with self-pairs excluded.

If $\xi'(r)$ is defined for $r < D$, then it could be thought of as the \emph{hypothetical} correlation function in the absence of halo exclusion.
From the observed correlation function $\xi(r)$, as given by \cref{eq:exclusion_xi_1R}, it is not possible to infer what the correlation function would have looked like below the distance $D$ in the absence of halo exclusion.

Fourier transforming \cref{eq:exclusion_xi_1R} gives the power spectrum that takes into account halo exclusion for the case of a single exclusion distance $D$:
\begin{align} 
	P_{2h}(k) &= \int_0^\infty  \xi(r) \exp (-i\bm{k}\cdot \bm{r}) {\rm d}^3 r \nonumber \\
            & =  4\pi\int_0^\infty  \xi(r) \frac{\sin(kr)}{kr} r^2 {\rm d}r \nonumber \\
            &=  -4\pi\int_0^D \, \frac{\sin(kr)}{kr} r^2 {\rm d}r
            + 4\pi\int_D^\infty\, \xi'(r) \frac{\sin(kr)}{kr} r^2 {\rm d}r,
\end{align}
where the simplification to the second line assumes spherical symmetry.

\section{Relating the 1-halo term to the halo profile}

\subsection{Derivation of the 1-halo term}
\label{app:halo_profile}

\Cref{eq:1halo_discrete_u} is the 1-halo term in terms of the measured halo profile, $\hat{v}^{(s)}_h(k)$, i.e. the Fourier transform of the actual (discrete) galaxy distribution. It is common to assume that galaxies are sampled from an underlying smooth density distribution, such as the halo matter distribution with Fourier transform ${\hat u}(k)$. The functions 
$\hat{v}^{(s)}_h(k)$ and ${\hat u}(k)$ differ by the galaxy shot noise. To see how this comes about,
we separate the 1-halo term into self-pairs and distinct pairs, where the self-pairs term gives rise to the shot noise:
\begin{align}\label{eq:1halo_divide}
	&P_{1h,w}(k) \nonumber\\
 &= \frac{1}{V} \braket{\sum_{\h } \sum_{\gj\in \h } \sum_{\gl \in \h } W\subj W\subl  \, \exp[-i\bm{k}\cdot(\bm{r}\subj -\bm{r}\subl )]} \nonumber\\
  &= \frac{1}{V} \Braket{\sum_{\h } \sum_{\gj \in \h } \left(\sum_{\substack{\gl \in \h  \\ \gl\neq \gj}}+ \sum_{\substack{\gl \in \h  \\ \gl= \gj}} \right)W\subj W\subl  \, \exp[-i\bm{k}\cdot(\bm{r}\subj -\bm{r}\subl )]} \nonumber\\
	&= \frac{1}{V} \Braket{\sum_{\h } \sum_{\gj \in \h } \sum_{\substack{\gl \in \h  \\ \gl\neq \gj}} W\subj W\subl  \, \exp[-i\bm{k}\cdot(\bm{r}\subj -\bm{r}\subl )]} \nonumber\\
	& + \frac{1}{V} \Braket{\sum_{\h } \sum_{\gj \in \h } W\subj ^2},
\end{align}
where the outer sum with index $h$ is over haloes, and the product of sums with indices $g$ and $g'$
are over galaxies in halo $h$. The second term in the last line arises from the case where $g=g'$ and is the shot noise term. To describe the $g\neq g'$ term, we define
\begin{align}\label{eq:little_u_p}
    |\uparent_h (k)|^2
    &= \frac{\Braket{\sum_{\gj \in \h } \sum_{\gl \neq \gj \in \h } W\subj W\subl  \, \exp[-i\bm{k}\cdot(\bm{r}\subj -\bm{r}\subl )]}}{\sum_{\gj \in \h }\sum_{\gl \neq \gj \in \h } W\subj W\subl}\,.
\end{align}
The relation between this term and the halo profile is explained in \cref{app:profile_sampled_from}.
This allows us to rewrite the first term of Eq.~(\ref{eq:1halo_divide}) as
\begin{align} \label{eq:1halo_divide2}
    &\frac{1}{V} \sum_{\h } \sum_{\gj \in \h } \sum_{\substack{\gl \in \h  \\ \gl\neq \gj}} W\subj W\subl  |\uparent (k)|^2 \nonumber \\
    &=\frac{1}{V} \sum_{\h } \sum_{\gj \in \h }\left(\sum_{\gl \in \h } - \sum_{\substack{\gl \in \h  \\ \gl= \gj}}\right) W\subj W\subl  |\uparent (k)|^2 \nonumber \\
    &=\frac{1}{V} \sum_{\h } \left(\sum_{\gj \in \h }W\subj \sum_{\gl \in \h }W\subl  - \sum_{\gj \in \h }W\subj ^2\right)  |\uparent (k)|^2 \nonumber \\
    &=\frac{1}{V} \sum_{\h } \left(W_{\h }^2 - \sum_{\gj \in \h } W\subj ^2\right)  |\uparent (k)|^2.
\end{align}
This result motivates us to define
\begin{align}
    U(k)^2 &\equiv \frac{\sum_{\h } (W_{\h }^2 - \sum_{\gj \in \h } W\subj ^2) |\uparent (k)|^2}{\sum_{\h } (W_{\h }^2 - \sum_{\gj \in \h } W\subj ^2)}.
\end{align}
Substituting this into \cref{eq:1halo_divide2} gives
\begin{align} \label{eq:1halo_distinct_derivation}
    &\frac{1}{V} \sum_{\h } \left(W_{\h }^2 - \sum_{\gj \in \h } W\subj ^2\right)  U(k)^2 \nonumber \\
    &=\frac{1}{V} \left(\sum_{\h } W_{\h }^2 - \sum_{\gj \in G}W\subj ^2\right)  U(k)^2 \nonumber \\
    &= (P_{\text{shot},I}^{\rm halo} - P_{\text{shot},I}^{\mathrm{gal}}) U(k)^2,
\end{align}
where
\begin{align}
    P_{\text{shot},I}^{\rm halo} &= \frac{1}{V} \sum_h W_{\h }^2
\end{align}
is the {\it halo} shot noise.
\Cref{eq:1halo_distinct_derivation} gives the first term in \cref{eq:galaxy_ps_1halo}.

\subsection{Effect of weighting on the halo profile contribution to the 1-halo term}
\label{app:profile_sampled_from}

We stated in \cref{app:halo_profile} that the function $\uparent (k)$ given by \cref{eq:little_u_p} can be related to the halo profile. Here, we investigate in what way they are related.

First, we assume that satellites in the same halo are uncorrelated with each other. This simplifies Eq.~(\ref{eq:little_u_p}) to
\begin{align}\label{eq:independent_uk}
    &|\uparent_h (k)|^2 \nonumber \\
    &= \frac{\sum_{\gj \in \h } \sum_{\gl \neq \gj \in \h } \Braket{W\subj  \, \exp[-i\bm{k}\cdot \bm{r}\subj ]} \Braket{W\subl \exp[i\bm{k}\cdot \bm{r}\subl ]}}{\sum_{\gj \in \h }\sum_{\gl \neq \gj \in \h } W\subj W\subl }.
\end{align}
Next, we assume that the weight and position of a satellite are uncorrelated, yielding
\begin{align}
    |\uparent_h (k)|^2 \nonumber
    =& \frac{\sum_{\gj \in \h } \Braket{W\subj}\Braket{\exp[-i\bm{k}\cdot \bm{r}\subj ]}}{\sum_{\gj \in \h }\sum_{\gl \neq \gj \in \h } W\subj W\subl} \nonumber \\
    &\times \sum_{\gl \neq \gj \in \h }\Braket{W\subl}\Braket{\exp[i\bm{k}\cdot \bm{r}\subl ]}\nonumber\\
    =& \frac{\sum_{\gj \in \h } \sum_{\gl \neq \gj \in \h } \Braket{\exp[-i\bm{k}\cdot \bm{r}\subj ]} \Braket{\exp[i\bm{k}\cdot \bm{r}\subl ]}}{N_g(N_g - 1)}\,,
\end{align}
where $N_g$ is the number of galaxies in halo $h$. Not surprisingly, $\uparent (k)$ no longer depends on the weights.
Under these assumptions, $\uparent (k)$ should be the Fourier transform of the mean satellite distribution.

We find that weighting the galaxies by their \sfr\ changes the shape of $\uparent (k)$ in \tng:
the weight and position {\em are} correlated in \tng. In that case, we can rewrite \cref{eq:independent_uk} by partitioning the sum into bins split by position as follows:
\begin{align}\label{eq:weight_binned}
    |\uparent_h(k)|^2 &= \frac{\sum_{R_p} \sum_{\gj \in R_p} \Braket{W_{R_p}} \Braket{\exp[-i\bm{k}\cdot \bm{r}\subj ]}}{\sum_{\gj}\sum_{\gl \neq \gj} W\subj W\subl} \nonumber \\
    &\quad \times \sum_{R_q}\sum_{\gl \neq \gj \in R_q} \Braket{W_{R_q}} \Braket{\exp[i\bm{k}\cdot \bm{r}\subl ]}.
\end{align}
where $R_p$ denotes the radial bin $R_{p} = [r_p, r_p+dr]$.  We have assumed that within each bin, the weights are sampled independently. In this case, the weighted $\uparent (k)$ can be considered to be the same as multiplying the number density profile by the mean weight distribution. 

Our motivation for rewriting $\uparent (k)$ is that we want to separate how satellites are distributed spatially from how their weights are distributed. For example, the spatial distribution might follow an \nfw\ profile, but their weights are position dependent -- so that $\uparent (k)$ is {\em not} (the Fourier transform of) an \nfw\ profile. 

We want to test whether multiplying an \nfw\ profile by the mean weight distribution as a function of radius can properly predict the weighted $\uparent (k)$. In the tests in this appendix, we consider \tng\ but with galaxies outside the virial radii of the haloes removed, so that there is no additional complication due to the {\sc fof} halo being slightly different from the spherical overdensity halo.

\begin{figure*}
\centering
\begin{subfigure}[t]{.5\textwidth}
    \centering
    \includegraphics[width=\linewidth]{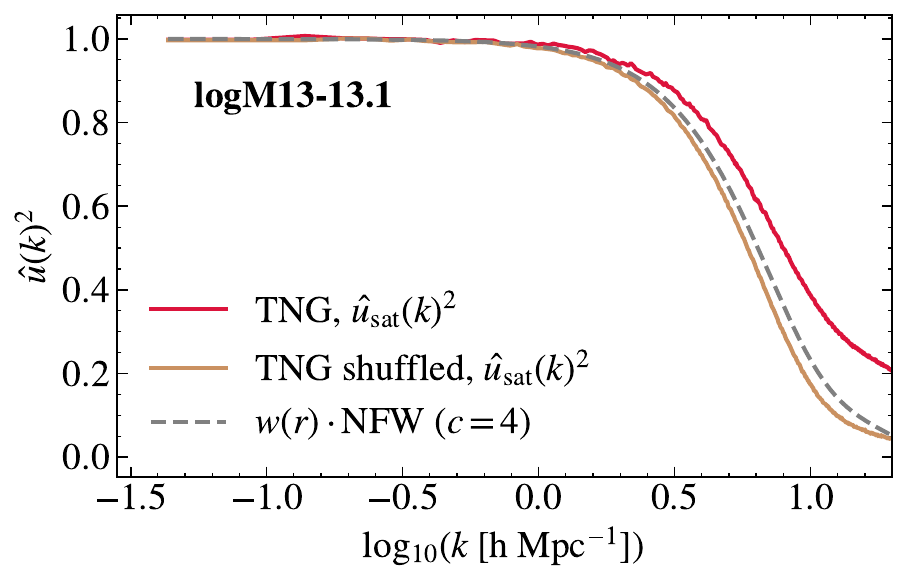}
\end{subfigure}%
\begin{subfigure}[t]{.5\textwidth}
    \centering
    \includegraphics[width=\linewidth]{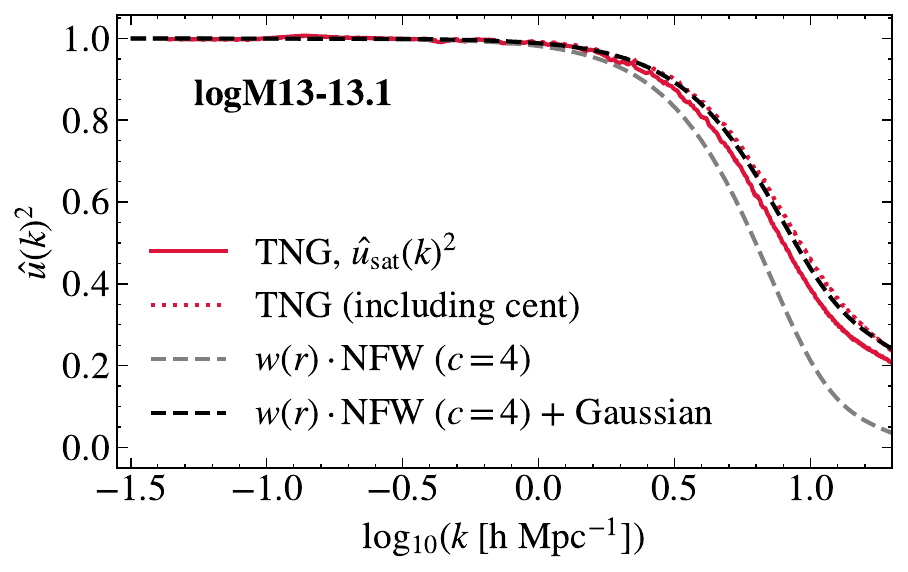}
\end{subfigure}
\caption{\textbf{\textit{Left panel}}: The {\em solid red line} is the weighted halo profile, $\hat{u}(k)^2$, computed using the satellite galaxies in haloes with virial mass in a narrow range around $\sim 10^{13}M_\odot$ in the \tng\ simulation. The {\em brown line} is also from the simulation, but we have randomly reassigned satellites to these same haloes. The {\em grey line} is the model, in which satellites are distributed according to an \nfw\ profile with concentration $c=4$, and assigned the position-dependent average weight as measured in \tng. The {\em red} and {\em brown lines} differ substantially, signalling that there are correlations between the \sfr s of satellites in the same halo.
The model describes the shuffled case very well. 
\textbf{\textit{Right panel}}: The {\em solid red} and {\em dashed grey lines} are the same as in the left panel. The {\em dotted red line} shows the $\hat{u}(k)^2$ for the original \tng\ simulation where the central galaxy has been included. This is fitted well when adding a Gaussian function to the centre of the halo in addition to the weighted \nfw\ profile represented by the {\em dashed grey line}.
}
\label{fig:u_k_physical}
\end{figure*}

The $\uparent (k)^2$ corresponding to this weighted \nfw\ profile is shown by the dashed grey line in \cref{fig:u_k_physical}. We use the same \nfw\ parameters as that corresponding to the dashed blue line in the left panel of \cref{fig:u_k_logM13} ($c=4$, $r_{\mathrm{vir}}=0.5$ \lenunit, $r_{\mathrm{max}}=1$ \lenunit). This \nfw\ profile describes the number density profile, while the weight distribution takes into account the correlation of the \sfr s with distance from the centre of the halo. The weighted $\uparent (k)$ measured in TNG is shown by the red line, indicating that it still deviates from the prediction (dashed grey line). However, we find that by shuffling the satellites between haloes, we can reproduce the prediction. The shuffled case is shown by the brown line in the left panel of \cref{fig:u_k_physical}, which is close to the dashed grey line, which represents the prediction given by the \nfw\ profile multiplied by the mean weight distribution.
The fact that the original \tng\ $\uparent (k)$ (the solid red line in the left panel of \cref{fig:u_k_physical}) deviates from the prediction suggests that the \sfr s of satellites within the same halo are correlated with each other, indicating that the assumption of independent sampling does not fully hold. Closer examination shows that the difference between the unshuffled and shuffled cases is mostly caused by a few haloes with highly star-forming satellites close to the centre.

In the right panel of \cref{fig:u_k_physical}, we investigate the effect of the central galaxy: the red solid line is \tng\ without central galaxies, while the dotted line is \tng\ with the central galaxy included.
The difference between the two red lines is not very large, implying that the central galaxy has little impact on the profile, at least for this halo mass. The dashed grey line is repeated from the left panel and is an \nfw\ profile with $c=4$, multiplied by the average weight of galaxies as a function of radius - the function $w(r)$. This profile deviates a lot from the measured profile. We try to account for central star formation -- either from a central, or from highly star-forming satellites near the centre -- by adding a Gaussian profile to the \sfr-weighted model: this yields the black dashed line.
This improved model reproduces \tng\ quite well.
\Cref{table:physical_fit} summarises the model parameters.

In \cref{sec:halo_profile}, we found that the unweighted power spectrum resembles the Fourier transform of an {\sc nfw} profile with concentration typical of the dark matter profile for haloes of that mass, but applying weighting changes $\uparent (k)$. To fit the weighted $\uparent (k)$, we used an {\sc nfw} profile with high concentration, which seems somewhat unphysical. In this appendix, we have provided an alternative fit which can be decomposed into the contribution from the number density distribution, the correlation of weight with distance from centre, and the contribution from the central galaxies. Using this fit may allow a more physical interpretation of how the star formation is distributed in haloes.

\renewcommand{\arraystretch}{1.5}
\begin{table*}
\centering
\begin{tabular}{ |p{3cm}|p{6cm}|p{1.5cm}|p{5.5cm}| }
 \hline
 \multicolumn{4}{|c|}{Model parameters of 1-halo term} \\
 \hline
 Model component & Equation of component & Parameter & Value at $z\sim 1.5$ \\
\hline
\multirow{2}{\textwidth}{Number density profile} & \multirow{2}{\textwidth}{NFW $u_{\mathrm{NFW}}(r) \propto $\LARGE{$\frac{1}{(r/r_s)(1+r/r_s)^2}$}}& $r_s$  & 0.125 \lenunit \\
     &    &  & ($c = r_{\mathrm{vir}}/r_s = 4$ for $r_{\mathrm{vir}} = 0.5$ \lenunit) \\
         \hline
   &    &  $a$   & 4.8\\
  Weight profile  &   \large{$w(r) = a \exp (-b r) + c$}                            &  $b$   & $4.5$ \unit{h\, Mpc^{-1}}\\
   &                               &  $c$  & 1 \\
    \hline
 \multirow{2}{\textwidth}{Central profile}  &        \multirow{2}{\textwidth}{Gaussian  $G(r) = $ \large{$\frac{1}{\sqrt{2\pi}\sigma}\exp\left[-\frac{1}{2}(\frac{r-\mu}{\sigma})^2\right]$}}       &  $\sigma$  & 0.001 \lenunit \\
  &                                   & $\mu = 0$ & \\
       \hhline{|=|=|=|=|}
 Total profile &        $u(r) = A \cdot w(r) \cdot u_{\mathrm{NFW}}(r) + (1-A) \cdot G(r)$                              & $A$ & $2 \times 10^{-4}$\\         
  &        \large{$\hat{u}(k) = \int_0^{r_\mathrm{max}} u(r) \exp(-i\bm{k}\cdot \bm{r})\ \mathrm{d}^3r$}          & $r_{\mathrm{max}}$ & 0.5 \lenunit \\   
 \hline
\end{tabular}
\caption{Halo profile, $\hat{u}(k)$, broken down into the contribution from the number density distribution of satellites, given by $u_{\mathrm{NFW}}(r)$, the average weight distribution as a function of radius, $w(r)$, and the contribution from the central galaxy or satellites near the centre, modelled using a Gaussian function, $G(r)$. While the simulated $\hat{u}(k)$ may be approximated using a single \nfw\ profile (but with very high concentration), the fit for $\hat{u}(k)$ presented in this table allows for a more physical interpretation of the distribution of \sfr\ in the halo.}
\label{table:physical_fit}
\end{table*}

\renewcommand{\arraystretch}{1}

\section{Separating centrals from satellites}
\label{app:cent_sat}

The weights distribution separated into the contribution from central and satellite galaxies is given by
\begin{align}
w(\bm{r}) = \sum_{\h=1}^{N_h}\,\Big[W_{\h,c}\,\delta^{\mathrm{(D)}}&(\bm{r}-{\bm{r}}_{\h,c})  \nonumber\\
&+ \sum_{s=1}^{N_{\mathrm{sat},\h}} W_{\h,s}\,\delta^{\mathrm{(D)}}({\bm{r}}-{\bm{r}}_{\h,s})\Big]\,,
\end{align}
where the outer summation is over all $N_h$ haloes in the volume $V$, the inner summation is over the $N_{\mathrm{sat},\h}$ satellites of halo $\h$. The power spectrum, $P_{\rm tot}=V\langle\hat w({\bm{k}})\hat w({-\bm{k})})\rangle$, can be separated into the case where these haloes are the same, $h=h'$, and the case where they are different, $h\neq h'$, yielding the 1-halo term, $P_{1h}$, and 2-halo term, $P_{2h}$, respectively. The 1-halo term consists of a shot noise term due to centrals only, a central-satellite term, and a satellite-satellite term:
\begin{widetext}
\begin{align}\label{1h_cent_sat}
    P_{1h,w}(k)
    = \frac{1}{V}
    \underbrace{\sum^{}_{\h} W_{\h, c}^2}_{\hbox{\rm central}}\:
    +\: &\frac{1}{V}\underbrace{\Braket{\sum^{}_{\h}  W_{\h,c}\sum_{s} W_{\h,s}\Big(\exp[-i{\bm{k}} \cdot({\bm{r}}_{\h,s}-{\bm{r}}_{h,c})]\: +\: \exp[i{\bm{k}}\cdot({\bm{r}}_{\h,s}-{\bm{r}}_{h,c})] \Big)}}_{\hbox{\rm central-satellite (same halo)}}\nonumber\\
    +&\frac{1}{V}\underbrace{\left(\sum^{}_{\h}\sum_{s} W_{\h,s}^2 \quad + \quad
    \Braket{\sum^{}_{\h} \sum_{\substack{\sj,\sprime \\ \sj \neq \sprime}}\,W_{\h,\sj }W_{\h,\sprime  }\exp[-i{\bm{k}}\cdot({\bm{r}}_{\h,\sj }-{\bm{r}}_{\h,\sprime})]}\right)}_{\hbox{\rm satellite-satellite (same halo)}}.
\end{align}
Let 
\begin{align}
    \hat{u}_{\mathrm{sat},h}(k) = \frac{\sum_s \langle W_{h,s}\exp[-i{\bm{k}}\cdot({\bm{r}}_{h,s}-\bm{r}_{h,c})]\rangle}{\sum_{s} W_{h,s}}.
\end{align}
Assuming the positions and weights of satellites are uncorrelated, such that $\langle\,W_\sj \,W_\sprime \,\exp[-i{\bm{k}}\cdot ({\bm{r}}_\sj -{\bm{r}}_\sprime)]\rangle 
    = \langle\,W_\sj \,\exp[-i{\bm{k}}\cdot({\bm{r}}_\sj -\bm{r}_c)]\rangle\,
    \langle\,W_\sprime\exp[i{\bm{k}}\cdot({\bm{r}}_\sprime-{\bf r}_c)]\rangle$ when $s \neq s'$, we can substitute $\hat{u}_{\mathrm{sat},h}(k)$ into \cref{1h_cent_sat}, giving
\begin{align}\label{eq:1halo_usat}
    P_{1h,w}(k) =
    \underbrace{\frac{1}{V}\left\{
    \sum^{}_{\h}\left(
    W_{\h,c}^2+\sum_{s}W_{\h,s}^2
    \right)
    \right\}}_{\hbox{$P_{\mathrm{shot},w}^{\mathrm{gal}}$}}
    &+\frac{1}{V}
    \left\{
   \sum^{}_{\h} W_{\h,c}\sum_{s}\,W_{\h,s}\,(\usatp(k)+\usatp(-k))
    \right\}\, \nonumber\\
    &+\frac{1}{V}\left\{\sum^{}_{\h} \sum_{\substack{\sj,\sprime \\ \sj \neq \sprime}}\,W_{\h,\sj }W_{\h,\sprime}\,\usatp(k) \usatp(-k)\right\}.
\end{align}
In the limit where $k \to \infty$, such that $\usatp(k) \to 0$, this tends to $P_{\mathrm{shot},w}^{\mathrm{gal}}$.
In the limit where $k \to 0$, such that $\usatp(k) \to 1$, we have 
\begin{align}
    P_{1h,w}(k \to 0) &= \frac{1}{V}\,
    \sum_{\h} \left( W^2_{\h,c}+\sum_{s}W^2_{\h,s} 
    + 2 W_{\h,c}\sum_{s} W_{\h,s}
    + \sum_{\substack{\sj,\sprime \\ \sj \neq \sprime} } W_{\h,\sj }W_{\h,\sprime  }
    \right) \nonumber \\
    &= \frac{1}{V}\,
    \sum_{\h} \left( W_{\h,c} + \sum_{s} W_{\h,s}
    \right)^2 \nonumber\\
    &= P_{\mathrm{shot},w}^{\mathrm{halo}}.
\end{align}
%\end{widetext}

The 2-halo term similarly consists of a term due to the clustering of centrals, the clustering of centrals with satellites in another halo, and the clustering of satellites in one halo with satellites in another halo.
%\begin{widetext}
\begin{align}
    P_{2h,w}(k) &= \frac{1}{V} 
    \underbrace{\Braket{\sum_{\substack{\h,\hprime  \\ \h \neq \hprime}} W_{\h, c} W_{\hprime , c}\exp[-i{\bm{k}}\cdot({\bm{r}}_{\h, c}-{\bm{r}}_{\hprime , c})]}}_{\hbox{\rm central-central (different haloes)}} \nonumber\\
    &+\frac{1}{V} \underbrace{\Braket{\sum_{\substack{\h,\hprime  \\ \h \neq \hprime}}  W_{\h,c}\,\exp[-i{\bm{k}}\cdot({\bm{r}}_{\h,c}-{\bm{r}}_{\hprime ,c})]\sum_{\sprime \in \hprime} W_{\hprime ,\sprime}\,\exp[i{\bm{k}}\cdot({\bm{r}}_{\hprime ,\sprime}-{\bm{r}}_{\hprime ,c})]+cc}}_{\hbox{\rm central-satellite (different haloes)}}\nonumber\\
    &+ \frac{1}{V} 
    \underbrace{\Braket{\sum_{\substack{\h,\hprime \\\h \neq \hprime}} 
    \exp[i{\bm{k}}\cdot({\bm{r}}_{\hprime , c}-{\bm{r}}_{\h, c})]
    \sum_{\sj \in \h} W_{\h,\sj }\,\exp[-i{\bm{k}}\cdot({\bm{r}}_{\h,\sj }-{\bm{r}}_{\h, c})]\,\sum_{\sprime \in \hprime } W_{\hprime ,\sprime  }\,\exp[i{\bm{k}}\cdot({\bm{r}}_{\hprime ,\sprime }-{\bm{r}}_{\hprime ,c})]}}_{\hbox{\rm satellite-satellite (different haloes)}} \nonumber\\
    &= \frac{1}{V} \sum_{\substack{\h,\hprime  \\ \h \neq \hprime}}\Braket{ W_{\h, c} W_{\hprime , c}\exp[-i{\bm{k}}\cdot({\bm{r}}_{\h, c}-{\bm{r}}_{\hprime , c})]}\nonumber\\
    &+ \frac{1}{V} \sum_{\substack{\h,\hprime  \\ \h \neq \hprime}} \Braket{ W_{\h,c}\,\exp[-i{\bm{k}}\cdot({\bm{r}}_{\h,c}-{\bm{r}}_{\hprime ,c})]}\sum_{\sprime \in \hprime} W_{\hprime ,\sprime} \usatpprime(-k) \nonumber\\
    &+ \frac{1}{V} \sum_{\substack{\h,\hprime  \\ \h \neq \hprime}}  \Braket{ W_{\hprime ,c}\,\exp[-i{\bm{k}}\cdot({\bm{r}}_{\h,c}-{\bm{r}}_{\hprime ,c})]}\sum_{s \in \h} W_{\h,s} \hat{u}_{\mathrm{sat},h}(k) \nonumber\\
    &+\frac{1}{V} \sum_{\substack{\h,\hprime  \\ \h \neq \hprime}} \Braket{
    \exp[i{\bm{k}}\cdot({\bm{r}}_{\hprime , c}-{\bm{r}}_{\h, c})]}
    \sum_{\sj \in \h} W_{\h,\sj }\,\hat{u}_{\mathrm{sat},h}(k)\,\sum_{\sprime \in \hprime } W_{\hprime ,\sprime  }\,\usatpprime(-k).
\end{align}
Note that $cc$ refers to the complex conjugate.
On large scales such that $k \to 0$, $\hat{u}_{\mathrm{sat},h}(k) \to 1$, we have
\begin{align}
    P_{2h,w}(k \to 0) &= \frac{1}{V}\Braket{\sum_{\substack{\h,\hprime  \\ \h \neq \hprime}}(W_{\h, c} +\sum_{\sj \in \h } W_{\h,\sj })(W_{\hprime , c} +\sum_{\sprime \in \hprime} W_{\hprime ,\sprime  }) \exp[-i{\bm{k}}\cdot({\bm{r}}_{\h, c}-{\bm{r}}_{\hprime , c})]} \nonumber\\
    &= \frac{1}{V}\Braket{\sum_{\substack{\h,\hprime  \\ \h \neq \hprime}} W_{\h} W_{\hprime } \exp[-i{\bm{k}}\cdot({\bm{r}}_{\h, c}-{\bm{r}}_{\hprime , c})]} \nonumber\\
    &= P^{\mathrm{halo}}_{2h,w},
\end{align}
\end{widetext}
such that the power spectrum is equivalent to the 2-halo term of the halo power spectrum.
In the opposite limit of $k \to \infty$, the 2-halo term $P_{2h}\to 0$, since $P_{hh'}\to 0$ due to halo exclusion.

\section{Comparison between IllustrisTNG and EAGLE} \label{app:eagle}

\subsection{SFR fraction in central galaxy}\label{app:cent_frac_eagle}

In \cref{sec:sfr}, we plotted the average \sfr\ contributed by central and satellite galaxies as a function of the host halo mass for \tng. Here we compare this relation to that found in the \eagle\ simulation.

\begin{figure}
\centering
\includegraphics[width=\linewidth]{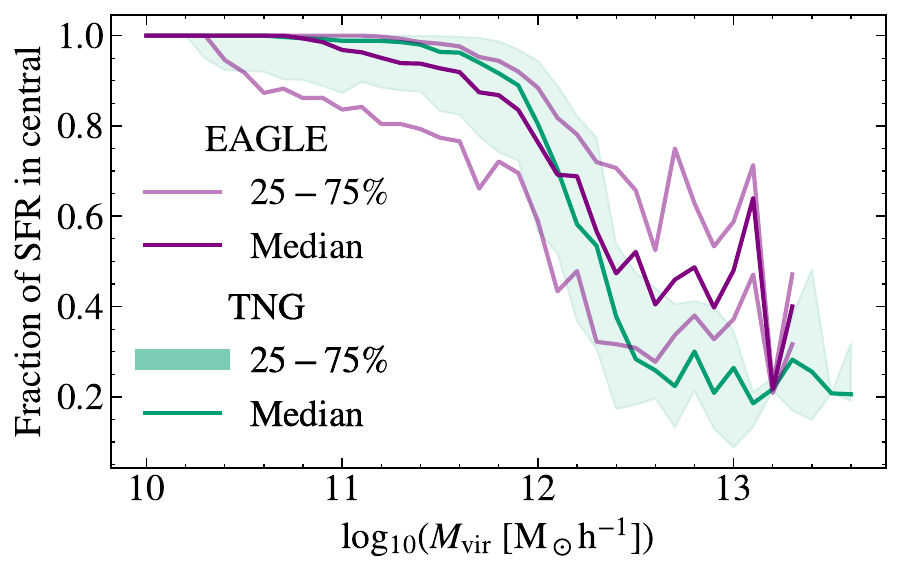}
\caption{Fraction of the halo's total \sfr\ that occurs in the central galaxy as a function of the halo's virial mass at $z \sim 1.5$. The {\em solid lines} are the median relation, with \eagle\ shown in {\em purple}, and \tng\ in {\em green}.
The {\em faint purple lines} include the 25th-75th percentiles for \eagle; the {\em green shading} is the same but for \tng. In both simulations, central galaxies dominate the halo \sfr\ in haloes with $\log M_{\mathrm{vir}} \lesssim 11.5$, but contribute only of order 50 percent to the \sfr\ for haloes with mass $\log M_{\mathrm{vir}} \gtrsim 12.5$.
In these more massive haloes, the fraction of \sfr\ in the central galaxy is higher in \eagle\ 
than in \tng. The scatter around the median relation is quite large in both cases. The transition from central-dominated \sfr\ to a larger contribution from satellites is more sudden in \tng.}
\label{fig:cent_frac_eagle}
\end{figure}

\Cref{fig:cent_frac_eagle} compares the fraction of \sfr\ in the central galaxy as a function of halo mass for TNG100-1 and \eagle\ RefL100N1504. For \eagle, the fraction also decreases with increasing halo mass, as \agn\ feedback quenches the central galaxy's \sfr, while satellite subhaloes become more massive and more numerous, thereby contributing a higher \sfr. The reduction in the central's \sfr\ contribution with increasing \mvir\ is more sudden in \tng\ than in \eagle.

\tng\ implements two modes of feedback, a thermal mode that mimics quasar feedback, and a kinetic mode that mimics radio-mode feedback. The kinetic mode is more efficient and switches on at a black hole mass of $\sim 10^8\,\mathrm{M}_\odot$, causing stronger quenching of central galaxies. The sudden increase in \agn\ feedback efficiency as the black hole passes this mass threshold may contribute to the sudden decrease in the \sfr\ of central galaxies. The \eagle\ \agn\ feedback scheme is always thermal. At higher masses, the reduction in the contribution of the central galaxy to \sfr\ is lower in \eagle\ than in \tng.
These results agree with the findings by \citet{Piotrowska_2022} (see their fig.~3). 

\subsection{Fit of 2-halo term in EAGLE}\label{app:2halo_eagle}

\begin{figure*}
\centering
\begin{subfigure}[t]{.5\textwidth}
    \centering
    \includegraphics[width=\linewidth]{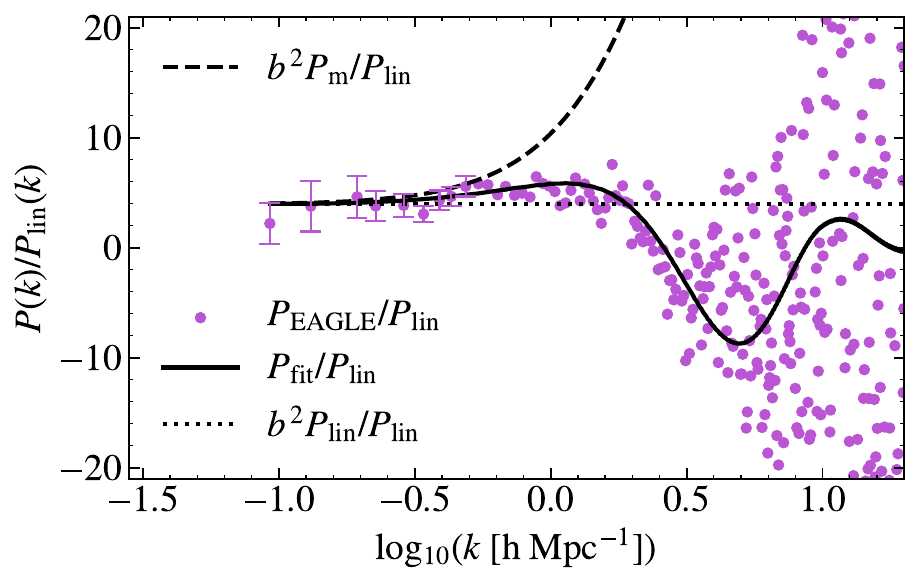}
\end{subfigure}%
\begin{subfigure}[t]{.5\textwidth}
    \centering
    \includegraphics[width=\linewidth]{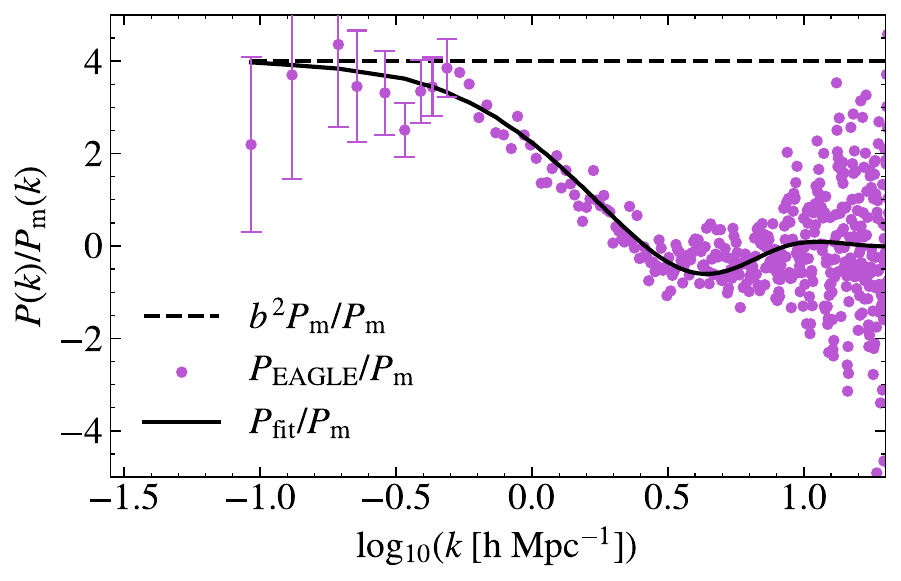}
\end{subfigure}
\caption{As Fig.~\ref{fig:bias2}, but for \eagle. The fit to the 2-halo term ({\em solid line}) developed for \tng\ works equally well for \eagle.
}
\label{fig:bias2_eagle}
\end{figure*}

\Cref{fig:bias2_eagle} is the equivalent of \cref{fig:bias2} but with the dots representing the 2-halo term measured in the \eagle\ simulation. The fit we proposed in \cref{sec:2halo_fit}, with the same parameters, also works reasonably well to reproduce the power spectrum observed in \eagle.

\ifSubfilesClassLoaded{%
  \bibliography{bibliography}%
}{}

\end{document}

\end{document}